 \documentclass[letterpaper]{JHEP3}

\usepackage{epsf,epsfig}
\usepackage{amssymb}
\usepackage{graphicx}
\usepackage{amsmath}
\usepackage{amsfonts}
 
\usepackage{graphicx}

\newcounter{fig}

\newcommand{\beq}{\begin{equation}}
\newcommand{\eeq}{\end{equation}}
\newcommand{\beqs}{\begin{eqnarray}}
\newcommand{\eeqs}{\end{eqnarray}}

\newcommand{\be}{\begin{equation}}
\newcommand{\ee}{\end{equation}}
\newcommand{\bea}{\begin{eqnarray}}
\newcommand{\eea}{\end{eqnarray}}

\numberwithin{equation}{section}

\abstract{ 
We propose a general framework
for the study of asymptotically flat black objects
with $k+1$ equal magnitude angular momenta
in $d\geq 5$ spacetime dimensions  (with $0\leq k\leq \big[\frac{d-5}{2} \big]$).
In this approach, the dependence on all angular coordinates but one is factorized, 
which leads to a codimension-two problem.
This framework can describe black holes with spherical horizon topology,
the simplest solutions corresponding to a class of Myers-Perry black holes. 
A different set of solutions describes   
balanced black objects with 
$ S^{n+1} \times S^{2k+1}$
horizon topology. 
The simplest members of this family are the black rings $(k=0)$.
The solutions with  $k>0$ are dubbed {\it black ringoids}.
Based on the nonperturbative numerical results found for several 
values of $(n,k)$,
we propose a general picture 
for the properties and the phase diagram of these solutions
and the associated black holes with spherical horizon topology:
$n=1$ black ringoids repeat the $k=0$ pattern of black rings
and Myers-Perry black holes in 5 dimensions, whereas
$n>1$ black ringoids follow the pattern of
higher dimensional black rings associated with `pinched' black holes
and Myers-Perry black holes.
}

\keywords{ black holes, numerical solutions}\preprint{ }

\title{   
Black ringoids: spinning balanced black objects in $d\geq 5$ dimensions --
the codimension-two case  
} 
  
 \author{
 {\large Burkhard Kleihaus}$^{\dagger}$, {\large Jutta Kunz}$^{\dagger}$  
  and 
 {\large Eugen Radu}$^{\ddagger}$
\\ 
\\
$^{\dagger}$
{\small Institut f\"ur Physik, Universit\"at Oldenburg, Postfach 2503
D-26111 Oldenburg, Germany} 
\\
$^{\ddagger}$   
{\small  Departamento de F\'\i sica da Universidade de Aveiro and I3N, 
  Campus de Santiago, 3810-183 Aveiro, Portugal}
 
 }
 \begin{document}

\section{ Introduction}

The physics of the black hole event horizon 
has proven a fruitful field of research in gravitational physics.  
Following Hawking's black hole topology theorem
\cite{Hawking:1973uf},
for many decades the focus was on 
asymptotically flat black holes in four dimensions 
with horizons of spherical topology.
Also 
the Tangerlini \cite{Tangherlini:1963bw}  
and the Myers-Perry (MP) \cite{Myers:1986un}
black hole solutions,
which provide natural higher dimensional
generalizations of the  $d=4$
Schwarzschild and Kerr solutions, respectively,
possess horizons of spherical topology.
Nevertheless, already Myers and Perry argued
that black rings with a horizon topology $S^2\times S^1$
should exist \cite{Myers:1986un},
making Emparan and Reall's discovery  
of the black ring (BR) in $d=5$ spacetime dimensions 
\cite{Emparan:2001wn,Emparan:2001wk}
a celebrated and long awaited result. 
 
The discovery of the BRs
made clear that a number of well known results in $d=4$ gravity
do not have a simple extension  to higher dimensions.  
For example, the uniqueness of vacuum black holes is violated in $d=5$,
since three distinct solutions may exist for the same 
global charges (two BRs and a MP black hole).
The rapid progress following the discovery in 
\cite{Emparan:2001wn,Emparan:2001wk}
 provided a rather extensive picture of the solutions'
landscape for the five dimensional case,
with a large variety of physically interesting solutions 
(for a review, see  
\cite{Emparan:2008eg},
\cite{maeda},
\cite{horowitz},
\cite{Reall:2012it}).

However, despite the presence of several partial results in the literature, 
the $d > 5$ case has remained largely unexplored. 
At the same time, 
there is overwhelming evidence 
that as the dimension increases, the phase structure
of the solutions
becomes increasingly intricate and diverse, 
with a variety of other horizon topologies apart from the spherical  one \cite{Emparan:2008eg}.
The main obstacle stopping the progress in this field seems to be the absence of
closed form solutions (apart from the MP black holes), 
since the Weyl formalism and various solution generation techniques
(which were very useful in $d=4,5$) do not apply for the $d>5$  asymptotically flat case.
 
Most of our knowledge in this area is based on results
 found by using the method of matched asymptotic expansions 
\cite{Emparan:2007wm},
\cite{Emparan:2009cs}, 
\cite{Emparan:2009at},
\cite{Emparan:2009vd}.
Here the central assumption  
 is that some black objects
can be approximated by a certain very thin
black brane curved into a given shape.
In a remarkable development,
this has led to the development of the {\it blackfold effective worldvolume theory}.
This theory provides
a general formalism leading to quantitative predictions for the behaviour
of various $d>4$ general relativity solutions in the ultraspinning regime\footnote{   
This approach 
is an extension of the theory of classical brane dynamics originally developed by
Carter to provide an effective description of some field theory solitons in flat space
(see $e.g.$ the recent
review \cite{Carter:2011ab}).
}. 
In this way, it was possible to achieve a partial description of 
a plethora of higher dimensional black objects with various event horizon topologies.

However, this theory has some clear limitations; for
example it is supposed to work only if the length scales involved are
widely separated.
Also, the blackfold approximation cannot say anything about
 the issue of the limiting behaviour of the black objects with
 a nonspherical horizon topology, which is supposed to occur in the 
region of relatively small angular momenta.
Moreover,  black holes without a black membrane limiting behavior 
cannot be described by the blackfold approach 
\cite{Emparan:2009vd}.
 
Therefore the construction of higher dimensional black objects
 with a non-spherical horizon topology
 within a nonperturbative approach 
 remains a pertinent task.
 In the absence of exact solutions,
this task has been approached recently  
by employing numerical methods,
see $e.g.$
the work in 
\cite{Kudoh:2006xd},
\cite{Kleihaus:2009wh},
\cite{Kleihaus:2010pr},
\cite{Kleihaus:2012xh},
\cite{Kleihaus:2013zpa},
\cite{Dias:2014cia}.
Such an approach can be considered as complementary to
the analytical one in
\cite{Emparan:2007wm}-\cite{Emparan:2009vd}.
For example, the numerical results
may provide evidence for the existence 
of the solutions beyond the various approximations 
employed in the blackfold effective worldvolume theory.
  At the same time, the analytical predictions there 
can be used to cross-check the numerical
results in some region of the parameter space.

For example, the work 
\cite{Kleihaus:2012xh},
\cite{Dias:2014cia}
   has given numerical evidence for the
existence of balanced spinning vacuum BRs in 
$d>5$ dimensions 
 and analyzed their basic properties.
The results there show that the analytical results from the blackfold
approximation work very well for thin BRs.
However, a rather complicated picture, 
which cannot be captured within the blackfold formalism,
is found for `fat' BRs.
There a different class of solutions starts playing a role -- {\it the `pinched' black holes}.
Their existence results from the fact that the ultraspinning MP black
holes exhibit a Gregory-Laflamme-type of instability \cite{Dias:2009iu,Dias:2010maa}. 
The `pinched' black holes (which are not yet known in closed form)
connect the MP solutions
with the branch of `fat' BRs, via
a topology changing merger solution \cite{Dias:2014cia}.

However, apart from  
the BRs,
 relatively little is known about the nonperturbative 
 behaviour of
 other $d>5$ solutions
 with a non-spherical horizon topology.
Solutions with an 
$S^2\times S^{d-4}$
horizon topology have been studied in 
 \cite{Kleihaus:2009wh},
\cite{Kleihaus:2010pr}.
However, these solutions are static,
and supported against collapse by conical singularities.

The main purpose of this paper is to present a general 
 nonperturbative 
framework capable to describe  
a class of
balanced black object  
 with  $S^{n+1}\times S^{2k+1}$  horizon topology,
 in 
$d\geq 5$ spacetime dimensions,
 \begin{eqnarray}
d=2k+n+4,~~~{\rm with}~~~~n\geq 1.
\end{eqnarray}
In this case,
the rotation provides a centrifugal repulsion that allows 
regular solutions to exist.
The study here is restricted to the special case 
of $k+1$ equal magnitude angular momenta,
with
\begin{eqnarray}
0\leq k\leq \big[ \frac{d-5}{2}\big],
\end{eqnarray}
an assumption which leads to a treatable codimension-2 numerical problem.

For $k=0$, the framework proposed here reduces to that used in  \cite{Kleihaus:2012xh}
to construct higher dimensional BR solutions.
One of the purposes of this work is to 
present a more detailed discussion of the BRs in \cite{Kleihaus:2012xh},
together with the properties of the coordinate system introduced there.
Apart from that, we shall consider $d>6$ solutions with
$k>0$, which are dubbed {\it black ringoids}.
Numerical results are reported for the simplest case $d=7,~k=1$.

However, apart from these black objects with a non-spherical
horizon topology,
the proposed framework can describe also 
a class of MP black holes, whose properties
we review in this work\footnote{Moreover, the `pinched' black holes (which also possess a horizon of spherical topology)
can be studied as well within the proposed framework, although we do not consider them here.}. 
These MP black holes can also be characterized by the 
integers $n$ and $k$, associated to the non-rotating and rotating  
parts of the metric.
In particular, they possess $k+1$ equal magnitude angular momenta,
which is less than the maximally possible number 
$N= \left[ \frac{d-1}{2} \right]$ for MP black holes.

When compiling the results for these two different horizon topologies,
we are led to conjecture
that the basic properties 
of the $d>5$ BRs still hold for $n>1$
black ringoids, in particular, for their behaviour in the 
nonperturbative region, not covered by the
blackfold approach.
However, we suggest that the solutions with  $n=1$, $i.e.$ black ringoids with
$S^2\times S^{d-4}$ horizon topology are special,
since they share the
basic 
 properties of 
the $d=5$ BRs.
This behaviour is related to that of the corresponding MP black holes,
which possess an ultraspinning regime for $n>1$ only. 

This paper is organized as follows:
in the next  Section we present a  discussion of the coordinate system used to impose
a non-spherical topology of the event horizon.
The general framework is introduced in Section  3.
In Section 4 we review the basic properties of the known exact solutions:
the $d\geq 5$ MP black holes
with $k+1$
equal magnitude angular momenta and the $d=5$
Emparan-Reall BRs.
We continue with Section 5, where we exhibit the numerical 
results for several values of $(d,k)$.  
We give our conclusions and remarks in the final Section.
The Appendix A contains
an approximate form of the solutions
on the boundaries of the domain of integration.
The expression of the $d=5$
balanced BR in the coordinate system introduced in this work
is given in Appendix B.

\section{ A special coordinate system}

All solutions in this work approach at infinity
the Minkowski spacetime background in $d=D+1$ dimensions,
with a line element
\begin{eqnarray}
\label{Mink}
ds^2=-dt^2+d\sigma_D^2,~~{\rm where}~~
d\sigma_D^2=d\rho^2+ \rho^2 d\Omega_{d-2}^2,
\end{eqnarray}
and a parametrization of the $(d-2)$-dimensional sphere
\begin{eqnarray}
\label{sphere}
d\Omega_{d-2}^2=d\Theta^2+ \cos ^2 \Theta d\Omega_n^2+\sin^2\Theta  d\Omega_{p}^2,~~~
{\rm with}~~D=n+p+2.
\end{eqnarray}

In the above relations, $\rho$ and $t$ are a radial
and a time coordinate, respectively, while $\Theta$ is an angular
coordinate, with 
$0\leq \Theta \leq \pi/2$.
Also, $d\Omega_n^2$ is the metric on the $n$-dimensional sphere.
For $d=5$, these are the usual bi-azimuthal coordinates, with $n=p=1$
and
\begin{eqnarray}
\label{omega4}
d\sigma_4^2=d\rho^2+\rho^2(d\Theta^2+ \cos ^2 \Theta d\phi^2+\sin^2\Theta  d\psi^2),
 ~~{\rm with}~~0\leq (\phi,\psi)<2 \pi.
\end{eqnarray}

The numerical  scheme used in this work requires 
a rectangular boundary for the coordinates, 
such that both the event horizon and the spacelike infinity are located at a constant value of 
one of the coordinates.
As we shall see in Section 4, this is possible for MP black holes,
where a surface of constant radial coordinate in a general line element based on (\ref{Mink}) is 
topologically  a sphere.

In what follows, we show the existence of a parametrization of the flat space
 with the property that
(\ref{Mink}) is approached only asymptotically,
while a surface of constant (new) radial coordinate possesses, for some of its range,
a $S^{n+1}\times S^{p}$ topology\footnote{It is interesting to notice the formal analogy with the
Kaluza-Klein caged black holes in $d$-dimensions.
In some sense, those solutions are the opposite of the BRs,
possessing a spherical horizon topology,
and approaching,
however, a background which is
the product of the Minkowski
 spacetime with a circle.  
 The numerical problem of constructing 
 solutions with this behaviour has been solved 
 in \cite{Kudoh:2003ki} by using a special coordinate system
 in the spirit of the one introduced in Section (\ref{sec-s1}) 
(see also \cite{Harmark:2002tr}).
 For the coordinate system in \cite{Kudoh:2003ki},
  a surface of constant radial coordinate has the topology $S^{d-2}$
 close to the horizon and $S^{d-3}\times S^1$ in the asymptotic region.
 }.

\subsection{The new coordinates in $D=4$}
\label{sec-s1}

The coordinates usually used in the study of $d=5$
BRs naturally occur
when considering a foliation of the $D=4$ flat space
in terms of the equipotential surfaces of a two form potential 
sourced by a ring \cite{Emparan:2006mm}.
In these coordinates, the flat space metric reads
\begin{eqnarray}
\label{ring-coords}
 ds^2=\frac{R^2}{(x-y)^2}
 \left[
 \frac{dx^2}{1-x^2}
 +\frac{dy^2}{y^2-1}
 +(1-x^2)d\phi^2
 +(y^2-1)d\psi^2
 \right],
\end{eqnarray} 
with $R>0$ an arbitrary parameter and  
\begin{eqnarray}
\label{ring1}
 -\infty<y<-1,~~-1\leq x\leq 1.
\end{eqnarray}  
Although these coordinates are physically rather opaque,
they result in a simple and compact form of the $d=5$
BR solution.
However, in a numerical approach, their disadvantage is that
 the asymptotic infinity is approached
at a single point, $x\to-1,~y \to -1$.
Therefore, the imposition of the boundary conditions and the extraction of 
the mass and the angular momenta of the solutions 
is problematic, at least for the scheme used in this work,
and represents an obstacle which we could not overcome so far.

We solve this problem by working with a different coordinate system,
with a foliation of the flat space in terms of equipotential
surfaces of a scalar field $\Psi$ solving the Laplace equation
\begin{eqnarray}
\label{eq1}
\nabla^2 \Psi=0,
\end{eqnarray}
outside of a ring source at $\rho=R>0$, $\Theta=0$.
The corresponding solution reads
\begin{eqnarray}
\label{eq2}
 \Psi(\rho,\Theta)=\frac{1}{\sqrt{(R^2+\rho^2)^2-4 R^2\rho^2 \cos^2\Theta}}~.
\end{eqnarray}
\newpage
\setlength{\unitlength}{1cm}
\begin{picture}(8,7.5)
\put(1,0.0){\epsfig{file=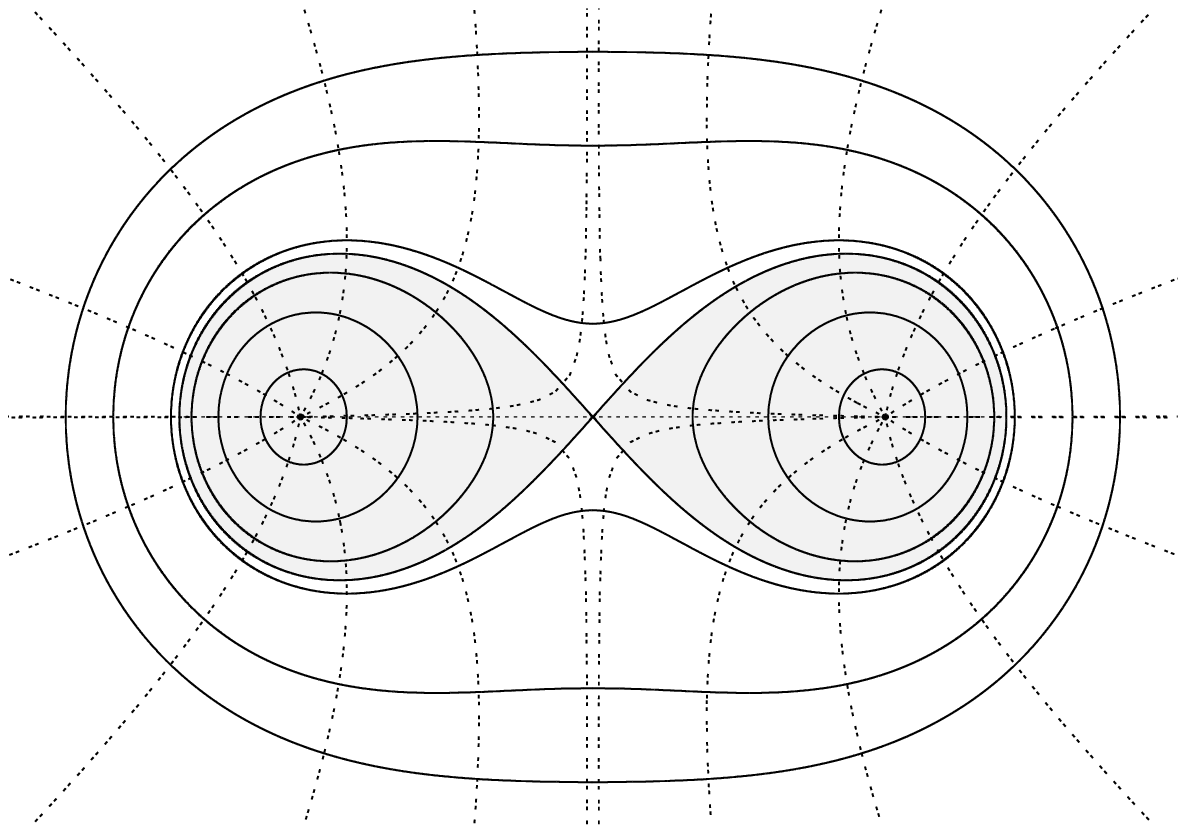,width=12cm}}
\end{picture}
\\
\\
{\small {\bf Figure 1.}
The  new coordinates for the $D=4$
flat space metric on a section at constant $\phi$ and $\psi$
(and  $\phi+\pi$ and $\psi+\pi$).
The solid lines shown here have constant values of $r$,
while the dotted lines have constant $\theta$.
Also, the shaded gray region covers the domain $r<R$.
}
\vspace{0.5cm}

Then, following the corresponding approach  in  \cite{Emparan:2006mm} for 
a two form potential,
we introduce the new coordinates
($r,\theta$)
that correspond to surfaces of constant $\Psi$
and their gradients  surfaces.

The coordinate transformation between $(\rho,\Theta)$ in (\ref{omega4}) and $(r,\theta)$ is
\begin{eqnarray}
\label{transf}
\rho=r\sqrt{U(r,\theta)},~~\tan \Theta=(\frac{r^2+\rho^2+R^2}{r^2+\rho^2-R^2})\tan \theta ,
\end{eqnarray}
the relation with the   usual ring coordinates $(x,y)$
being
\begin{eqnarray}
\label{transf-ring}
x=\frac{R^2}{r^2}-U(r,\theta),~~y=-\frac{R^2}{r^2}-U(r,\theta).
\end{eqnarray}
In the above relations we note
 \begin{eqnarray}
\label{U}
U(r,\theta)=\sqrt{1+\frac{R^4}{r^4}-\frac{2R^2}{r^2}\cos2\theta}.
\end{eqnarray} 
The coordinate range here is $0\leq r<\infty$, $0\leq \theta \leq \pi/2$.
 
A straightforward computation leads to the following
 expression of the $D=4$ flat space line element as written in $(r,\theta)$-coordinates 
\begin{eqnarray}
\label{ER-solution}
 d\sigma_4^2=F_1(r,\theta)(dr^2+r^2 d\theta^2)+F_2(r,\theta) d\psi^2+F_3(r,\theta) d\phi^2,
\end{eqnarray} 
where
\begin{eqnarray}
\label{Fi} 
&F_1(r,\theta)=\frac{1}{U },~
F_2(r,\theta)=r^2\left(\cos^2\theta -\frac{1}{2}(1+\frac{R^2}{r^2}-U )\right),~
F_3(r,\theta)=r^2\left(\sin^2\theta -\frac{1}{2}(1-\frac{R^2}{r^2}-U )\right),~~~{~~~}
\end{eqnarray} 
with $U$ given by (\ref{U}). 

Since $r$ runs from zero to infinity, one can think of it as a sort of radial coordinate.
As $r\to 0$, the behaviour of the metric functions is
\begin{eqnarray}
\label{r0}
F_1=\frac{r^2}{R^2}+O(r^4),~~F_2=\frac{r^4}{R^2}\sin^2\theta \cos^2\theta+O(r^6),~~
F_3=R^2+O(r^2).
\end{eqnarray}
By defining $r=\sqrt{2R \bar r}$, $\theta=\bar \theta/2$, one can show that $r=0$ is a regular origin,
with
\begin{eqnarray}
\label{le-r0}
ds^2=d\bar r^2+ \bar r^2 d\bar \theta^2+\bar r^2 \sin^2 \bar\theta d\psi^2+R^2 d\phi^2,
\end{eqnarray}
in the vicinity of that point.

In fact, one can see that for $0<r<R$,
a surface of constant $r$ has a $S^2\times S^1$ topology,
where the $S^2$ is parametrized by $(\theta,\psi)$
and the $S^1$ by $\phi$.
For $r>R$, one recovers the usual $S^3$ topology of an $r=const.$ foliation.
 $r=R$, $\theta=0$ is a special point with a coordinate system singularity.
These features are shown in Fig. 1, where we present a section at constant $\phi$ and $\psi$.
(Note that,  for greater clarity,
  the antipodal sections at  $\phi+\pi$ and $\psi+\pi$ are also shown there.)

It is also useful to consider the expansion of the functions $F_i$
at $\theta=0,\pi/2$. Starting with $\theta=0$, one finds
 \begin{eqnarray}
\label{t01i}
\nonumber
F_1=\frac{r^2}{R^2-r^2}+O(\theta^2),~~F_2=\frac{r^4}{R^2-r^2}\theta^2+O(\theta^4),
~~
F_3=R^2-r^2+O(\theta^2),
\end{eqnarray}
 for $r<R$, and 
  \begin{eqnarray}
\label{t02i}
\nonumber
F_1=\frac{r^2}{r^2-R^2}+O(\theta^2),~~F_2=r^2-R^2+O(\theta^2),
~~
F_3=\frac{r^4}{R^2-r^2}\theta^2+O(\theta^4),
\end{eqnarray}
 for $r>R$.
The corresponding expansion for $\theta=\pi/2$ is
  \begin{eqnarray}
\label{tpi2i}
\nonumber
F_1=\frac{r^2}{r^2+R^2}+O(\theta-\frac{\pi}{2})^2,~~F_2=\frac{r^4}{r^2+R^2}(\theta-\frac{\pi}{2})^2+O(\theta^4),
~~
F_3=(r^2+R^2)+O(\theta-\frac{\pi}{2})^2.
\end{eqnarray}
For completeness, we give also the asymptotic form of the functions valid for large $r$
 \begin{eqnarray}
\label{inf}
\nonumber
 F_1= 1+\frac{R^2}{r^2}\cos2\theta+O(1/r^4),~~F_2=r^2\cos^2\theta (1-\frac{R^2}{r^2})+O(1/r^2),
~~
F_3=r^2\sin^2\theta (1+\frac{R^2}{r^2})+O(1/r^2),
\end{eqnarray}
such that asymptotically $(r,\theta)$
correspond to the usual bi-azimuthal coordinates.

\subsection{The $D>4$ case and the issue of the metric ansatz  }
The above coordinates generalize straightforwardly 
to $D>4$ dimensions\footnote{Note that the interpretation of the $(r,\theta)$ coordinates as corresponding to
equipotential surfaces of a
scalar field is lost for $D>4$. 
Although one can devise such a coordinate system,
the resulting expressions are too complicated to use in practice.}.
By using the same transformation (\ref{transf}),
the flat space line element $d\sigma_D^2$ in (\ref{Mink}), (\ref{sphere})
becomes
\begin{eqnarray}
\label{new-form}
 ds^2=F_1(r,\theta)(dr^2+r^2 d\theta^2)+F_2(r,\theta) d\Omega_n^2+F_3(r,\theta)  d\Omega_p^2,
\end{eqnarray} 
with the same expression (\ref{Fi}) for the $F_i$
and the same coordinate range for $(r,\theta)$. 
Then, for $0<r<R$,
a surface of constant $r$ has a $S^{n+1}\times S^p$ topology,
while for $r>R$,  an $r=const.$ surface is topologically a sphere.

It is now obvious that this parametrization of flat space
can be used  to describe black objects with a non-spherical horizon topology.
The corresponding line element in $d=D+1$ dimensions
should preserve the basic structure of (\ref{new-form}) ($e.g.$ the behaviour at $\theta=0,\pi/2$),
containing, however, additional
terms that encode the gravity effects.   
The event horizon will be located at a constant (positive) value of $r<R$,
and so the black objects will inherit the  $S^{n+1}\times S^p$ topology.
For values larger than $R$, the coordinate $r$
would  correspond to the usual radial coordinate.

However, the metric ansatz should also be general enough to allow for rotation\footnote{
Unfortunately, the only way to achieve balance for a non-spherical
horizon topology seems to be to rotate the solutions.
To our knowledge,  no other mechanism
is known at this moment.
For example, the results in \cite{Kleihaus:2009dm} show that the Gauss-Bonnet corrections
to Einstein gravity cannot eliminate the conical singularity of a $d = 5$ static BR.
A similar result is likely to hold also for the Einstein-Gauss-Bonnet 
generalizations of the higher dimensional configurations discussed
in this work.
}.
Then the centrifugal force would  
prevent the collapse of such black objects with a non-spherical horizon topology,
and balance them.
A generic metric ansatz based on (\ref{new-form})
which describes a rotating spacetime, would contain metric functions
 with a nontrivial dependence of
 at least one more coordinate apart from $r,\theta$.
 However, this is a very hard numerical problem
 which we have not yet solved.

However, the problem is greatly simplified
 for the special case
\begin{eqnarray}
p=2k+1,~~~{\rm with}~~k\geq 0,
\end{eqnarray}
by assuming that all angular momenta on the $S^p$
have equal magnitude.  
At the same time, all other possible angular momenta vanish.
(We recall that in $d$ spacetime dimensions, there are 
$N=\big[\frac{d-1}{2}\big]$ independent angular momenta.)
This would factorize the dependence of the coordinates
on $S^p$,
leading to a 
cohomogeneity-2 ansatz, 
the resulting equations of motion forming a set of coupled 
nonlinear PDEs in terms of $(r,\theta)$ only.

The inclusion of rotation on the $S^{2k+1}$ is based on
the simple observation that one can always 
write the metric of an odd-dimensional 
(round) sphere
as an $S^1$ fibration over the complex projective
space $\mathbb C \mathbb P^k$,
\begin{eqnarray}
\label{s1}
d\Omega_{2k+1}^2=(d\psi+{\cal A})^2+d\Sigma_k^2,
\end{eqnarray}
where $d\Sigma_k^2$
is the metric on the unit $\mathbb C \mathbb P^k$ space and
${\cal A}=A_i dx^i $ is its K\"ahler form. 
 The fibre is parameterized by
the coordinate $\psi$, which has period $2\pi$.

A simple explicit form for (\ref{s1})
is found by introducing
$k+1$
complex coordinates $z_i$ (with $\displaystyle\sum_{i}^{k+1} z_i  \bar{z}_i=1$),
 such that
$d\Omega_{2k+1}^2=\displaystyle\sum_{i}dz_id\bar{z}_i$.
A simple expression of $z_i$ is (see $e.g.$  \cite{Stotyn:2013yka}):
\begin{eqnarray}
z_i=   e^{i(\psi+\phi_i)}\cos\theta_i\displaystyle\prod_{j<i}\sin\theta_j,
~~
{\rm for}~~i=1,\dots,k,~~ 
{\rm and}~~~
z_{k+1}= e^{i\psi} \prod_{j=1}^{\frac{n-1}{2}}\sin\theta_j.
\end{eqnarray}
(Note that the coordinates $\phi_i$ have period $2\pi$ while the $\theta_i$ have period $\pi/2$.)
%
The corresponding expression of the  K\"ahler form ${\cal A}$ is
\begin{eqnarray}
{\cal A}=A_i dx^i=\sum_{i=1}^{k}{\cos^2\theta_i\left[\prod_{j<i}\sin^2\theta_j\right]d\phi_i}~.
\end{eqnarray}

In this approach\footnote{Note that a similar approach has been used 
in the literature to numerically construct $d\geq 5$ spinning black holes with a spherical horizon
topology,
for various theories
where an exact solution is missing, see $e.g.$ 
\cite{Kunz:2006eh},
\cite{Kunz:2006yp},
\cite{Brihaye:2010wx},
\cite{Stotyn:2013yka}
(as well as in the perturbative construction of exact solutions
\cite{Allahverdizadeh:2010xx},\cite{Allahverdizadeh:2010fn}).
Note   that in all these cases  it was possible to reduce the problem
to solving a set of ordinary differential equations.
However, a non-spherical horizon topology prevents this possibility.
}, the rotation will be introduced by adding an extra term $W dt$
to the form $d\psi+{\cal A}$.
Also, the rotation will deform the sphere $S^{2k+1}$,
with different factors for the two parts in (\ref{s1}).

\section{A general framework}

\subsection{The line element and special cases}

The above considerations lead to the following metric ansatz: 
 \begin{eqnarray}
   \label{metric} 
 &&
 ds^2=f_1(r,\theta)\left(dr^2+\Delta(r) d\theta^2 \right)
 +f_2(r,\theta) d\Omega_{n}^2  -f_0(r,\theta) dt^2
  \\
  \nonumber
  &&{~~~~~~~~~~~~~~~~~~~~~~~~~~~~}
  +f_3(r,\theta) \big(d\psi+{\cal A}-W(r,\theta) dt\big)^2+f_4(r,\theta) d\Sigma_{k}^2~,
\end{eqnarray} 
which can be used to describe a class of black objects  with a $S^{n+1}\times S^{2k+1}$ horizon topology.
However, as we shall see in the next Section,
the MP black holes
with $k+1$ equal angular momenta
can also be written in the above form.

In our approach,   the information on the solutions is encoded in the 
unknown functions $(f_i, W)$, ($i=0,\dots 4)$.
Note that  the dependence of the coordinates on the $S^{2k+1}$
factorizes, such that the problem is effectively codimension-2.
Also,  $\Delta(r)$ is a given
`background' function which is chosen for convenience
by using the residual metric gauge freedom.
In the numerical study of the solutions with non-spherical horizon topology, we set
\begin{eqnarray}
   \label{delta}
\Delta(r)=r^2 ,
\end{eqnarray}
without any loss of generality. 
However, as we shall see, the MP black holes take a 
simple form for a different choice of $\Delta(r)$.

\newpage
\vspace{0.5cm}
\begin{table}[t!]
\centering
\begin{tabular}{|c|c|c|c|c|c|} 

\hline
 $~$ & ${\it spherical~horizon}$ & {\it black~rings} & \multicolumn{3}{c|}{{\it black~ringoids}  } 
\\
\hline
$~$ & MP/`pinched'
  &  $k=0$ &  $k=1$ &  $k=2$ &  $k=3$
\\
\hline
\hline
$d=5$   & $S^3$ &  ${\mathbf S^2}\times {\mathbf S^1}$ & $~$  & $~$ & $~$  
\\ 
$d=6$   & $S^4$ & $S^3\times S^1$ & $~$  & $~$ & $~$  
\\ 
$d=7$   & $S^5$ & $S^4\times S^1$ & ${\mathbf S^2}\times {\mathbf S^3}$  & $~$ & $~$   
\\ 
$d=8$   & $S^6$ & $S^5\times S^1$ & $S^3\times S^3$  & $~$ & $~$       
\\ 
$d=9$   & $S^7$ & $S^6\times S^1$ & $S^4\times S^3$  & ${\mathbf S^2}\times {\mathbf S^5}$   & $~$    
\\ 
$d=10$  & $S^8$ & $S^7\times S^1$ & $S^5\times S^3$  & $S^3\times S^5$   & $~$
\\ 
$d=11$  & $S^9$ & $S^8\times S^1$ & $S^6\times S^3$  & $S^4\times S^5$ & ${\mathbf S^2}\times {\mathbf S^7}$     
\\ 
 \hline
\end{tabular}
\label{table1}
\end{table}
{\small {\bf Table 1.} A list of horizon topologies for spinning balanced 
black objects which can be described by the metric ansatz (\ref{metric}).}
\vspace{0.5cm}

The range of the radial coordinate is $r_H\leq r<\infty$, 
and $r=r_H>0$ corresponds to the event horizon,  
where $f_0(r_H,\theta)=0$.
Also, the angular coordinate $\theta$ has the usual range, 
$0\leq \theta \leq \pi/2$.
 Thus the domain of integration has a rectangular shape,
and is well suited for numerical 
calculations.

The case $k=0$ is special, since the  $d\Sigma_{k}^2$ term
is absent in this case (also ${\cal A}=0$),
with a line element
\begin{eqnarray}
   \label{m-s1}
ds^2=f_1(r,\theta)\left(dr^2+\Delta(r) d\theta^2 \right)
 +f_2(r,\theta) d\Omega_{d-3}^2
 +f_3(r,\theta) \big(d\psi-W(r,\theta) dt\big)^2 -f_0(r,\theta) dt^2,~~{~~}
\end{eqnarray}
describing black objects with $S^{d-3}\times S^1$
topology of the event horizon ($i.e.$ the BRs),
 as well as MP black holes rotating in a single plane.
(Note that the `pinched' black holes in \cite{Dias:2014cia}
can also be studied within this ansatz.)
The corresponding relations are found by taking formally 
$k=0,~f_4=1$ in all general equations exhibited below.

Another case of interest
 is $n=1$,
  with a line element
  \begin{eqnarray}
   \label{metric-m1} 
 &&
 ds^2=f_1(r,\theta)\left(dr^2+\Delta(r) d\theta^2 \right)
 +f_2(r,\theta) d\phi^2  -f_0(r,\theta) dt^2
  \\
  \nonumber
  &&{~~~~~~~~~~~~~~~~~~~~~~~~~~~~}
  +f_3(r,\theta) \big(d\psi+{\cal A}-W(r,\theta) dt\big)^2+f_4(r,\theta) d\Sigma_{k}^2~,
\end{eqnarray}
describing black objects with a $S^{2}\times S^{d-4}$
topology of the event horizon in 
 $d=2k+5$ dimensions. (Therefore, the
$d=5$ line-element (\ref{m-s1}) is the first member of this family.)
As we shall see, the properties of the solutions are special in this case\footnote{The static limit
of (\ref{metric-m1}) has $f_3=f_4$, $W=0$,
and exists for any $d\geq 5$.
The properties of the  static un-balanced black objects 
with
$S^{2}\times S^{d-4}$
topology of the event horizon 
are discussed in \cite{Kleihaus:2009wh}. }.


Finally, in Table 1 we give 
a list of possible horizon topologies which can be studied within this framework,
 for $5\leq d \leq 11$ (the special case $n=1$ is highlighted there).
 
\subsection{The equations }

A suitable combination of the Einstein equations 
$G_r^r+G_\theta^\theta=0$, 
$G_\Omega^\Omega=0$,
$G_\psi^\psi=0$,
$G_\Sigma^\Sigma=0$,
$G_\psi^t=0$,
$G_t^t=0$ (with $G_{\mu}^\nu$ the Einstein tensor),
yield for the functions $(f_i,~W)$ the following set of equations:
\begin{eqnarray}
\label{eqf0}
&&
\nabla^2f_0-\frac{1}{2f_0}(\nabla f_0)^2
+\frac{(d-2k-4)}{2f_2}(\nabla f_0)\cdot  (\nabla f_2)
+\frac{1}{2f_3}(\nabla f_0)\cdot  (\nabla f_3)
\\
&&
\nonumber
{~~~~~~~~~~~}
-f_3 (\nabla W)^2+\frac{k}{f_4}(\nabla f_0)\cdot  (\nabla f_4)=0.
\end{eqnarray}
\begin{eqnarray}
&&
\nonumber
\nabla^2f_1
-\frac{1}{ f_1}(\nabla f_1)^2
-\frac{(d-2k-4)f_1}{2f_0f_2}(\nabla f_0)\cdot  (\nabla f_2)
-\frac{(d-2k-4)(d-2k-5)f_1}{4f_2^2}(\nabla f_2)^2{~~~~~~}
\\
\nonumber
&&
-\frac{f_1}{2f_0f_3}(\nabla f_0)\cdot  (\nabla f_3)
-\frac{(d-2k-4)f_1}{2f_2f_3}(\nabla f_2)\cdot  (\nabla f_3)
-\frac{f_1f_3}{2f_0}(\nabla W)^2
-f_1\left(\frac{\Delta'^2}{2\Delta^2}-\frac{\Delta''}{\Delta}   \right)
\\
\nonumber
&&
+\frac{(d-2k-4)(d-2k-5)f_1^2}{f_2} 
+k
\bigg(
-\frac{f_1}{f_0f_4}(\nabla f_0)\cdot  (\nabla f_4)
-\frac{(d-2k-4)f_1}{f_2f_4}(\nabla f_2)\cdot  (\nabla f_4)
\\
\label{eqf1}
&&
-\frac{ f_1}{f_3f_4}(\nabla f_3)\cdot  (\nabla f_4)
-\frac{(2k-1)f_1}{2f_2^2}(\nabla f_4)^2
+\frac{2f_1^2}{f_4}(2(k+1)-\frac{f_3}{f_4}) 
\bigg)=0,
 \end{eqnarray}
\begin{eqnarray}
\nonumber
&&
\nabla^2f_2
+\frac{1}{2f_0}(\nabla f_2)\cdot  (\nabla f_0)
+(d-2k-6)\frac{1}{2f_2}(\nabla f_2)^2
+\frac{1}{2f_3}(\nabla f_2)\cdot  (\nabla f_3)
\\
&&
\label{eqf2}
+\frac{k}{f_4}(\nabla f_2)\cdot  (\nabla f_4)
-2(d-2k-5)f_1=0,
\end{eqnarray}
\begin{eqnarray}
\nonumber
&&
\nabla^2f_3
+\frac{1}{2f_0}(\nabla f_3)\cdot  (\nabla f_0)
+(d-2k-4)\frac{1}{2f_2}(\nabla f_2)\cdot  (\nabla f_3)
-\frac{1}{2f_3}(\nabla f_3)^2
\\
&&
\label{eqf3}
+\frac{f_3^2}{ f_0}(\nabla W)^2
-\frac{4kf_1f_3^2}{f_4^2}
+k\frac{1}{f_4}(\nabla f_3)\cdot  (\nabla f_4)=0,
 \end{eqnarray}
\begin{eqnarray}
\nonumber
&&
\nabla^2f_4
+\frac{1}{2f_0}(\nabla f_4)\cdot  (\nabla f_0)
+(d-2k-4)\frac{1}{2f_2}(\nabla f_2)\cdot  (\nabla f_4)
+\frac{1}{2f_3}(\nabla f_3)\cdot  (\nabla f_4)
\\
\label{eqf4}
&&
+\frac{(k-1)}{f_4}(\nabla f_4)^2
-4(k+1) f_1+\frac{4f_1f_3}{f_4}=0,
 \end{eqnarray}
\begin{eqnarray}
\label{eqW}
&&
\nabla^2 W
-\frac{1}{2f_0}(\nabla W)\cdot  (\nabla f_0)
+(d-2k-4)\frac{1}{2f_2}(\nabla W)\cdot  (\nabla f_2)
\\
\nonumber
&&
{~~~~~~~~~~~~~~~~~~~~~~~~~~~~~~~}
+\frac{3}{2f_3}(\nabla W)\cdot  (\nabla f_3)
+\frac{k}{f_4}(\nabla W)\cdot  (\nabla f_4)=0.
 \end{eqnarray}
All other Einstein equations except for
$ G_r^\theta= 0$ and 
$G_r^r- G_\theta^\theta= 0$
 are linear combinations
of those used to derive the above equations or are identically zero.
The remaining equations
$ G_r^\theta= 0$ 
and 
$G_r^r- G_\theta^\theta= 0$
yield two constraints
\begin{eqnarray}
\label{eq11}
&&
-\frac{\Delta'}{4\Delta}
\bigg(
(d-2k-4)\frac{f_2'}{f_2}+\frac{f_3'}{f_3}+\frac{f_0'}{f_0}
\bigg)
-\frac{1}{4f_0^2}(f_0'^2-\frac{1}{\Delta}{\dot f_0^2})
-\frac{1}{2f_0 f_1}(f_0'f_1'-\frac{1}{\Delta}{\dot f_0}{\dot f_1})
\\
\nonumber
&&
-\frac{(d-2k-4)}{2f_1 f_2}(f_1'f_2'-\frac{1}{\Delta}{\dot f_1}{\dot f_2})
-\frac{(d-2k-4)}{4f_2^2}(f_2'^2-\frac{1}{\Delta}{\dot f_2}^2)
-\frac{1}{2f_1 f_3}(f_1'f_3'-\frac{1}{\Delta}{\dot f_1}{\dot f_3})
\\
\nonumber
&&
-\frac{f_3}{2f_0}(W'^2-\frac{1}{\Delta}{\dot W}^2)
-\frac{1}{4f_3^2}(f_3'^2-\frac{1}{\Delta}{\dot f_3}^2)
+\frac{1}{2f_0}(f_0''-\frac{1}{\Delta}{\ddot f_0})
+\frac{(d-2k-4)}{2f_2}(f_2''-\frac{1}{\Delta}{\ddot f_2})
\\
\nonumber
&&
+\frac{1}{2f_3}(f_3''-\frac{1}{\Delta}{\ddot f_3})
+k
\bigg(
f_4''-\frac{1}{\Delta}{\ddot f_4}
-\frac{\Delta'}{\Delta}\frac{f_4'}{2f_4}
-\frac{1}{f_1 f_4}(f_1'f_4'-\frac{1}{\Delta}{\dot f_1}{\dot f_4})
-\frac{1}{2f_4^2}(f_4'^2-\frac{1}{\Delta}{\dot f_1}^2 )
\bigg)
=0,
\end{eqnarray}
\begin{eqnarray}
\nonumber
&&
-\frac{\Delta'}{4\Delta}
\bigg (
\frac{{\dot f_0}}{f_0}+\frac{(d-2k-4){\dot f_2}}{f_2}+\frac{{\dot f_3}}{f_3}
\bigg)
-\frac{1}{4f_0f_1}({\dot f_1}f_0'+{\dot f_0}f_1')
-\frac{(d-2k-4)}{4f_1f_2}({\dot f_2}f_1'+{\dot f_1}f_2')
\\
\label{eq12}
&&
-\frac{1}{4f_1f_3}({\dot f_1}f_3'+{\dot f_3}f_1')
-\frac{1}{4f_0^2}{\dot f_0}f_0'
-\frac{(d-2k-4)}{4f_2^2}{\dot f_2}f_2'
-\frac{1}{4f_3^2}{\dot f_3}f_3'
-\frac{f_3}{2f_0}{\dot W}W'
\\
&&
\nonumber
+\frac{1}{2}
\bigg(
\frac{{\dot f_0}'}{f_0}
+\frac{(d-2k-4)){\dot f_2}'}{f_2}
+\frac{{\dot f_3}'}{f_3}
\bigg)
+k \bigg(
\frac{{\dot f_4}'}{f_4}
-\frac{{\dot f_4}f_4'}{2f_4^2}
-\frac{1}{2f_1f_4}
(
{\dot f_1}f_4'+{\dot f_4}f_1'
-\frac{\Delta'}{\Delta}\frac{{\dot f_4}}{2f_4}
)
\bigg)
=0.
\end{eqnarray}
In the above relations,
a prime denotes $\partial/\partial_r$, and a 
dot $\partial/\partial_\theta$.
Also, we have defined
\begin{eqnarray}
\nonumber
&&
(\nabla A)\cdot  (\nabla B)=A'B'+\frac{1}{\Delta}{\dot A}{\dot B},
\\
\nonumber
&&
\nabla^2 A =A''+\frac{1}{\Delta}{\ddot A}.
\end{eqnarray}
One can easily verify that the Minkowski spacetime background is 
recovered for 
\begin{eqnarray}
f_1=F_1,~~f_2=F_2,~~f_3=f_4=F_3,~f_0=1,~W=0,
\end{eqnarray}
with $F_i$, $\Delta$ given by (\ref{Fi}) and (\ref{delta}), respectively. 

The structure of these equations suggests that
the case $n=1$, $i.e.$ $d=2k+5$,
is special, since some source terms
associated with the curvature of the $S^n$-part of the metric
vanish in this case.
As we shall see,
the properties of the corresponding solutions
with $S^2\times S^{d-4}$
horizon topology are indeed different,
as well as those of the corresponding MP  black holes.

\subsection{The boundary conditions}
In Appendix A we give
an approximate form of the solutions on the boundaries 
of the domain of integration,
compatible with the $S^{n+1}\times S^{2k+1}$
and $S^{d-2}$ horizon topologies. 
The analysis there
leads to a natural set of boundary
conditions for the solutions in this work,
which are imposed in the numerics.
First, 
the boundary conditions satisfied at the horizon, $r=r_H$, are
\begin{eqnarray}
\label{bc-eh}
f_0=0,~~
r_H\partial_{r}f_1+2f_1=\partial_{r}f_2=\partial_{r}f_3=0,
~~W=\Omega_H.
\end{eqnarray}
As $r\to \infty$, the Minkowski spacetime background is recovered,
which implies
 \begin{eqnarray}
f_0=f_1= 1,~~f_2=r^2\cos^2 \theta, ~~f_3=f_4=r^2\sin^2 \theta,~~W=0.
\end{eqnarray}
 At $\theta=\pi/2$, we impose 
\begin{eqnarray}
\label{tpi23}
\partial_\theta f_0 =
\partial_\theta f_1 =
f_2 =
\partial_\theta f_3 =\partial_\theta f_4=
\partial_\theta W=0.
\end{eqnarray}

The boundary conditions at $\theta=0$ are more complicated.
For solutions with a $S^{n+1}\times S^{2k+1}$ horizon topology, we impose
\begin{eqnarray} 
\label{tpi234}
\partial_\theta f_0 =
\partial_\theta f_1 =
   f_2 =
\partial_\theta f_3 =\partial_\theta f_4=
\partial_\theta W =0,
\end{eqnarray}
for $r_H< r\leq R$,
and
\begin{eqnarray} 
\label{tpi25}
\partial_\theta f_0 =
\partial_\theta f_1 =
\partial_\theta f_2 =
f_3 =f_4=
\partial_\theta W =0,
\end{eqnarray} 
for
$r_H> R$.
The solutions with a spherical horizon topology
are subject to the conditions (\ref{tpi25}) 
for any $r> r_H$. (We recall that $R$
does not appear in this case.)

Apart from that,
the solutions on the boundaries are subject
to a number of extra-conditions,
originating mainly in the constraint equations 
($e.g.$  the constancy of the Hawking temperature  on the horizon,
see the analysis in 
Appendix A).
 However, these conditions are not imposed in the numerics,
but used to verify the accuracy of the results.

 In describing the  boundary conditions (\ref{bc-eh})-(\ref{tpi234}),
 we have found it useful\footnote{These diagrams should also be viewed together
 with the plots of the metric functions in the Figures 3, 4, 8, 12.
 } to introduce 
 the diagrams shown in Figure 2.
There, the domain of integration 
is shown together with the boundary conditions satisfied by some metric functions  which enter  the 
angular part of the metric 
(with $g_{\Omega \Omega}=f_{2}$ and $g_{\Sigma \Sigma}=f_3,f_4$).
In our conventions,
a wavy line indicates a horizon, a doted line represents infinity,
 a  thick line  means  that the coefficient $g_{\Omega \Omega}$
 vanishes
 and a double thin line stands for $g_{\Sigma \Sigma}=0$. 
 Thus,
the horizon topology can easily be read from such diagrams:
a spherical horizon continues with thick and double thin lines, while for a $S^{n+1}\times S^{2k+1}$ horizon topology, the horizon
continues with thick lines only ($i.e.$ the coefficient of the
$d\Omega_n^2$ part of the metric vanishes both at $\theta=0$ and $\theta=\pi/2$).

Finally, let us mention that the diagrams in Figure 2
encode also the generalized rod-structure of the solutions;
moreover,  for $d=5$
they help to make contact with the usual Weyl coordinates.
 A discussion of these aspects can be found in 
 \cite{Kleihaus:2010pr}.
As shown there, similar diagrams can be drawn to describe
composite black objects, $e.g.$
black Saturns or dirings.

\setlength{\unitlength}{1cm}
\begin{picture}(8,6)
\put(-1.2,0.0){\epsfig{file=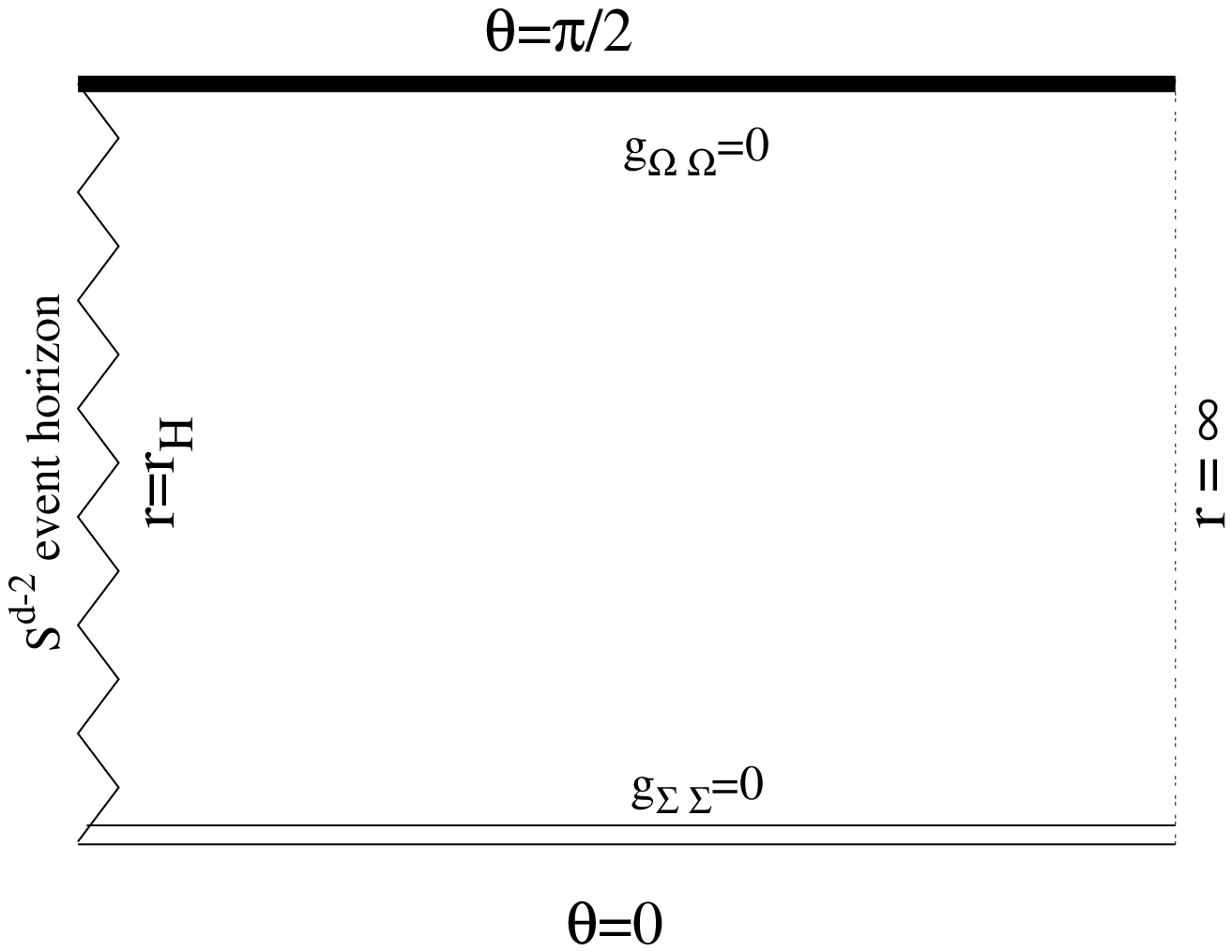,width=7.5cm}}
\put(7.6,0.0){\epsfig{file=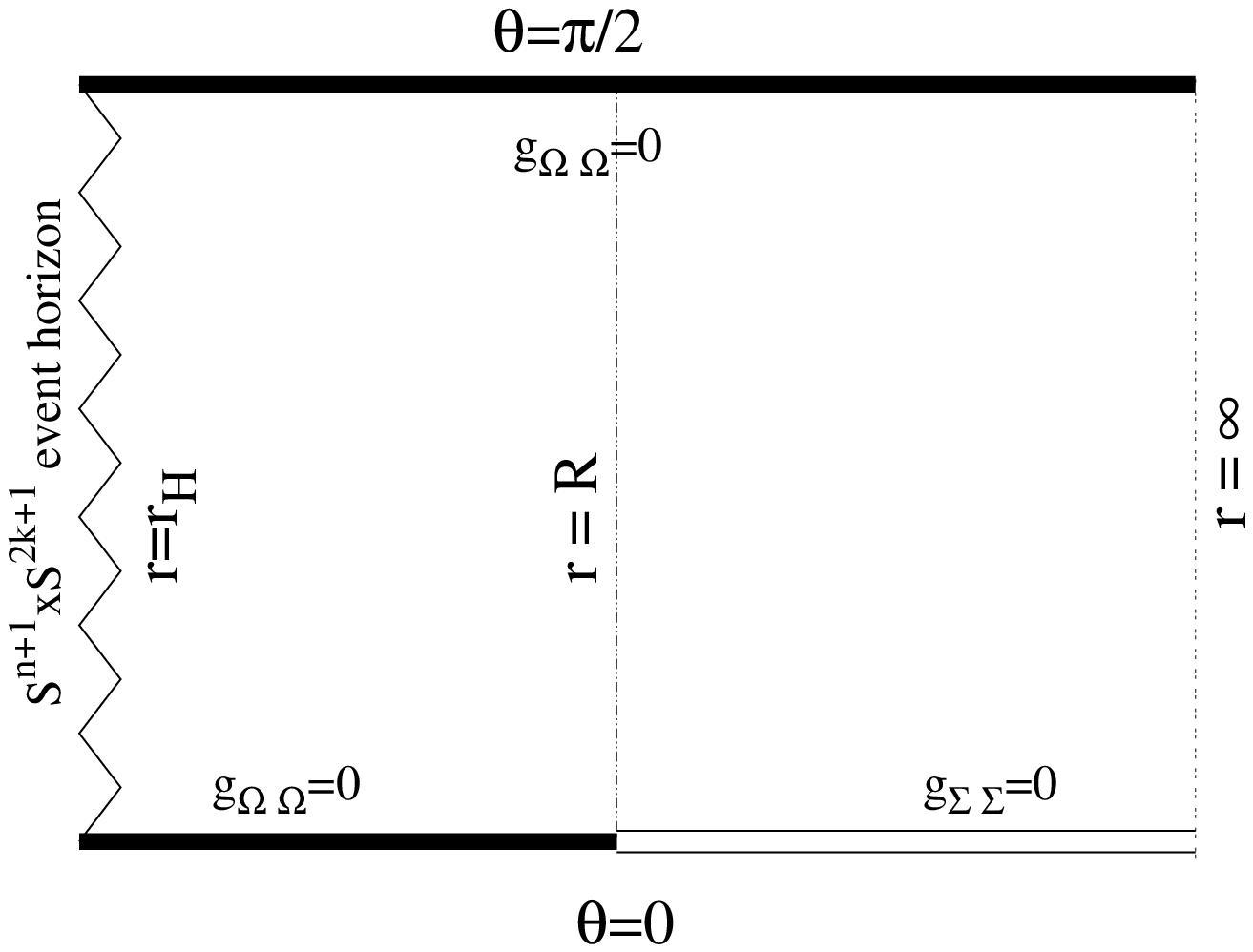,width=7.5cm}}
\end{picture}
\\
\\
{\small {\bf Figure 2.}
The domain of integration for the solutions in this work is shown for
 a black hole with spherical horizon topology
 and a black object with  $S^{n+1}\times S^{2k+1}$ horizon topology.}
\vspace{0.5cm}

\subsection{Quantities of interest}

\subsubsection{Horizon properties}

As discussed above, for any topology,
 the horizon is located at a constant value of the radial 
 coordinate, $r=r_H.$
The  metric of a spatial cross-section of the horizon is
\begin{eqnarray}
\label{eh-m}
d\sigma^2=  
f_1(r_H,\theta)r_H^2 d\theta^2 
 +f_2(r_H,\theta) d\Omega_{n}^2
 +f_3(r_H,\theta) (d\psi+ {\cal A})^2
  +f_4(r_H,\theta) d\Sigma_{k}^2~.
\end{eqnarray}
From the above boundary conditions 
and the discussion in Section 3.3, 
it is clear that the topology
of the horizon
of the generic solutions is $S^{n+1}\times S^{2k+1}$ (although both 
$S^{n+1}$ and $S^{2k+1}$ are not round spheres).
The crucial point here is that the functions $f_3$, $f_4$ multiplying
the  $S^{2k+1}$ part
 are nonzero for any $r\leq R$, 
while $f_2$  vanishes as $\epsilon^2$
at both $\theta=0$ and $\theta=\pi/2$ 
(which will correspond to the poles of the $S^{n+1}$-sphere).

However, the same horizon metric is shared by black objects with an
$S^{d-2}$ horizon topology,
in which case $f_2$ vanishes at $\theta=\pi/2$ (with $f_3$, $f_4$ nonzero),
while $f_3$, $f_4$ are zero at $\theta=0$ (with $f_2$ nonvanishing there).

For any horizon topology, the event horizon area $A_H$,
Hawking temperature $T_H$ 
and event horizon velocity $\Omega_H$ of the solutions
are given by
\begin{eqnarray}
\label{eh-A} 
&&
A_H= r_H V_{(n)}V_{(2k+1)}
\int_0^{\pi/2}d\theta
\sqrt{f_1f_2^{n}f_3f_4^{2k}}\Bigg|_{r=r_H},\\
\nonumber
&&T_H= \frac{1}{2\pi}\lim_{r\to r_H} \frac{1}{(r-r_H) }\sqrt{\frac{f_{0} }{ f_{1}}},~~
~~
\Omega_H=W\big|_{r=r_H},
\end{eqnarray}
where $V_{(p)}$ is the area of the unit $S^p$ sphere. 
Also, one can see that the Killing vector
 \begin{eqnarray}
\label{rh5}
\xi=\partial/\partial_t+\Omega_H \partial/\partial_\psi
 \end{eqnarray}
 is orthogonal and null on the horizon.

For black holes  with a non-spherical horizon topology,
it is useful to get some estimates for the deformation of the two parts in the horizon metric (\ref{eh-m}).
To obtain a measure for the deformation of the $S^{d-3}$ sphere,
we compare the circumference at the equator, $L_e$
($\theta=\pi/4$, where the sphere is fattest),
with the circumference of the $S^{n+1}$ along the poles, $L_p$,
\begin{eqnarray}
\label{Lep}
L_e=2 \pi \sqrt{f_2(r_H,\pi/4)},~~L_p=2\int_0^{\pi/2}d\theta~ r_H\sqrt{f_1(r_H,\theta)},
\end{eqnarray}
and consider, in particular, their ratio $L_e/L_p$. 
The sphere $S^{2k+1}$  in (\ref{eh-m})
is also deformed;
a possible  estimate of its
deformation is given by the ratio
 $R_{2k+1}^{(in)}/R_{2k+1}^{(out)}$,
 where we define
\begin{eqnarray}
\label{def2}
R_{2k+1}^{(in)}=   \left( f_3(r_H,0) f_4^{2k}(r_H,0) \right)^\frac{1}{2(2k+1)}  , ~~
R_{2k+1}^{(out)}=  \left( f_3(r_H,\pi/2) f_4^{2k}(r_H,\pi/2) \right)^\frac{1}{2(2k+1)}~.~~~~{~~}
\end{eqnarray}
These expressions are found by introducing
an effective ($\theta-$dependent) radius of the  $S^{2k+1}$ via its area,
and taking its value inside the ring(oid) at $\theta=0$, and
outside at $\theta=\pi/2$.

\subsubsection{The global charges}

 The mass and angular momenta are read from the large$-r$ 
 asymptotics of the metric functions,
 $g_{tt}=-1+\frac{C_t}{r^{d-3}}+\dots,
~g_{\psi t}=-f_3W= \frac{ C_\psi}{r^{d-3}}\sin^2 \theta+\dots,$
 with ($G=1$):
\begin{eqnarray}
\label{MJ}
{\cal M}=\frac{(d-2)V_{(d-2)}}{16 \pi }C_t,~~J_1=\dots=J_{k+1}=
J,~~{\rm where}~~~J=\frac{V_{(d-2)}}{8\pi }C_{\psi}.
\end{eqnarray} 
Also, the solutions satisfy the Smarr relation  
\begin{eqnarray}
\label{Smarr}
\frac{d-3}{d-2}{\cal M}=T_H \frac{A_H}{4}+(k+1)\Omega_H J,
\end{eqnarray}
and the $1^{{\rm st}}$ law
\begin{eqnarray}
\label{1st}
d{\cal M}=\frac{1}{4}T_H dA_H+(k+1)\Omega_H dJ.
\end{eqnarray}
The black objects have an entropy
which is given by the area law, $S=\frac{A_H}{4}$.
  
  It is well-known that different thermodynamic ensembles are not exactly equivalent
  (for example they may not lead to the same conclusions regarding the thermodynamic stability as they
correspond to different physical situations).
  We study the solutions in a canonical ensemble by keeping the temperature $T_H$ 
   and the angular momentum fixed. 
   The associated thermodynamic
potential is the Helmholz free energy
\begin{eqnarray}
\label{F}
F={\cal M}-T_H \frac{A_H}{4}.
\end{eqnarray}
The situation of  black objects in a grand canonical ensemble is also of interest, in
which case we keep the temperature  and the angular velocity of the horizon fixed. In this
case, the thermodynamics is obtained from the Gibbs potential
\begin{eqnarray}
\label{W}
W={\cal M}-T_H \frac{A_H}{4}-(k+1)\Omega_H J.
\end{eqnarray}
Using the Smarr relation (\ref{Smarr}),
one finds
\begin{eqnarray}
\label{W1}
W=\frac{{\cal M}}{d-2}.
\end{eqnarray}

  
Following the usual convention in the BRs/blackfold
literature, we fix the overall scale 
of the solutions by fixing their mass ${\cal M}$.
Then the solutions are characterized by a set of
reduced dimensionless 
quantities, obtained by dividing out an appropriate power of ${\cal M}$:
\begin{eqnarray}
\label{dim1}
j=c_j \frac{J}{{\cal M}^{\frac{d-2}{d-3}}},~~
a_H=c_a \frac{A_H}{{\cal M}^{\frac{d-2}{d-3}}},~~
w_H=c_w\Omega_H {\cal M}^{\frac{1}{d-3}},~~
t_H=c_t T_H {\cal M}^{\frac{1}{d-3}},~~
\end{eqnarray}
with the coefficients\footnote{These coefficients are chosen such that to agree with those
 in  \cite{Emparan:2007wm}  for $k=0$.}
\begin{eqnarray}
\label{dim2}
&&
c_j=\frac{(d-2)^{\frac{d-2}{d-3}}}{(16\pi)^{\frac{1}{d-3}}2^{\frac{d-2}{d-3}}}\frac{1+k}{\sqrt{(d-3)(2k+1)}}(V_{(n+1)}V_{(2k+1)})^{\frac{1}{d-3}},
\\
&&
\nonumber
c_a=\frac{2^{\frac{2}{d-3}}}{(16\pi)^{\frac{d-2}{d-3}}}(d-2)^{\frac{d-2}{d-3}}\sqrt{\frac{d-2k-4}{d-3}}(V_{(n+1)}V_{(2k+1)})^{\frac{1}{d-3}},
\\
\nonumber
&&
c_w= \frac{2^{\frac{1}{d-3}}}{(d-2)^{\frac{1}{d-3}}}\sqrt{\frac{d-3}{2k+1}}\frac{(16\pi)^{\frac{1}{d-3}}}{(V_{(n+1)}V_{(2k+1)})^{\frac{1}{d-3}}},
\\
\nonumber
&&
c_t= \frac{(d-4)\sqrt{d-3}}{2^{\frac{2(d-2)}{d-3}}(d-2)^{\frac{1}{d-3}}}
\frac{(16\pi)^{\frac{d-2}{d-3}}}{(d-2k-4)^{\frac{3}{2}}(V_{(n+1)}V_{(2k+1)})^{\frac{1}{d-3}}}.
\end{eqnarray}

Finally, let us mention that
all solutions possess an ergo-region, 
defined as the domain in which the metric
function $g_{tt}$ is positive.
For the line element in this work,
this domain is bounded by the event horizon and by the surface where
\begin{eqnarray}
\label{ergo}
-f_0+f_3 W^2=0.
\end{eqnarray}

\subsection{Remarks on the numerics }
Given the above framework,
the only solutions of the Einstein equations
which are known in closed form correspond to MP black holes
with $k+1$ equal angular momenta, and to
the single spinning $d=5$ Emparan-Reall BR ($i.e.$ with $k=0$).
These configurations are discussed in the next Section.

All other solutions  
 in this work are found by solving numerically the eqs.~(\ref{eqf0})-(\ref{eqW})
within a nonperturbative approach.
In our scheme, for given $(d,k)$,
the only input parameters are $R,~r_H$ and the angular velocity $\Omega_H$.
(Note that although $R$ and $r_H$ have no invariant meaning,
they provide a rough measure of the  $S^{n+1}$ and $S^{2k+1}$  spheres, respectively,  on the horizon.)
Then all other quantities of interest, in particular the Hawking temperature $T_H$,
the horizon area $A_H$ and the global charges ${\cal M},~J$
are extracted  from the numerical output, 
being encoded in the values of   $(f_i,W)$.

To find these functions,
we employ a numerical 
algorithm developed in 
\cite{Kleihaus:2009wh},
\cite{Kleihaus:2010pr},
which uses a Newton-Raphson method 
whilst ensuring that all the Einstein equations are satisfied.
In this approach,
the functions $f_i$ are expressed  
as products of suitable background functions $f_i^{(0)}$  which 
possess the required behaviour on the boundaries,
 and
unknown functions ${\cal F}_i$. 
The simplest choice for the background functions\footnote{
However,
 we have found that a choice for $f_i^{(0)}$
corresponding to the functions which enter
the $d=5$ static BR
 leads to better results.
In this case, $f_1^{(0)}$, $f_2^{(0)}$ and $f_3^{(0)}$ 
are essentially $F_1$, $F_2$ and $F_3$ in  (\ref{metric}),
though with some $r_H$-dependent corrections
which are finite and nonzero everywhere.
Most of the numerical results
reported in this work have been found for this choice of 
$f_i^{(0)}$.
} of the solutions 
 with a non-spherical horizon topology
is given by $F_1$, $F_2$ and $F_3$ in (\ref{Fi}).

The advantage of this approach is that the coordinate singularities
are essentially subtracted,
while imposing  at the same time the event horizon topology
as well as the asymptotic structure of spacetime.
The crucial point here is that the functions  ${\cal F}_i$ stay non-zero and finite
everywhere.
 In particular, this holds on the boundaries,
 such that the behaviour of the solutions there remains as fixed by the background 
 functions.
The reader is referred to Ref.~\cite{Kleihaus:2010pr} 
for details of this procedure.

The equations for the ${\cal F}_i,W$ 
result directly from (\ref{eqf0})-(\ref{eqW})
and are solved 
by using a finite difference solver \cite{schoen}.
This professional software
provides an error estimate for each unknown function, which is
the maximum of the discretization error divided by the maximum
of the function. The typical numerical error
for the solutions here is estimated to be lower than $10^{-3}$.
(Note that 
we use an order six  for the discretization of derivatives.)
We have extensively
tested the numerical results, including the convergence of
the code for different resolutions of the mesh.
Also, we have been able to
recover $d = 5$ balanced BRs and  
$d = 5, 6$ MP black hole solutions
with a single angular momentum, 
starting with the corresponding static
configurations.

One should mention that we have constructed the $d=6$ BRs 
independently  
by using a multi-domain spectral solver. 
Here the functions are expanded in products of Chebychev polynomials. 
The resulting systems of algebraic equations for the expansion coefficients
are then solved with the Newton-Raphson method. 
The iteration matrix of the `linear problem' is no longer
sparse and is solved by Gaussian elimination. 
We have found a very good agreement for the results
obtained by these two different numerical schemes.

For both approaches, 
another kind of test of the numerics is provided by
 the Smarr relation (\ref{Smarr})
  and by the 1$^{st}$ law (\ref{1st}).
The typical  relative errors found in this way are
 $<10^{-3}$.
 A further numerical
test is provided by the constraint equations,
$ G_r^\theta= 0$
and 
$G_r^r- G_\theta^\theta= 0$,
which in our scheme are not solved directly.
  However, usually these constraints are satisfied with the same order of the relative error as the
Smarr relation.

\section{Exact solutions}

\subsection{A spherical horizon topology: the  Myers-Perry black holes }

The simplest solutions of the eqs.~(\ref{eqf0})-(\ref{eqW})
correspond to the MP solutions with $k+1$ equal angular momenta.
\setcounter{figure}{2}
\begin{figure}[t!]
\setlength{\unitlength}{1cm}
\begin{picture}(15,18)
\put(-1.,0){\epsfig{file=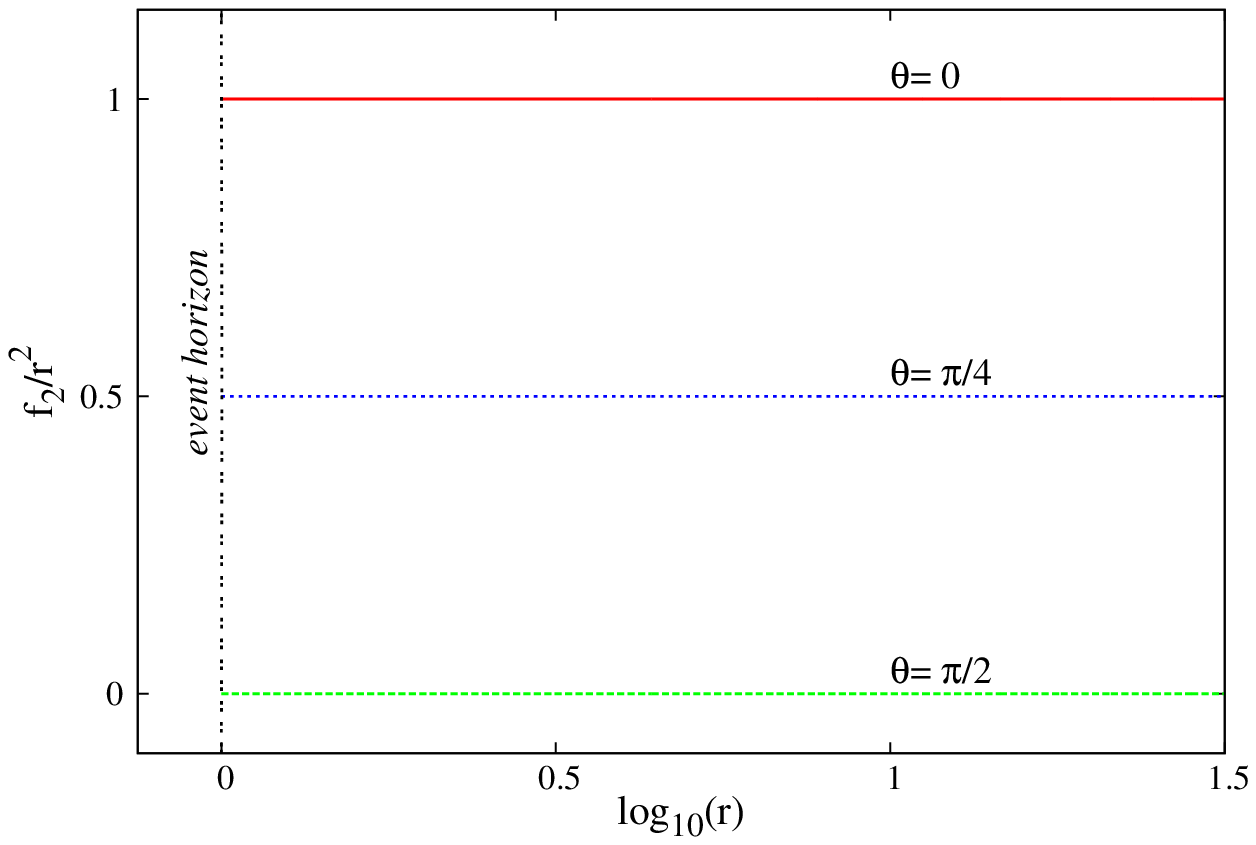,width=8cm}}
\put(7,-0.5){\epsfig{file=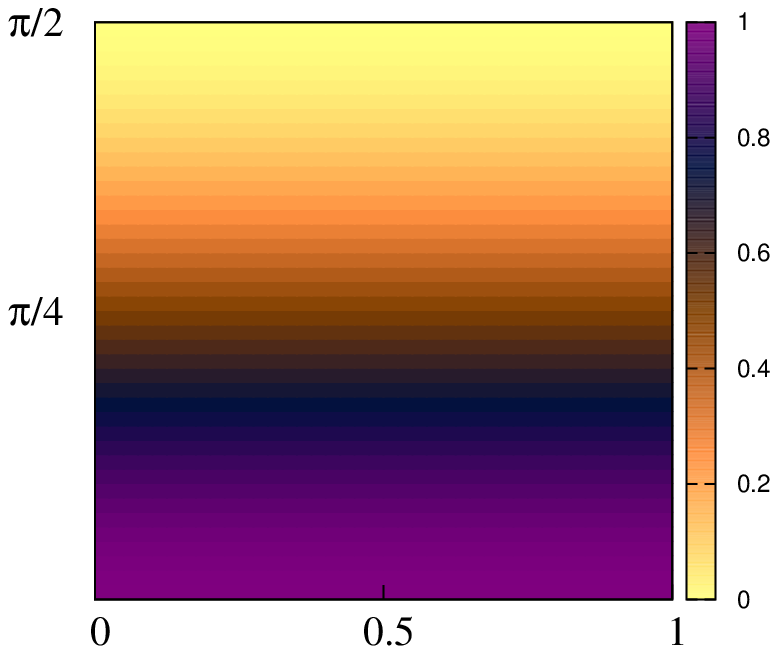,width=10cm}}
\put(-1,6){\epsfig{file=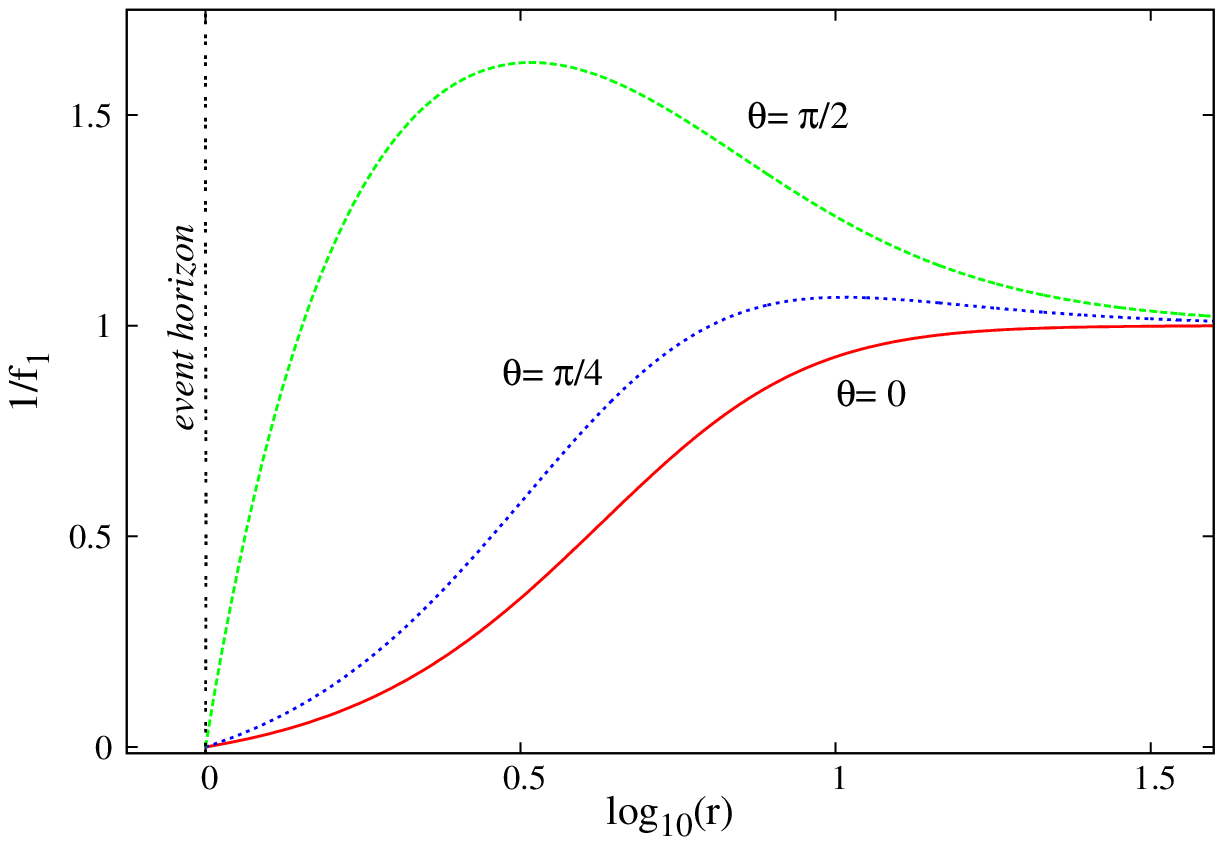,width=8cm}}
\put(7,5.5){\epsfig{file=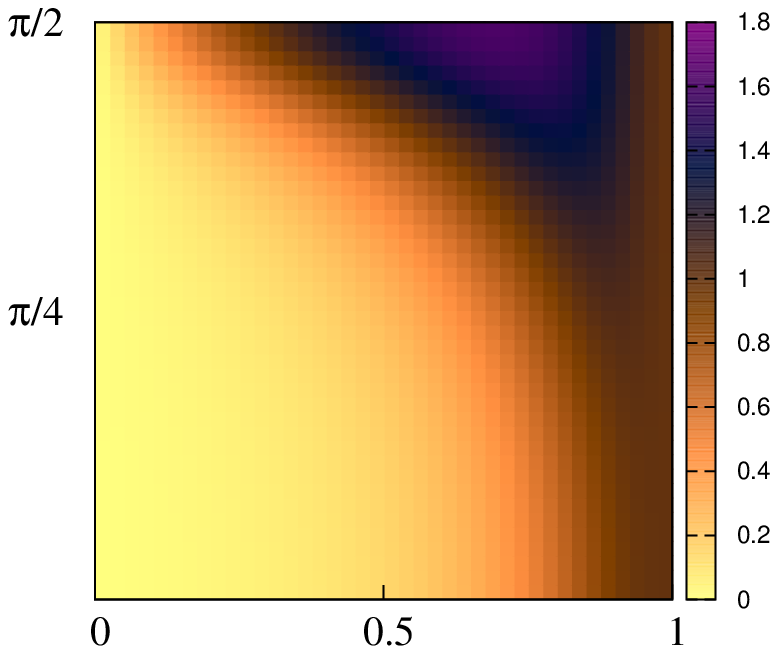,width=10cm}}
\put(-1,12){\epsfig{file=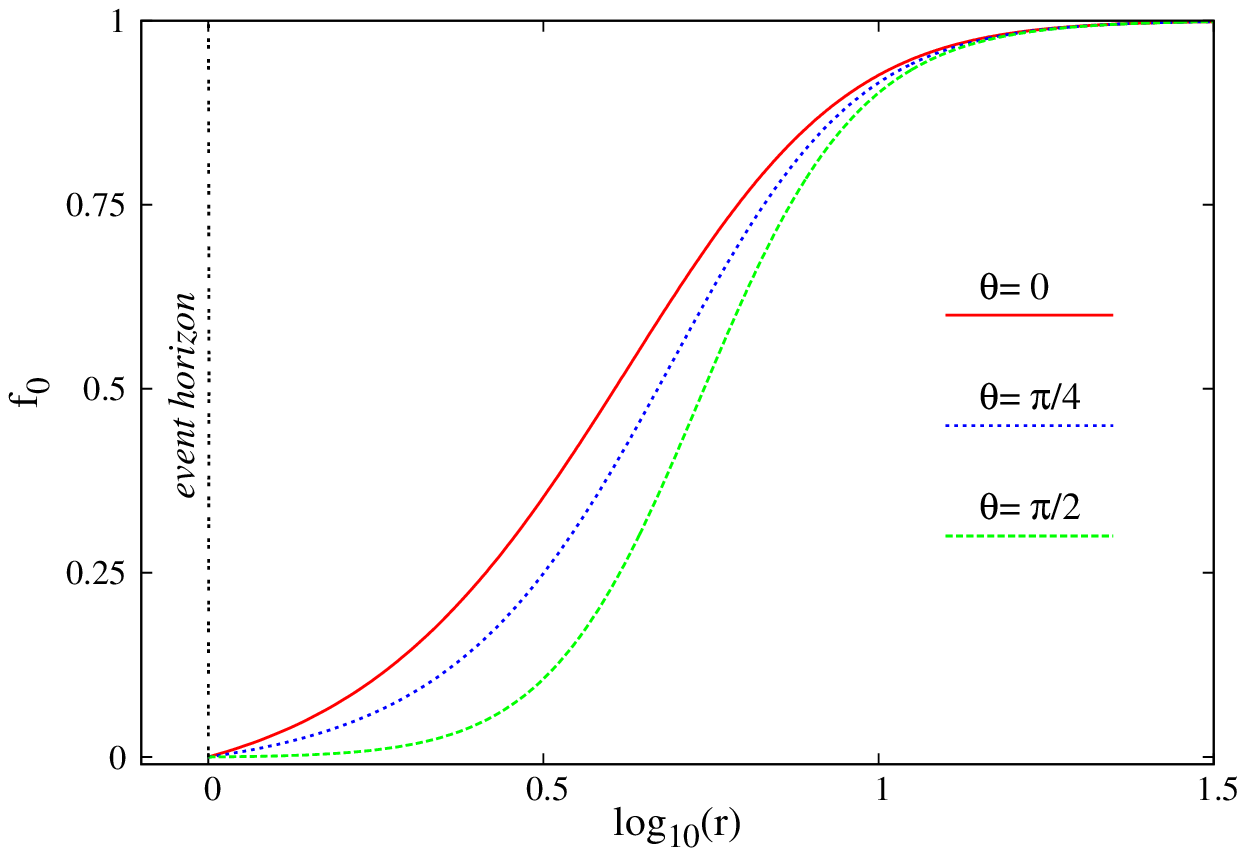,width=8cm}}
\put(7,11.5){\epsfig{file=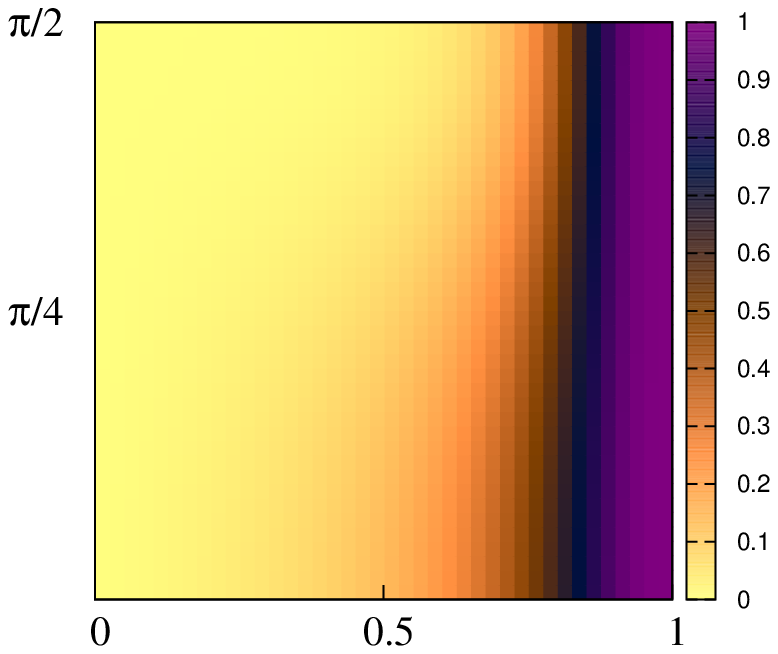,width=10cm}}
\end{picture}
\caption{
The metric functions  $f_i,W$ and the Kretschmann scalar $K=R_{\mu\nu\alpha \beta}R^{\mu\nu\alpha \beta}$ 
 are shown for a $d=7,\ k=1$ Myers-Perry black hole
with the input parameters
$r_H=1$
and 
$\Omega_H\simeq 0.162$.
Here and in Figures 4, 8 and 13
the left panels show the profiles for several
angles $\theta$;
the right panels are colour maps of the same functions in terms of $(x=1-r_H/r,~\theta)$,
which should be viewed together with the diagrams in Fig. 2.
}
 \end{figure}

\setcounter{figure}{2}
\begin{figure}
\setlength{\unitlength}{1cm}
\begin{picture}(15,18)
\put(-1.,0){\epsfig{file=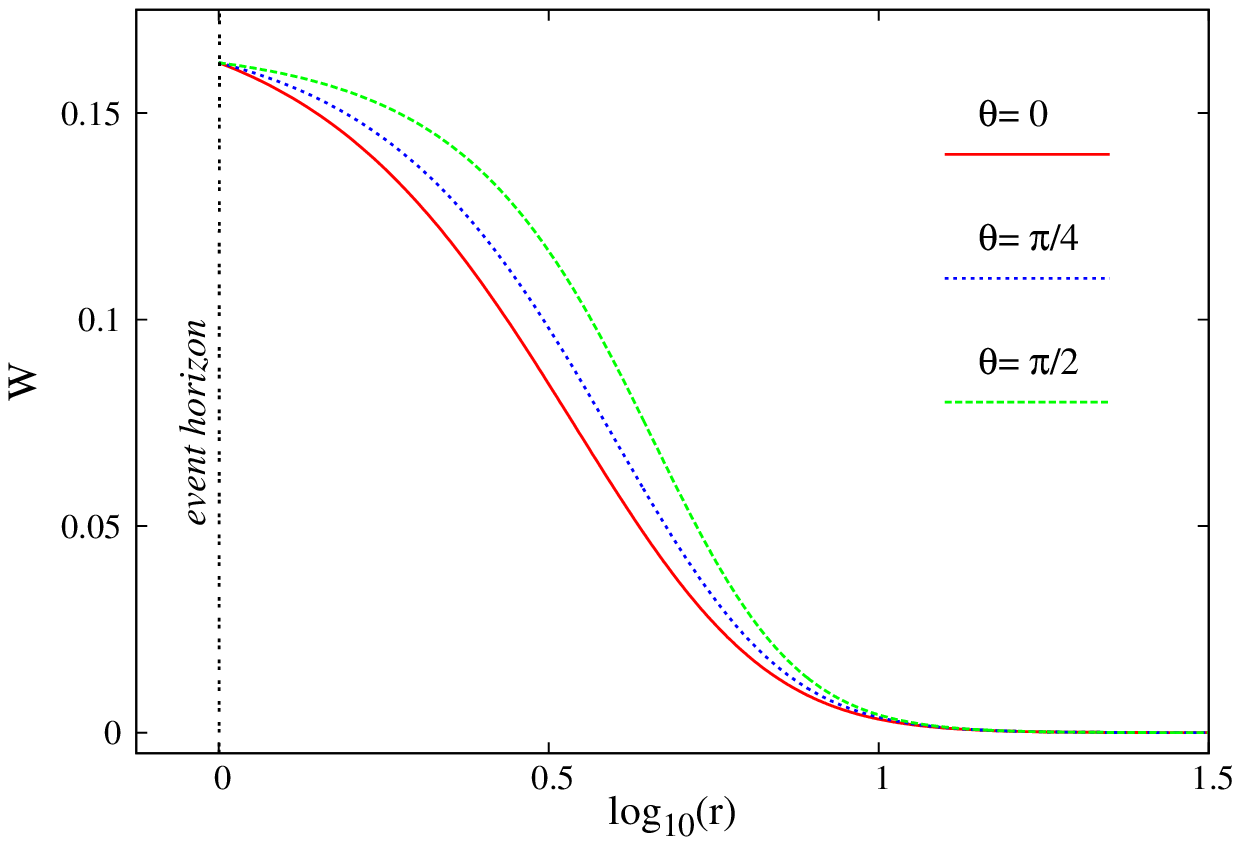,width=8cm}}
\put(7,-0.5){\epsfig{file=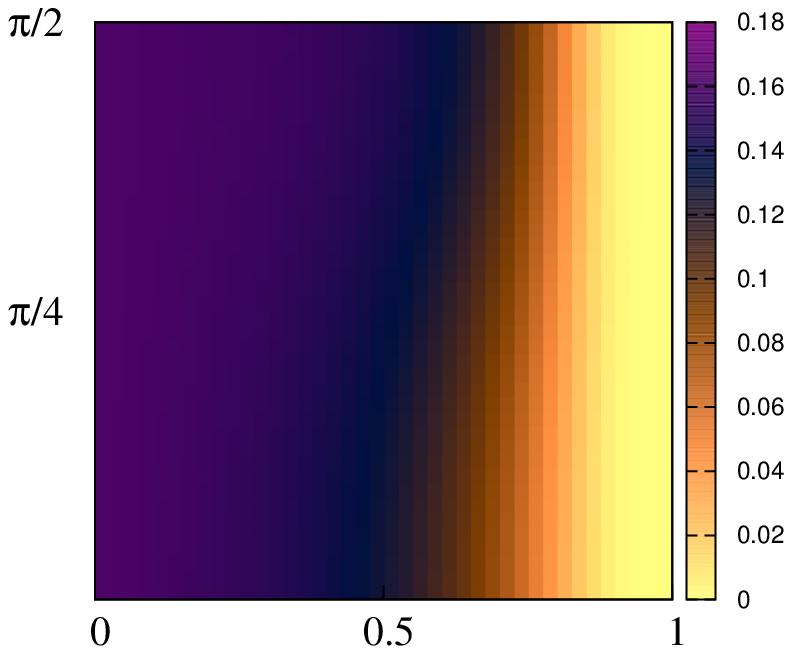,width=10cm}}
\put(-1,6){\epsfig{file=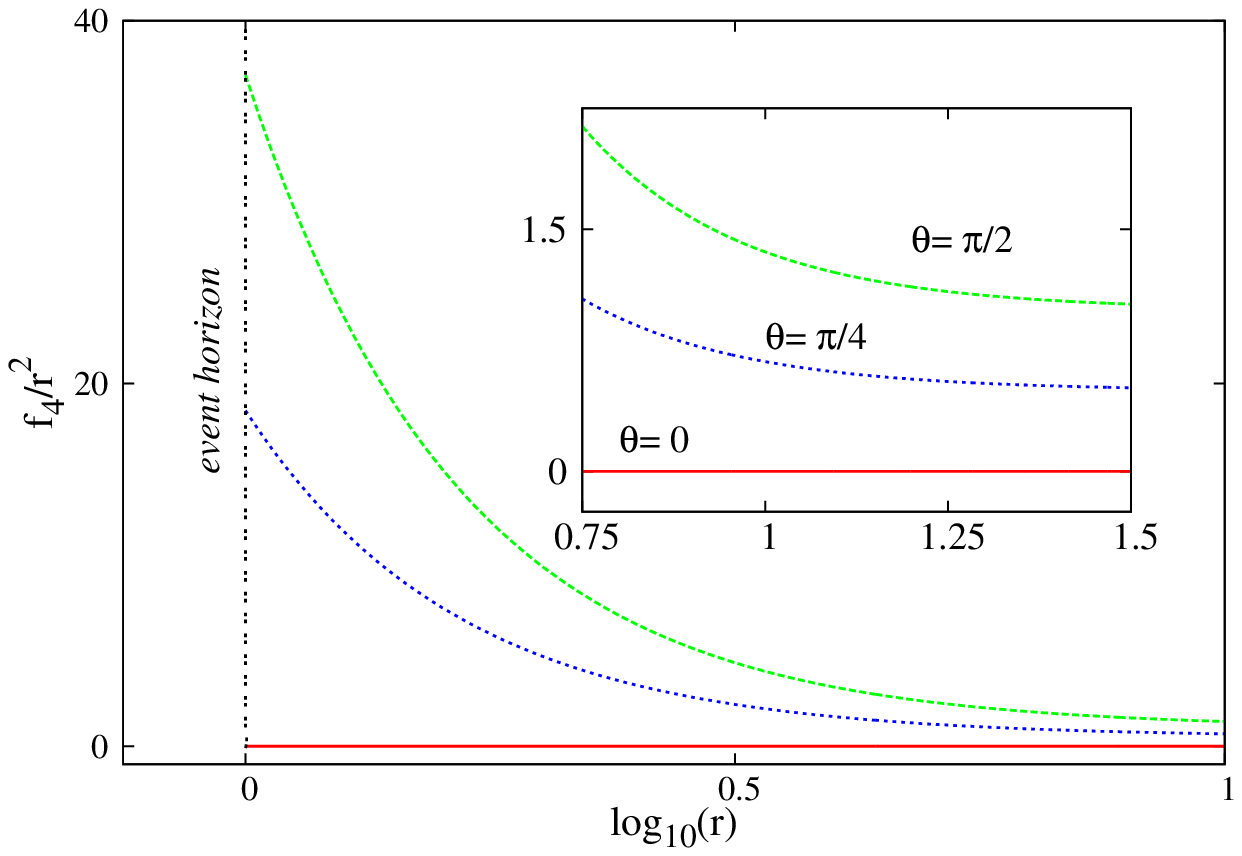,width=8cm}}
\put(7,5.5){\epsfig{file=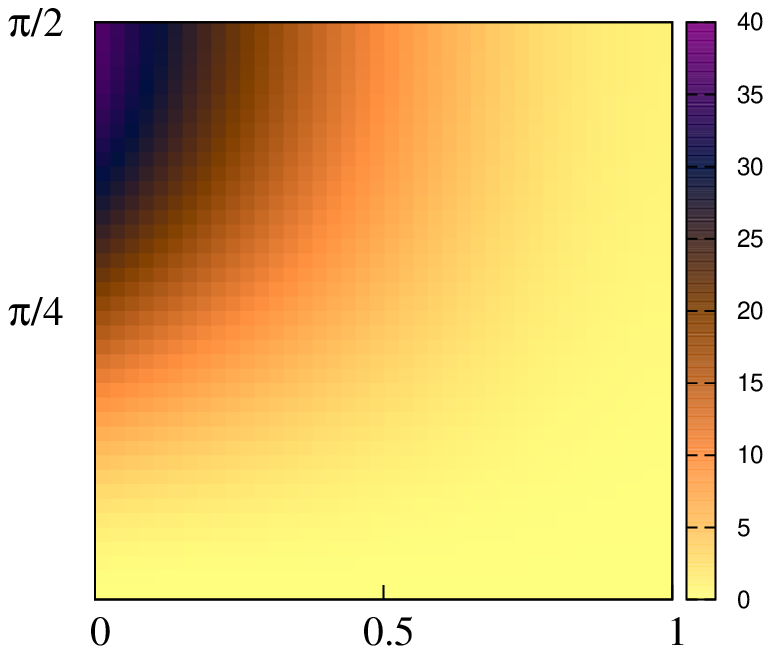,width=10cm}}
\put(-1,12){\epsfig{file=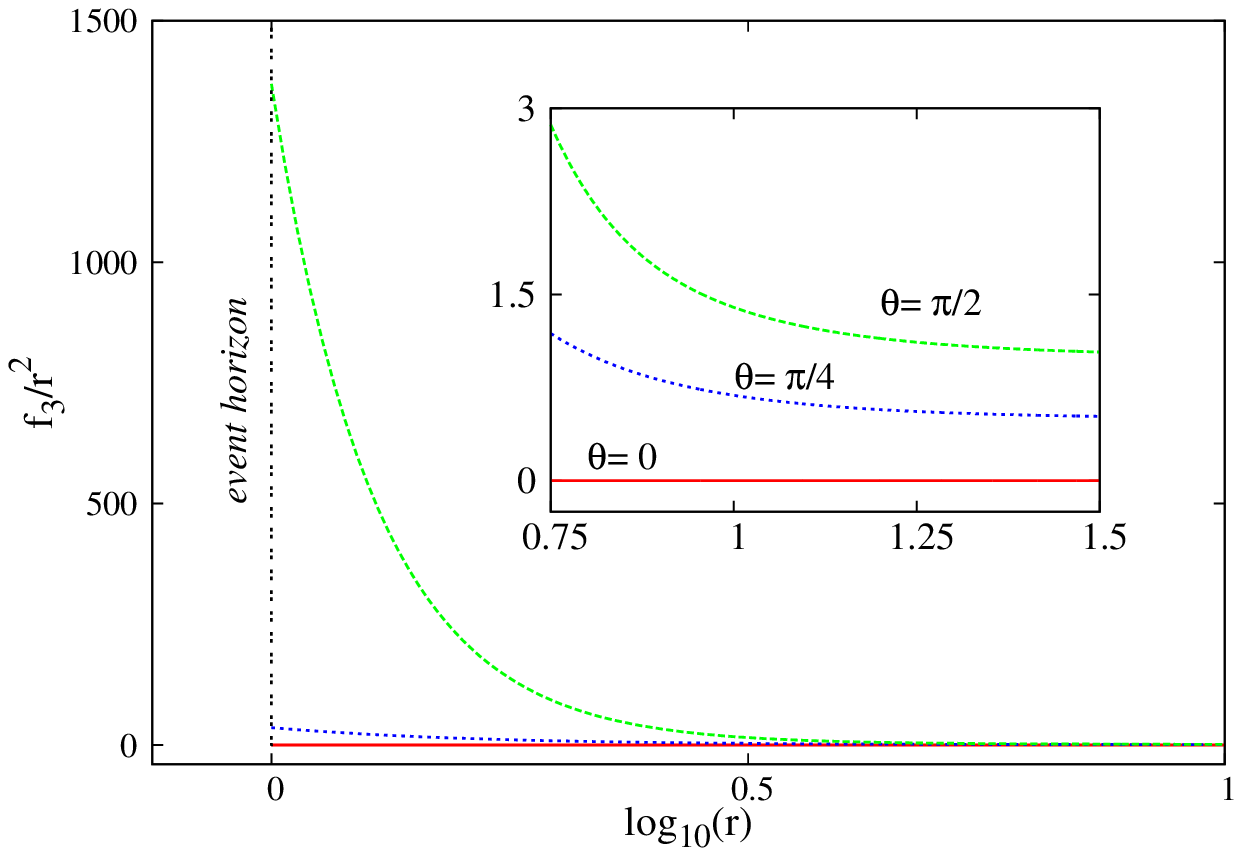,width=8cm}}
\put(7,11.5){\epsfig{file=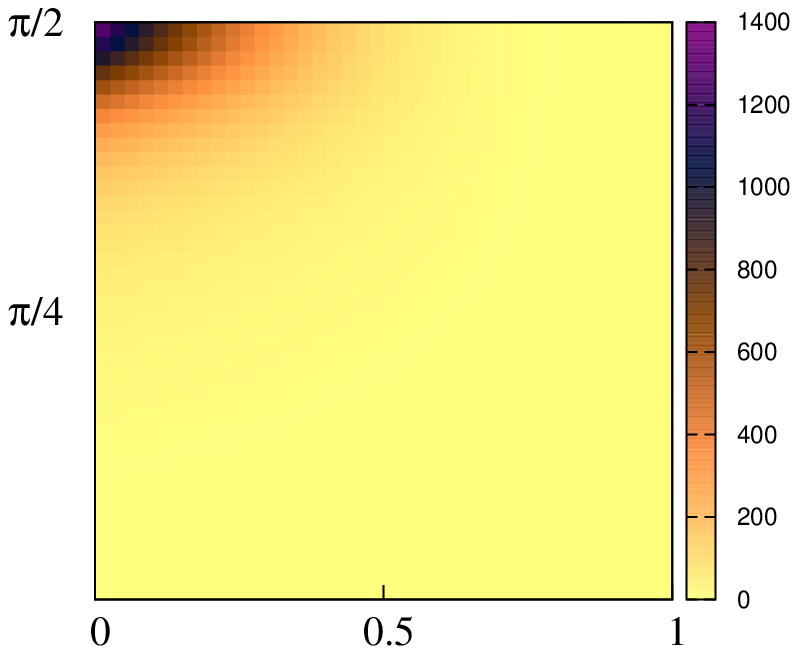,width=10cm}}
\end{picture}
\caption{
continued.
}
\end{figure}

\setcounter{figure}{2}
\begin{figure}
\setlength{\unitlength}{1cm}
\begin{picture}(8,6)
\put(-1,0.0){\epsfig{file=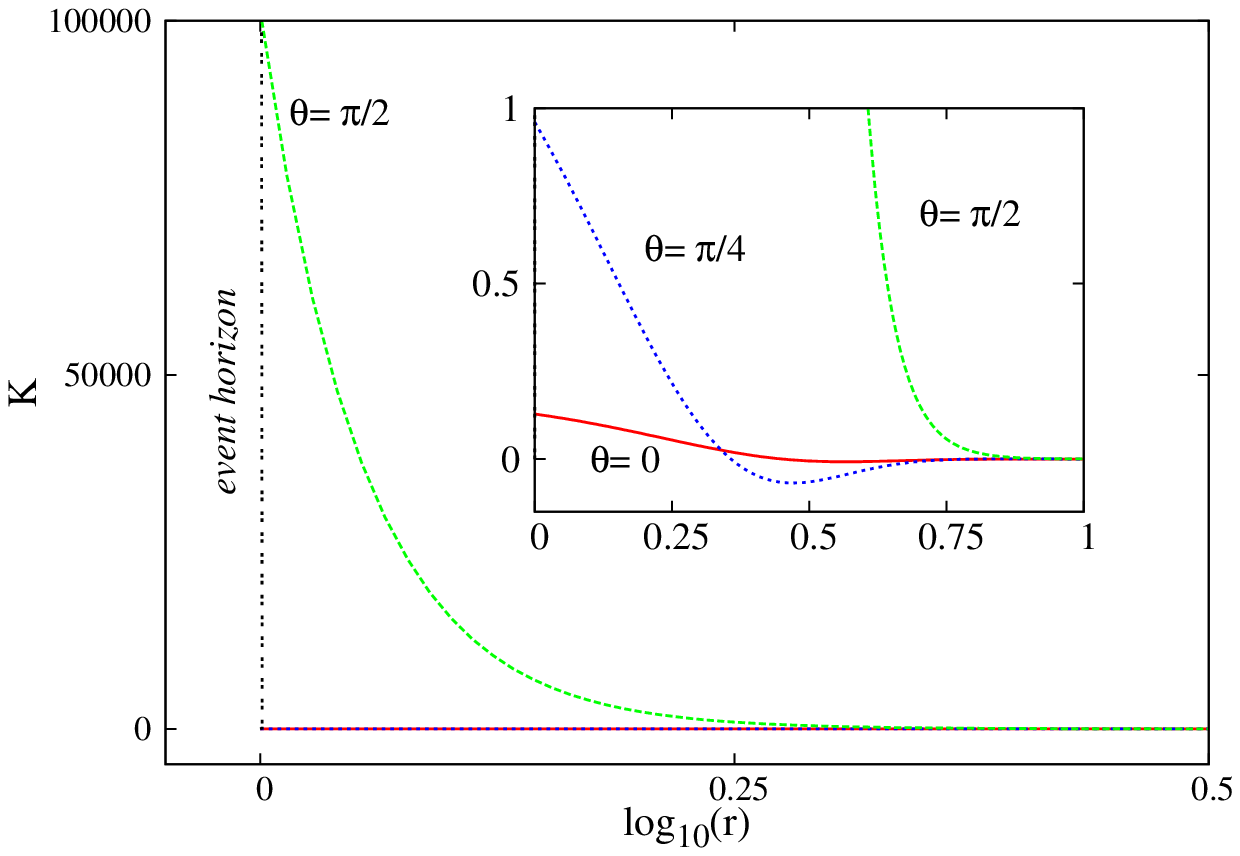,width=8cm}}
\put(7,-0.5){\epsfig{file=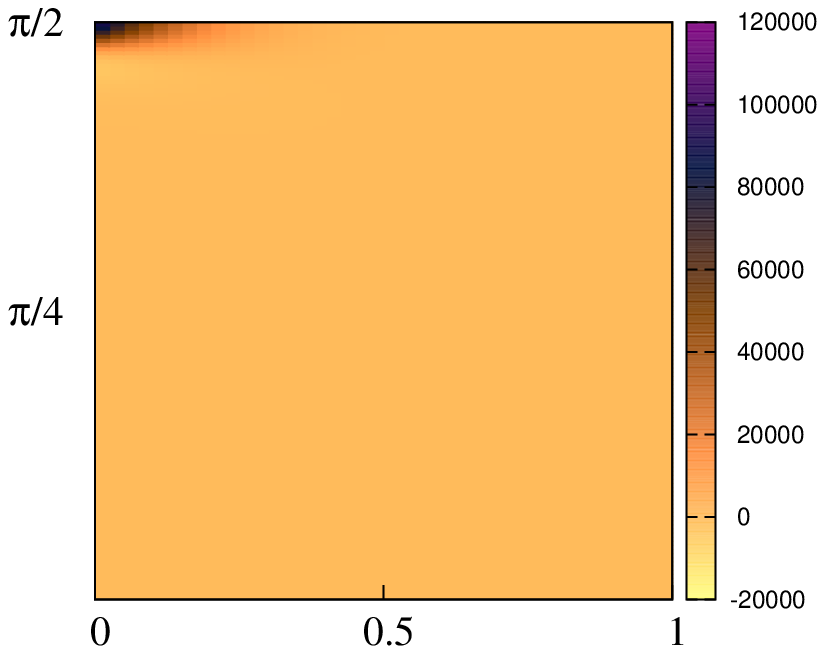,width=10cm}}
\end{picture}
\caption{
continued.
}
\end{figure}

A convenient expression of the metric functions $f_i$ which enter (\ref{metric}) is
\begin{eqnarray}
\nonumber
&&
f_0(r,\theta)=\frac{\Delta(r)}{(r^2+a^2)P(r,\theta)},~~
f_1(r,\theta)=\frac{r^2+a^2 \cos^2\theta}{\Delta(r)},~~
f_2(r,\theta)=r^2\cos^2\theta,
\\
&&
f_3(r,\theta)=(r^2+a^2)\sin^2\theta P(r,\theta),~~
f_4(r,\theta)=(r^2+a^2)\sin^2\theta,~~
\\
&&
\nonumber
W(r,\theta)=\frac{M}{r^{d-(2k+5)}}\frac{a}{(r^2+a^2)^{k+1}(r^2+a^2\cos^2\theta)P(r,\theta)}~,
\end{eqnarray}
where
\begin{eqnarray}
&&
\Delta(r)=(r^2+a^2)
\left(1-\frac{M}{r^{d-(2k+5)}(r^2+a^2)^{k+1}}
\right),~~
\\
\nonumber
&&
P(r,\theta)=1+\frac{M}{r^{d-(2k+5)}}\frac{a^2\sin^2\theta}{(r^2+a^2)^{k+1}(r^2+a^2\cos^2\theta) },
\end{eqnarray}
while $M,a$ are two input parameters\footnote{Note that,
when written in this form
the near horizon expression of the solutions differs from (\ref{rh}), (\ref{rh-sup}).
The relations  (\ref{rh}), (\ref{rh-sup})
are recovered by working with a different radial coordinate.
However, the expression  of the solution looks much more complicated 
in that case.}.
To give an idea about these functions in contrast to the ring(oid) case, we exhibit in Figure 3
the profiles of a typical  $d=7,~k=1$ configuration.
The Kretschmann scalar $K=R_{\mu\nu\alpha \beta}R^{\mu\nu\alpha \beta}$ 
is also shown there
(note that a straightforward computation shows that $K$
is finite everywhere, in particular at $r=r_H,\theta=\pi/2$).

These black holes have a horizon of spherical topology located
at 
$r=r_H$, where $\Delta(r_H)=0$, which implies
\begin{eqnarray}
M=(r_H^2+a^2)^{k+1}r_H^{d-(2k+5)}.
\end{eqnarray}

The quantities of interest which enter the thermodynamics of these solutions are
given by
\begin{eqnarray}
&&
{\cal M}=\frac{(d-2)V_{(d-2)}r_H^{d-2k-5}}{16\pi }(r_H^2+a^2)^{k+1},~~
J=\frac{V_{(d-2)}}{8\pi}a r_H^{d-2k-5}(r_H^2+a^2)^{k+1},
\\
\nonumber
&&
A_H=V_{(d-2)}r_H^{d-2k-4} (r_H^2+a^2)^{k+1},~~
T_H=\frac{1}{4\pi r_H}\left(d-3-\frac{2a^2(k+1)}{a^2+r_H^2} \right),~~
\Omega_H=\frac{a}{a^2+r_H^2}.
\end{eqnarray}

 This implies the following relations for the scaled
dimensionless quantities as defined by (\ref{dim1}): 
\begin{eqnarray}
\label{MP-rel1}
&&
j  =q_j \frac{x}{(1+x^2)^{\frac{k+1}{d-3}}},~
~~a_H=q_a \frac{1}{(1+x^2)^{\frac{k+1}{d-3}}},
\\
\nonumber
&&
t_H=q_t\frac{(d-2k-5)x^2+d-3}{(1+x^2)^{\frac{d-k-4}{d-3}}},
~~w_H=q_w \frac{x}{(1+x^2)^{\frac{d-k-4}{d-3}}},~~
\end{eqnarray} 
with
\begin{eqnarray}
x=\frac{a}{r_H},~~~~0\leq x<\infty,
\end{eqnarray}
and the coefficients:
\begin{eqnarray}
\label{MP-rel2}
&&
q_j=\frac{(k+1)\pi^{\frac{1}{2(d-3)}}}{\sqrt{(d-3)(2k+1)}}
\left(\frac{\Gamma(\frac{d-1}{2})}{\Gamma(\frac{d}{2}-(k+1))k!}\right)^{\frac{1}{d-3}},~
\\
\nonumber
&&
q_a=\frac{2^{\frac{2}{d-3}}\pi^{\frac{1}{2(d-3)}}}{\sqrt{\frac{d-3}{d-2k-4}}}
\left(\frac{\Gamma(\frac{d-1}{2})}{\Gamma(\frac{d}{2}-(k+1))k!}\right)^{\frac{1}{d-3}},
\end{eqnarray}
\begin{eqnarray}
\nonumber
&&
q_w=\frac{1}{\pi^{\frac{1}{2(d-3)}}}\sqrt{\frac{d-3}{2k+1}}
\left(\frac{\Gamma(\frac{d-1}{2})}{\Gamma(\frac{d}{2}-(k+1))k!}\right)^{\frac{1}{d-3}},~
\\
\nonumber
&&
q_t=\frac{1}{2^{\frac{3}{d-3}}}\frac{1}{\pi^{2(d-3)}}
\frac{(d-4)\sqrt{d-3}}{(d-2k-4)^{\frac{3}{2}}}
\left(\frac{\Gamma(\frac{d-1}{2})}{\Gamma(\frac{d}{2}-(k+1))k!}\right)^{\frac{1}{d-3}}.
\end{eqnarray}
From the above relations it is clear that the odd-dimensional case with $n=1$, $i.e.$
\begin{eqnarray}
\label{max-k}
d=2k+5
\end{eqnarray}
is special.
It is the only case where
extremality is possible (which is reached for $x\to \infty$) 
and the angular momentum is bounded from above.
The extremal solutions of this class
share the properties
of the $d=5$ extremal MP black holes with a single $J$ ($k=0$).
In particular, the event horizon  has a vanishing area.
The near horizon geometry of the extremal solutions 
is described  by the following line element\footnote{The case $d=5$, $k=0$ is discussed in \cite{Bardeen:1999px}.}
\begin{eqnarray}
\label{extr1}
&
ds^2=\cos^2 \theta \left(-\frac{r^2}{a^2}dt^2+\frac{a^2}{r^2}dr^2+r^2 d\phi^2 \right)
+\frac{(d-3)}{2}a^2 \left(\cos^2 \theta d\theta^2+\tan^2\theta (d\psi+{\cal A} )^2
+\sin^2\theta d\Sigma_k^2 \right),~~~{~~~}
\end{eqnarray} 
which solves the Einstein equations. 
Thus it turns out that the properties of the five dimensional solutions are generic,
with (\ref{extr1}) representing a singular geometry  for any value of $k$.

The situation is different for MP solutions with $n>1$, since 
in this case the
properties  are 
similar to those of the (better known) $d>5$
MP black holes with a single angular momentum.
Here the angular momenta 
do not possess an upper bound, while
$a_H$, $t_H$ are strictly positive quantities.
Thus these black holes possess an ultraspinning regime,
which is described by the corresponding
blackfolds, see the discussion in Section 5.1. 

\subsection{The $d=5$ Emparan-Reall black ring}
 
Despite various attempts,
the $d=5$ Emparan-Reall BR 
  (and its Pomeransky-Sen'kov generalization \cite{Pomeransky:2006bd}) 
 remains 
the only asymptotically flat  vacuum  (single) black object with a nonspherical
topology of the horizon which is known in closed form.
In all studies, this solution is written in (some version of) ring coordinates, 
or in Weyl coordinates.
However, the BR can also be studied by using the framework
introduced in the previous Section.

The line element is found by taking $k=0$, $d=5$ in (\ref{metric})
and reads 
\begin{eqnarray}
\label{ER1}
 ds^2=f_1(r,\theta)(dr^2+r^2d\theta^2)+f_2(r,\theta) d\phi^2+f_3(r,\theta)(d\psi-W(r,\theta) dt)^2-f_0(r,\theta) dt^2.
\end{eqnarray}
The expression of the metric functions $f_i,W$ is given in Appendix B,
together with the corresponding expansion at $r=r_H, \infty$ and $\theta=0,\pi/2$,
respectively.   
The profiles of a typical solution are given in Figure 4;  
for completeness and comparison with the higher dimensional case, we show there also 
the Kretschmann scalar $K=R_{\mu\nu\alpha \beta}R^{\mu\nu\alpha \beta}$ of the same solution.

The quantities which enter the thermodynamics of the 
$d=5$ BRs
exhibit a complicated dependence on the input parameters 
$(R,r_H)$: 
\begin{eqnarray}
\nonumber
&&
A_H=\frac{32\pi^2\sqrt{2}Rr_H^4\sqrt{R^4+r_H^4}}{(R^2-r_H)^2},~~
T_H=\frac{(R^2-r_H^2)^2}{8\pi \sqrt{2}Rr_H^2\sqrt{R^4+r_H^4}},~~
\Omega_H=\frac{R(R^2-r_H^2)}{\sqrt{2}(R^2+r_H^2)\sqrt{R^4+r_H^4}},
\\
&&
\label{cER1}
M=\frac{3\pi r_H^2(R^4+r_H^4)}{(r^2-r_H^2)^2},~~
J=\frac{\sqrt{2}\pi r_H^2(R^2+r_H^2)^3\sqrt{R^4+r_H^4} }{R(R^2-r_H^2)^3},
\end{eqnarray}
while for the quantities which encode the 
deformation of the 
horizon one finds
\begin{eqnarray}
&&
L_e=\frac{4\pi r_H^2 R}{\sqrt{R^4+r_H^4}},~~
L_p=\frac{8R r_H^2}{R^2-r_H^2}E(\frac{4R^2r_H^2}{(R^2+r_H^2)^2}),
\\
\nonumber
&&
R_{1}^{(in)}=\frac{\sqrt{2}\sqrt{R^4+r_H^4}}{R},~~
R_{1}^{(out)}=\frac{\sqrt{2}(R^2+r_H^2)^2\sqrt{R^4+r_H^4}}{R(R^2-r_H^2)^2}.
\end{eqnarray}
with $E(x)$ the complete elliptic integral of the second kind.

\setcounter{figure}{3}
\begin{figure}[t!]
\setlength{\unitlength}{1cm}
\begin{picture}(15,18)
\put(-1.,0){\epsfig{file=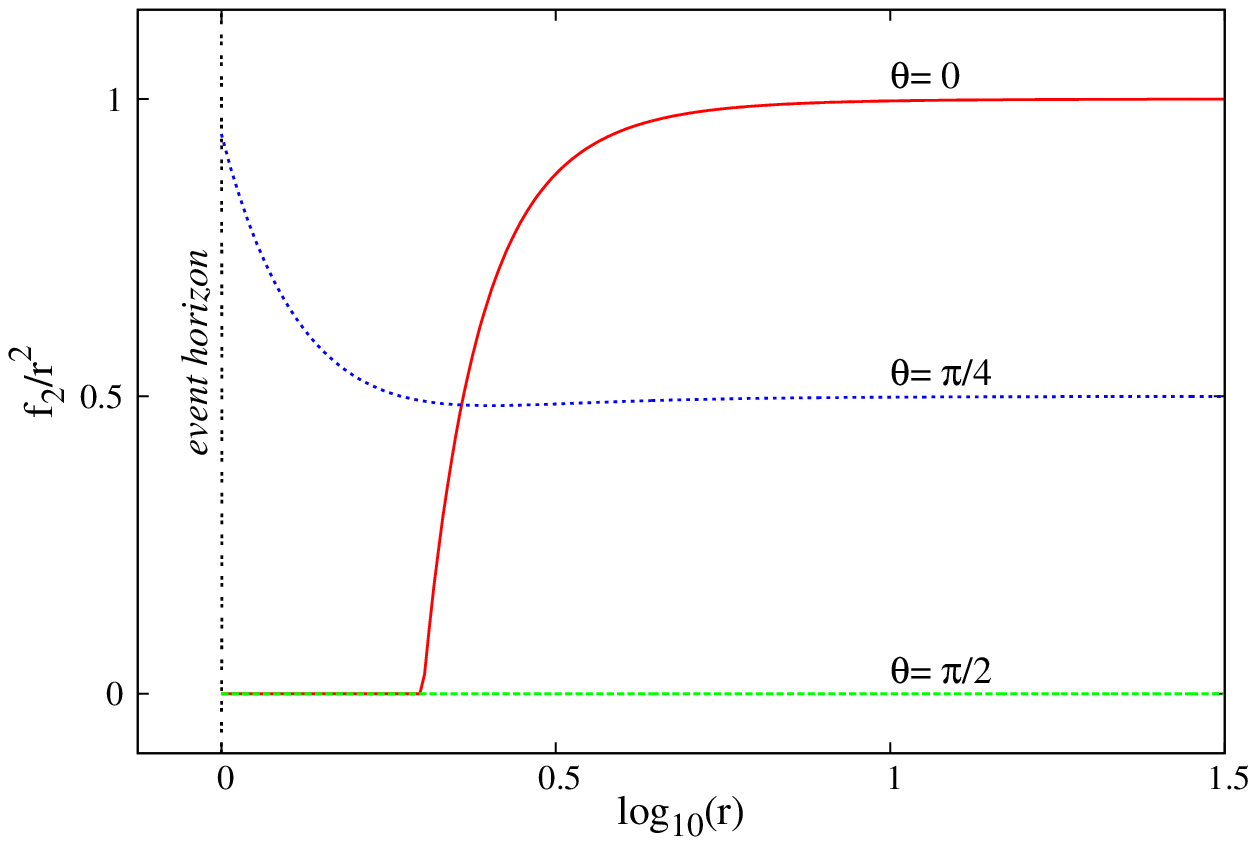,width=8cm}}
\put(7,-0.5){\epsfig{file=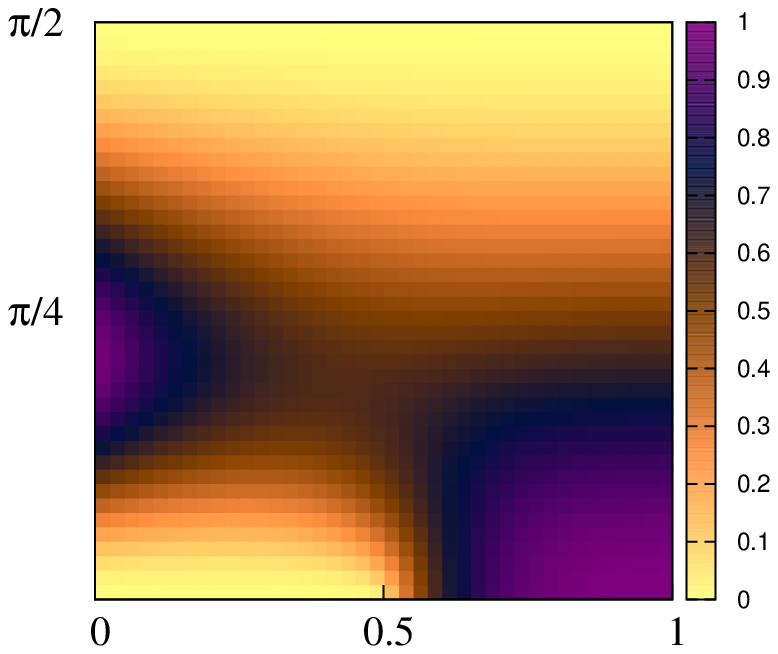,width=10cm}}
\put(-1,6){\epsfig{file=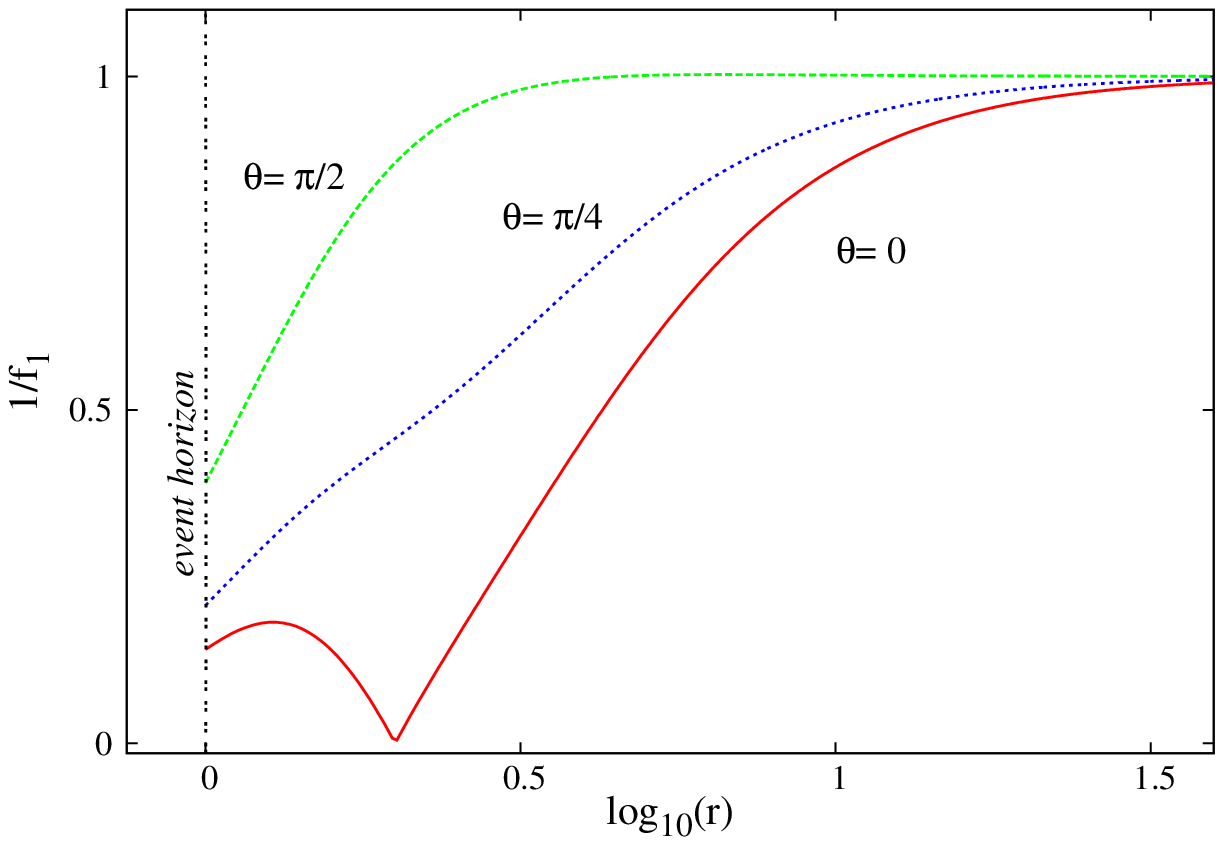,width=8cm}}
\put(7,5.5){\epsfig{file=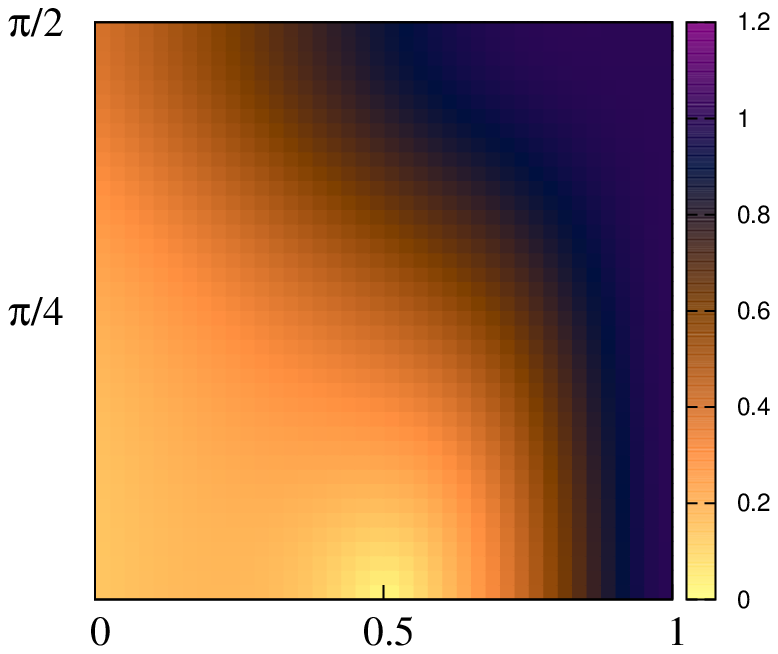,width=10cm}}
\put(-1,12){\epsfig{file=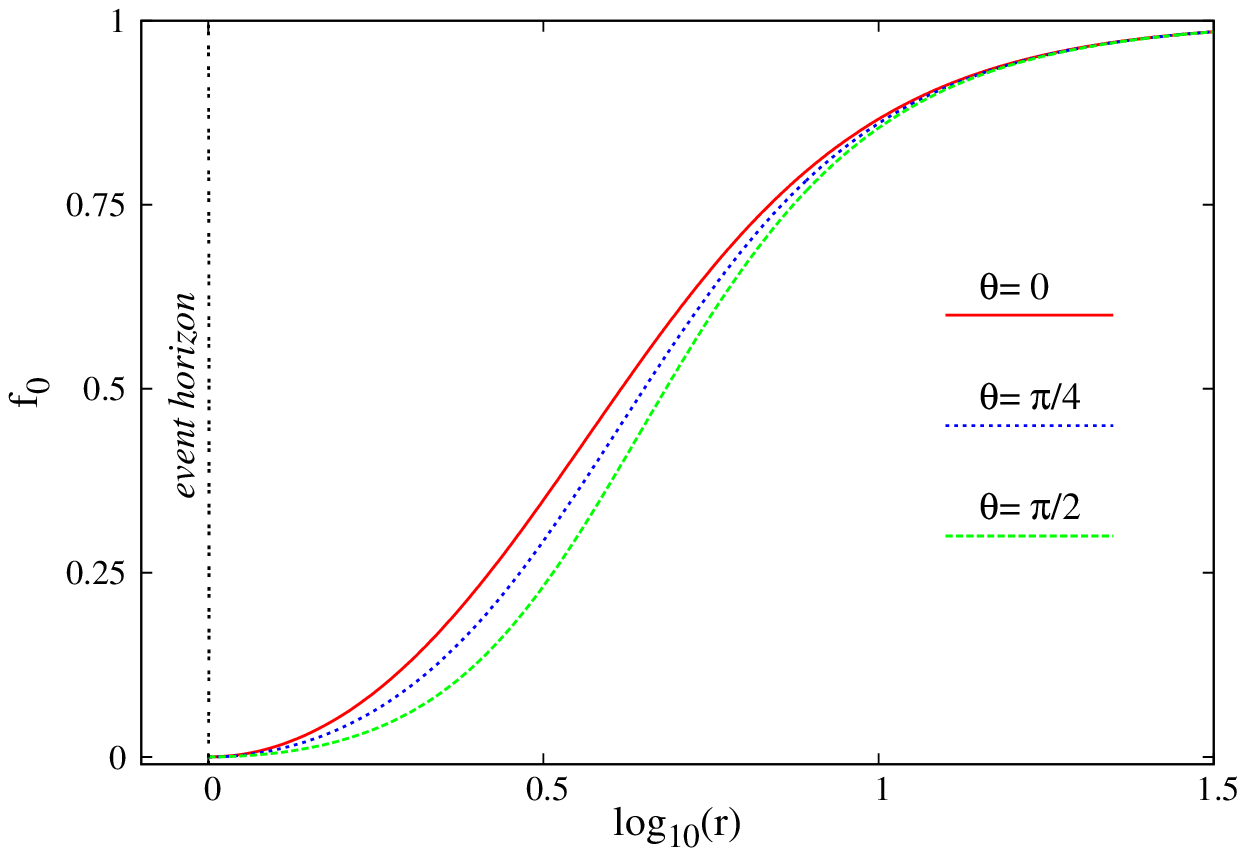,width=8cm}}
\put(7,11.5){\epsfig{file=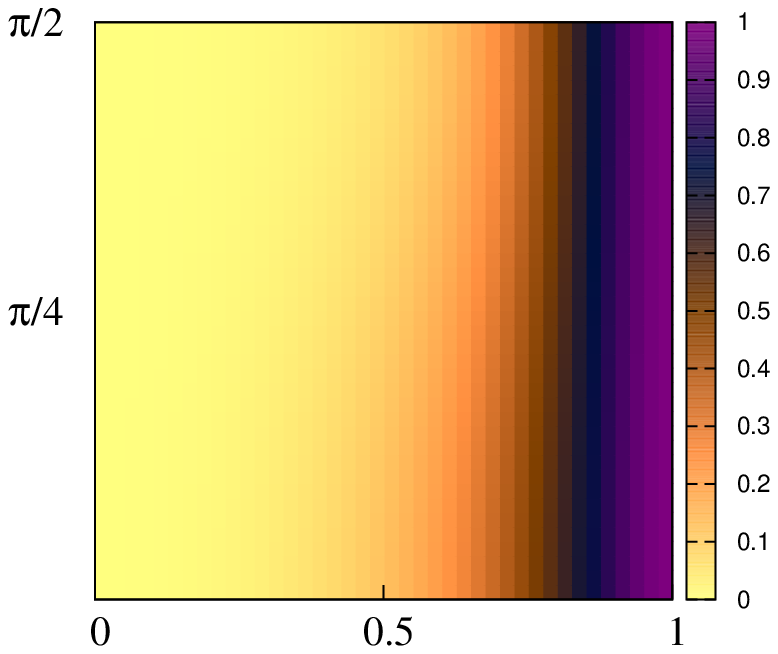,width=10cm}}
\end{picture}
\caption{
The metric functions  $f_i,W$ and the 
Kretschmann scalar $K=R_{\mu\nu\alpha \beta}R^{\mu\nu\alpha \beta}$  are shown for a $d=5$ Emparan-Reall black ring with 
$r_H=1$,
$R=2$
and
$\Omega_H\simeq 0.205$.
}
 \end{figure}

\setcounter{figure}{3}

\begin{figure}
\setlength{\unitlength}{1cm}
\begin{picture}(15,18)
\put(-1.,0){\epsfig{file=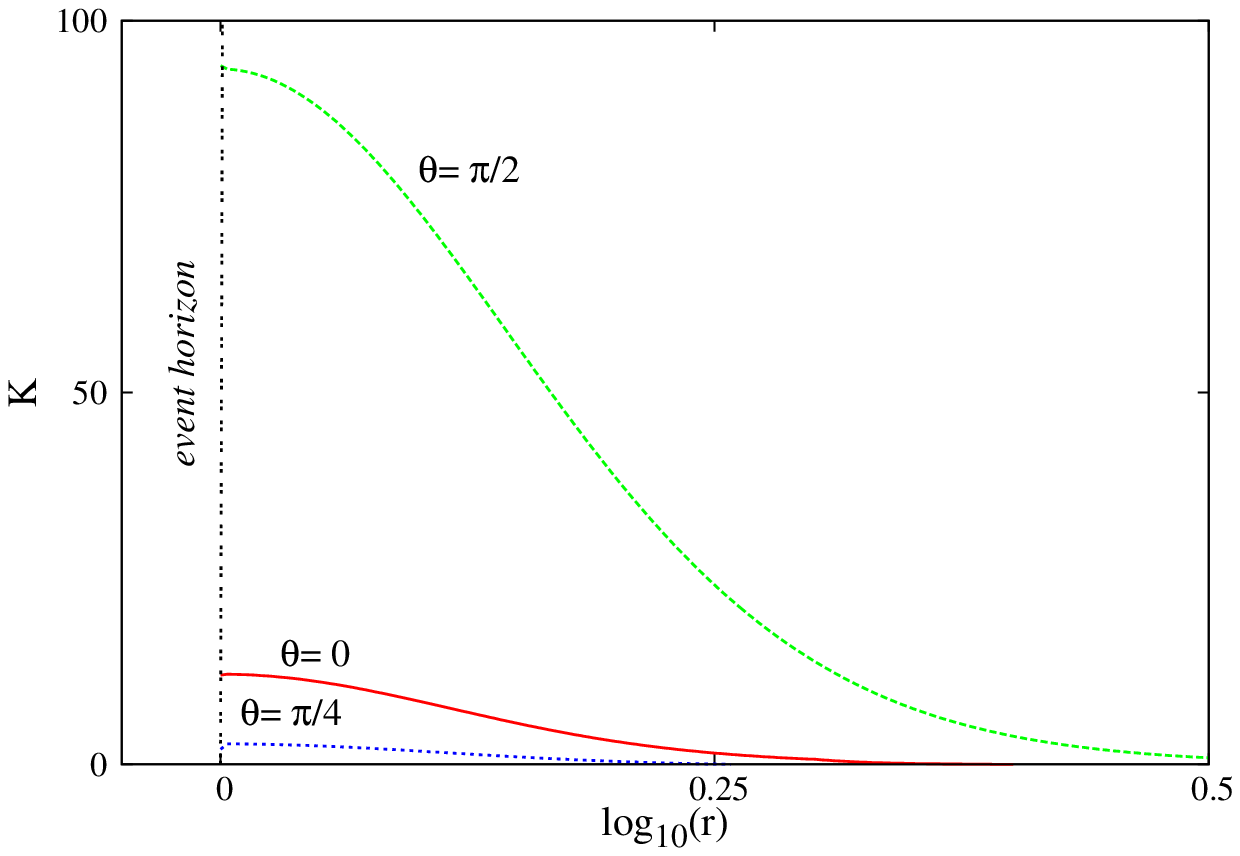,width=8cm}}
\put(7,-0.5){\epsfig{file=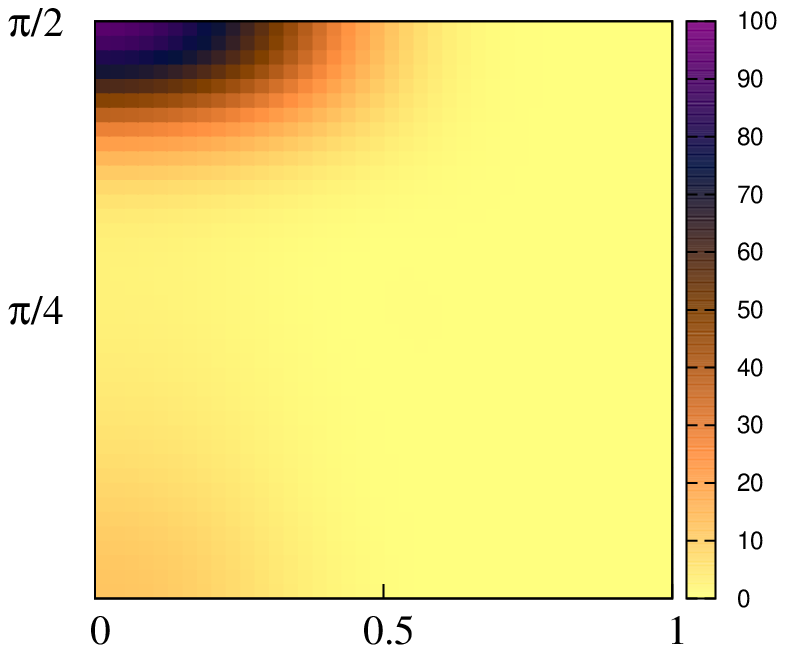,width=10cm}}
\put(-1,6){\epsfig{file=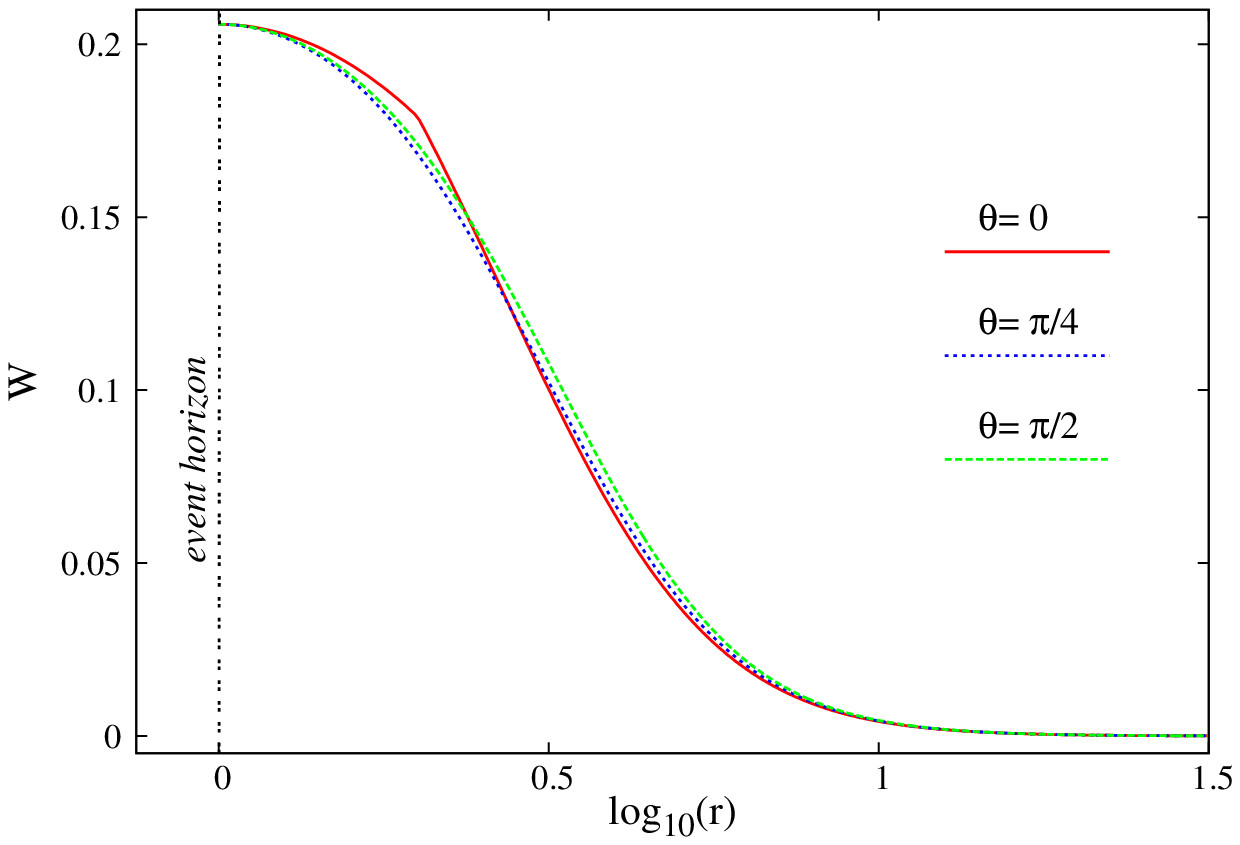,width=8cm}}
\put(7,5.5){\epsfig{file=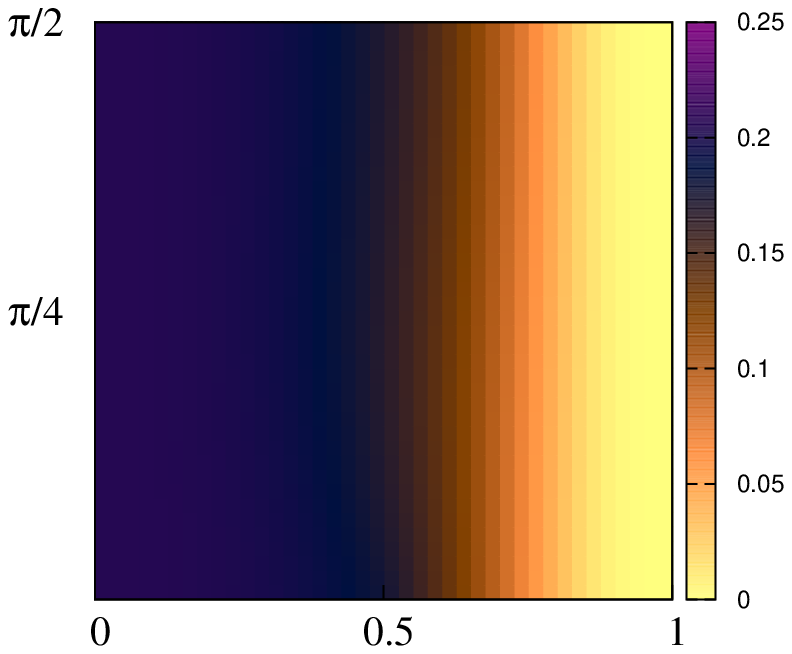,width=10cm}}
\put(-1,12){\epsfig{file=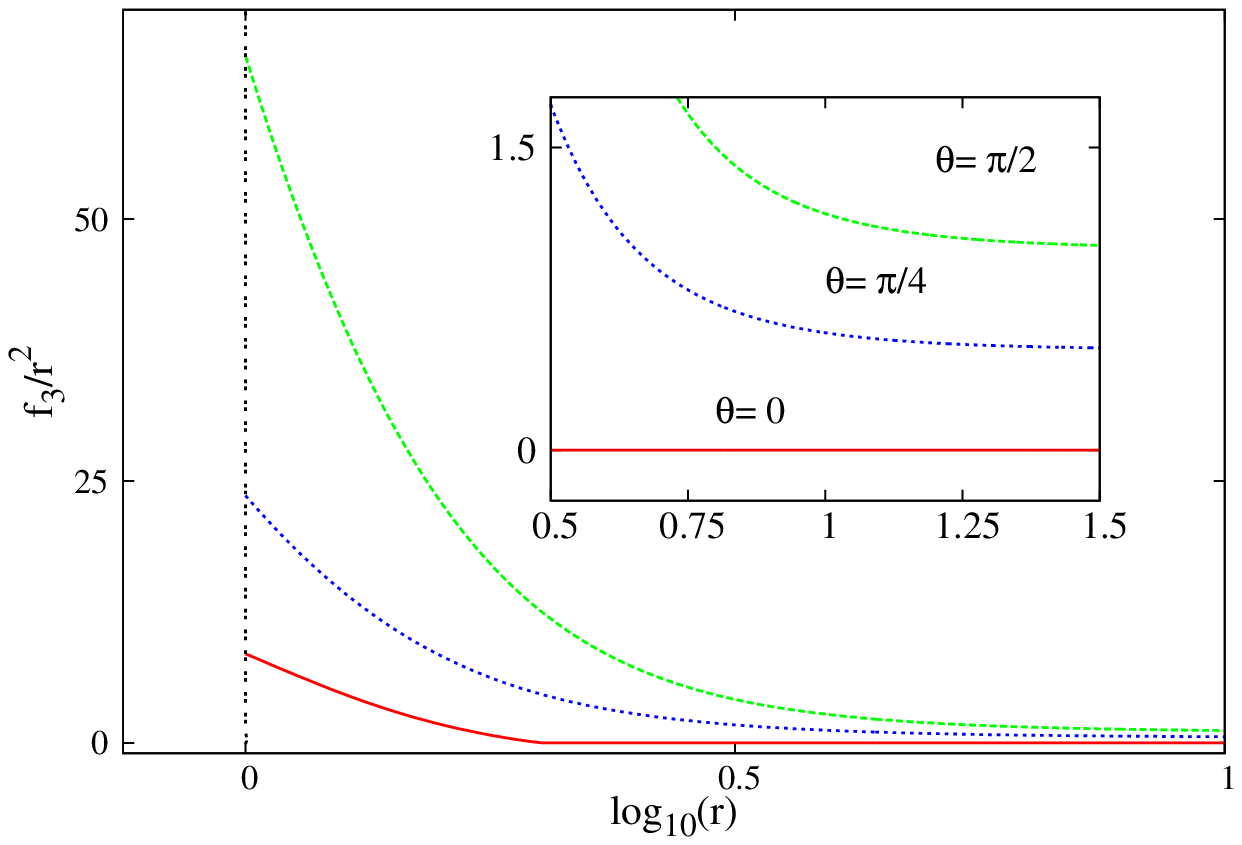,width=8cm}}
\put(7,11.5){\epsfig{file=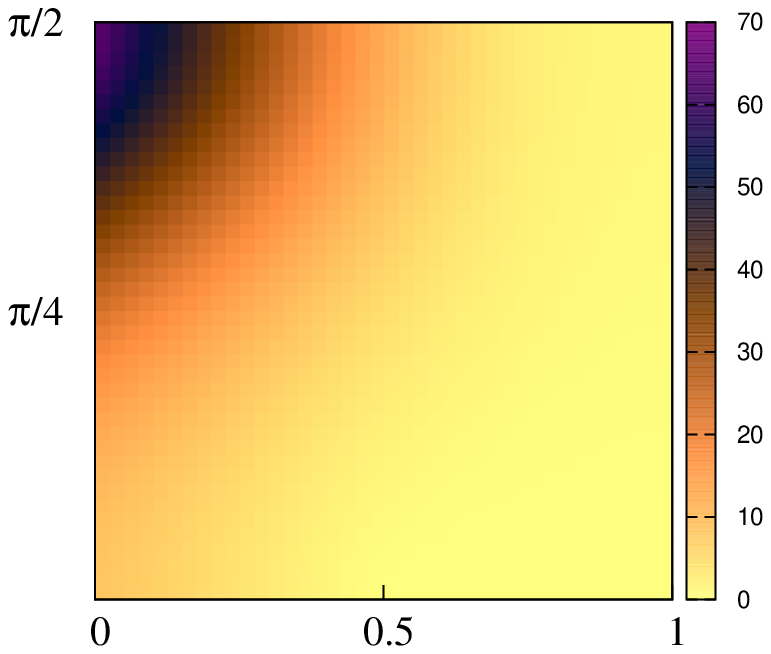,width=10cm}}
\end{picture}
\caption{
continued.
}
\end{figure}

Detailed discussions of 
the properties of this solution have appeared in various places in the literature,
see $e.g$ the review work \cite{Emparan:2006mm}.
Here we shall briefly mention only some features which occur later
when discussing the numerical solutions.

Let us start by observing that the expression above 
(including those in 
Appendix B) hold for a balanced BR.
However, unbalanced solutions exist as well, 
possessing one more free
 parameter.
That is, for given $(r_H,R)$,
BRs without conical singularities
are found for a single value of $\Omega_H$ only,
as given by (\ref{cER1}).
Also, 
the BRs with $R\gg r_H$ (and thus, from (\ref{cER1}), with large $J$)
 effectively become
boosted black strings,
being well described by the blackfold

\newpage
\setlength{\unitlength}{1cm}
\begin{picture}(8,6)
\put(-1,0.0){\epsfig{file=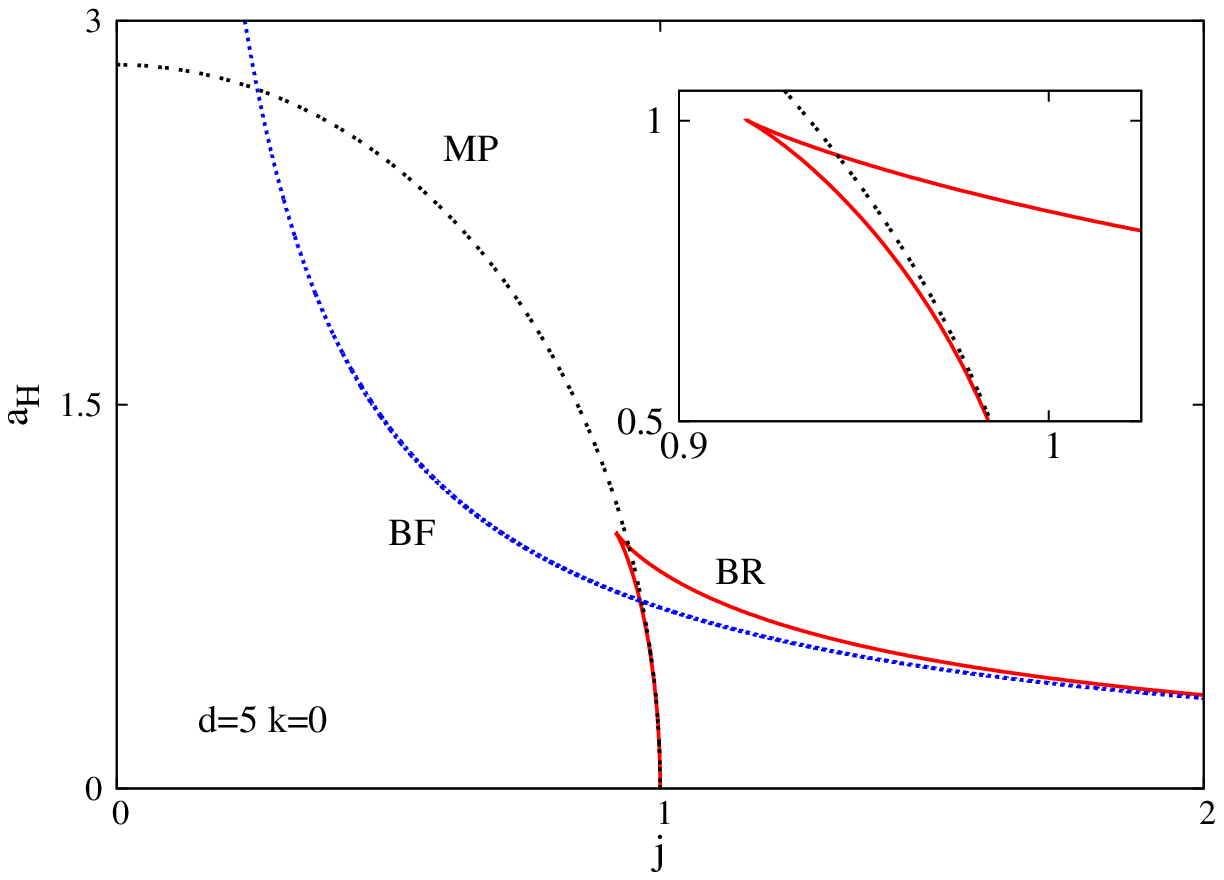,width=8cm}}
\put(7,0.0){\epsfig{file=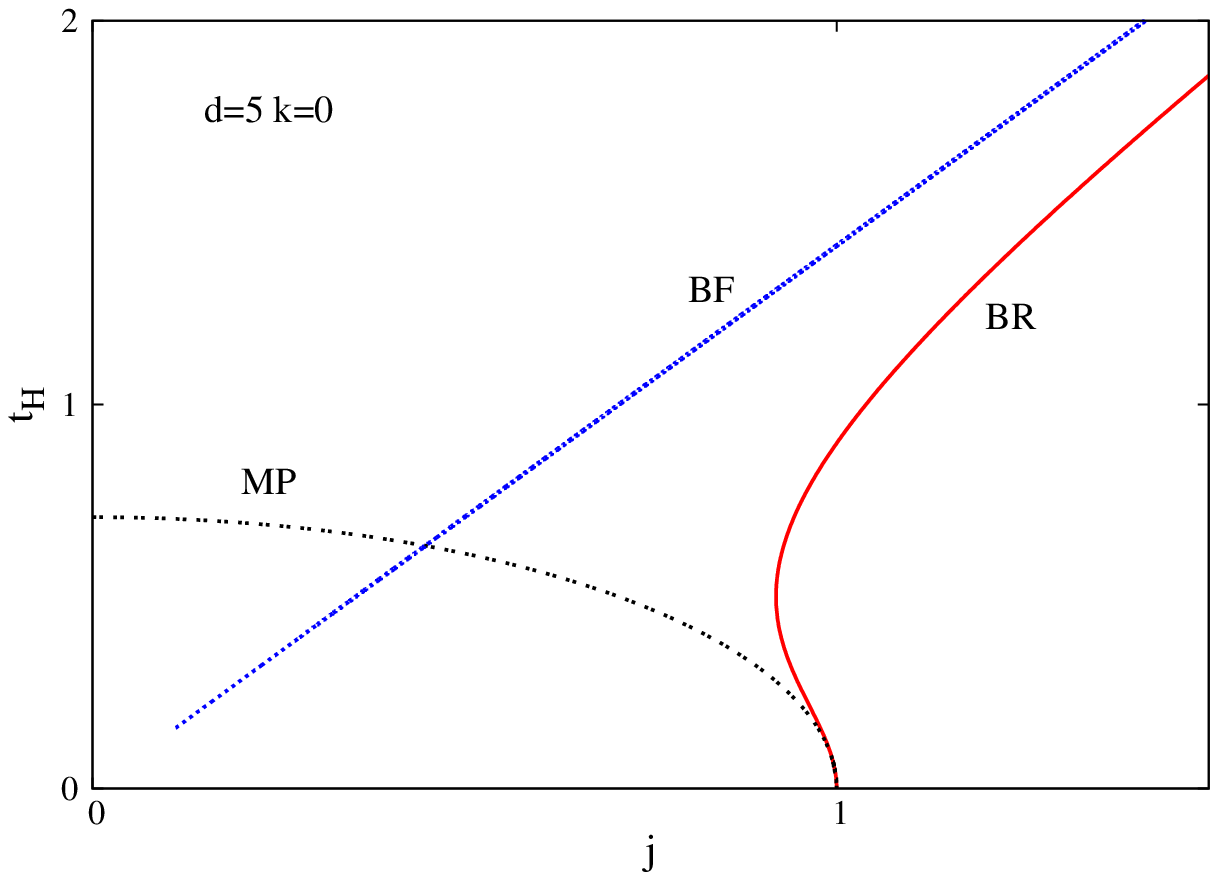,width=8cm}}
\end{picture}
\vspace{0.5cm}
  
\setlength{\unitlength}{1cm}
\begin{picture}(8,6)
\put(2,0.0){\epsfig{file=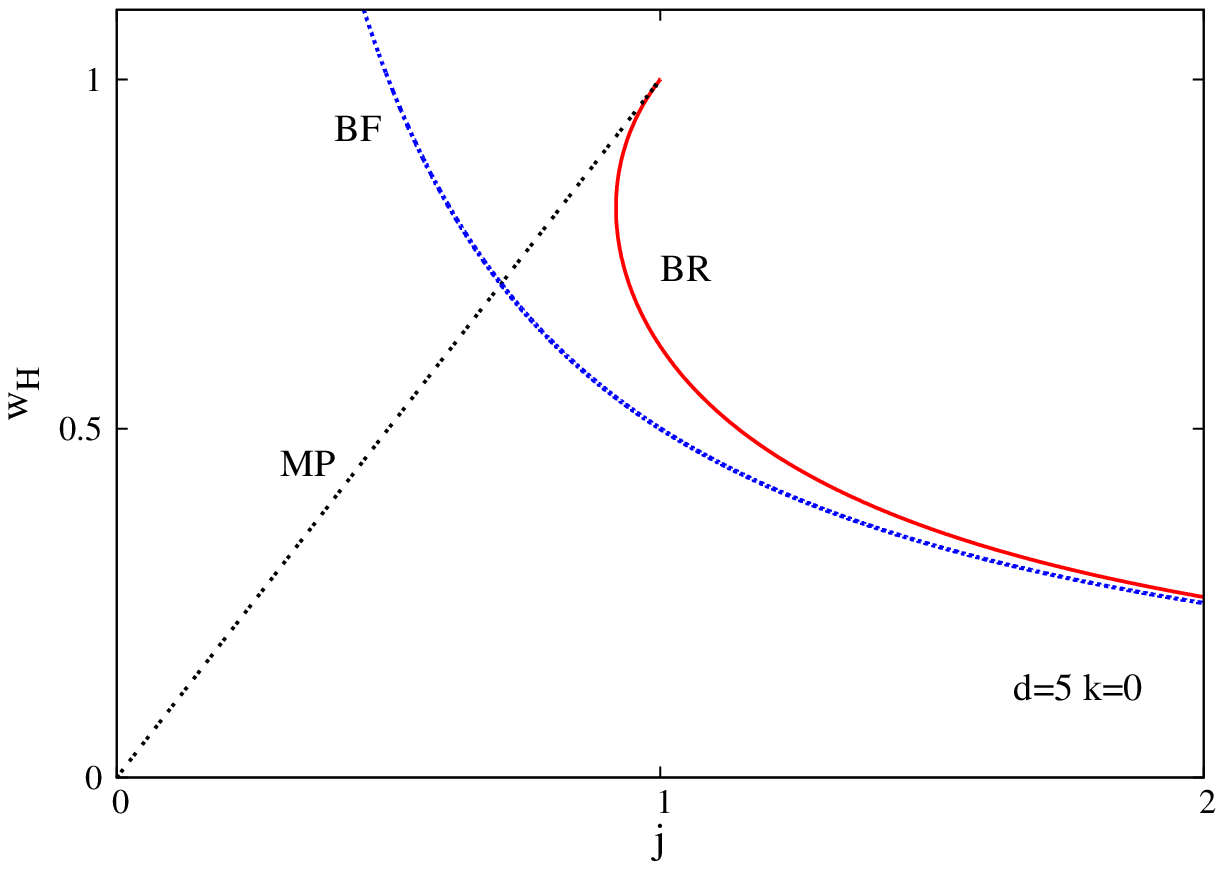,width=8cm}} 
\end{picture}
\\
\\
{\small {\bf Figure 5.}
The  reduced area $a_H$, the reduced temperature $t_H$ and the reduced angular velocity $w_H$  
are shown $vs.$
the reduced angular momentum $j$ for $d=5$
black rings (BR) and Myers-Perry (MP) black holes.
For comparison with the higher dimensional case,
we include here also the lowest order blackfold (BF) prediction.}
\vspace{0.5cm}
 \\
  formalism.
(Note that
the angular momentum of the
BR (for fixed mass) is bounded below, but not above.)

However, perhaps the most unexpected feature of the BRs is
the existence of
two branches, which branch off from a cusp at $(j^2, a_H) = (27/32, 1)$.
The existence and the properties of the 
branch of `fat' BRs cannot be predicted by
the blackfold approach.
It has a small extent, meeting at 
 $(j^2, a_H) = (1, 0)$ the
$k=0,~d=5$ singular MP solution.

These features are shown in Figure 5;
although these plots can be found in  the literature,
we have included them here since it is interesting to contrast the situation with the 
higher dimensional case.

\setlength{\unitlength}{1cm}
\begin{picture}(8,6) 
\put(-1.,0.0){\epsfig{file=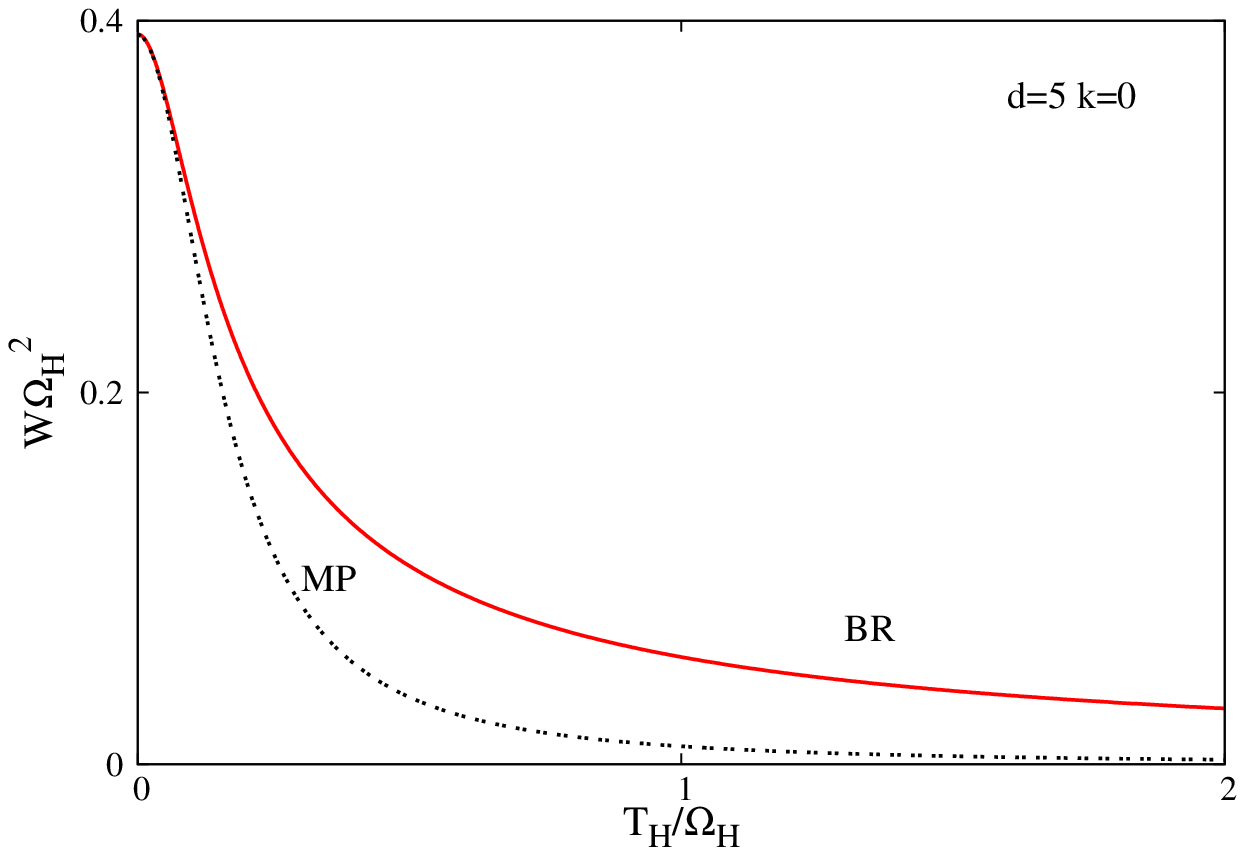,width=8cm}}
\put(7,0.0){\epsfig{file=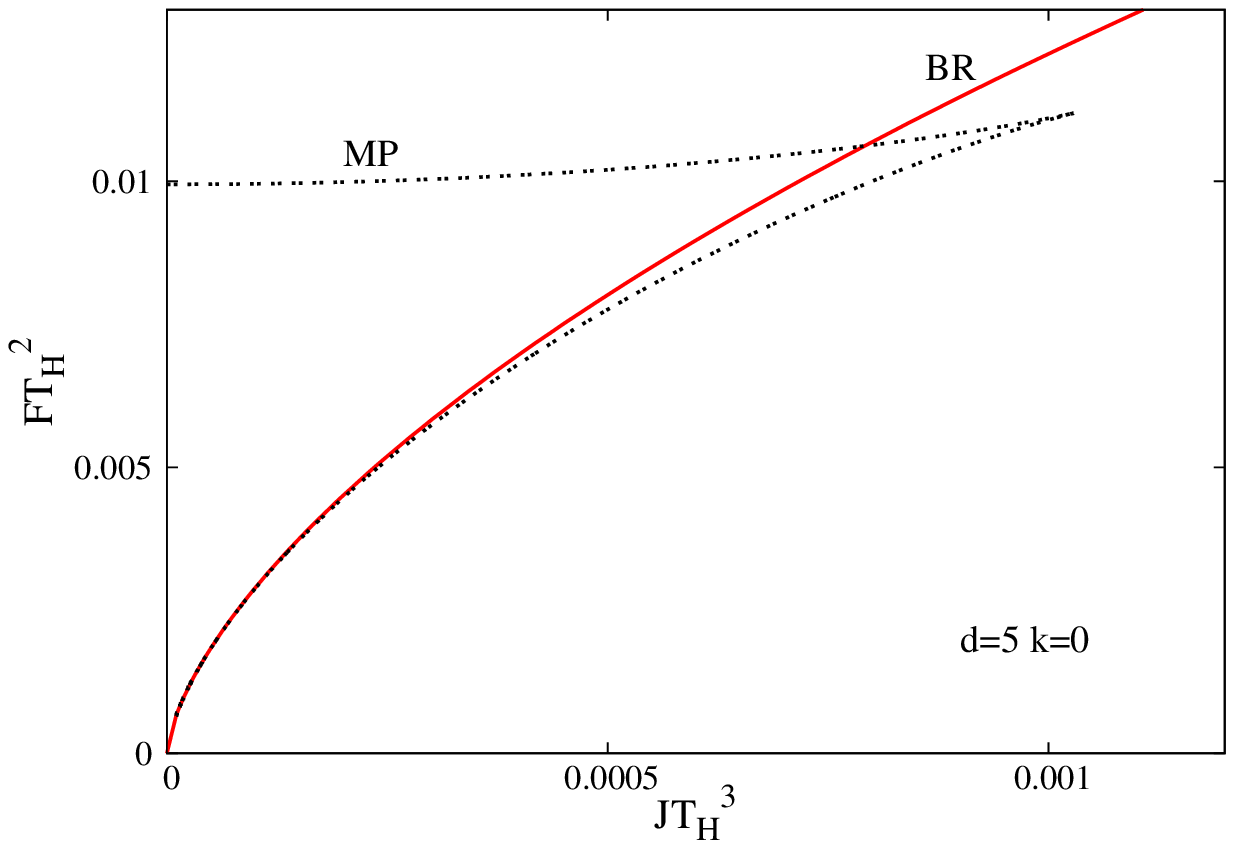,width=8cm}}
\end{picture}
\\
\\
{\small {\bf Figure 6.}
{\it Left}: The grand canonical potential $W$
is shown as a function of the Hawking temperature $T_H$
for $d=5$ black rings (BR) and Myers-Perry (MP)
black holes with fixed angular velocity of the horizon $\Omega_H$.
{\it Right}:
The canonical potential $F$
is shown as a function of the angular momentum $J$
for $d=5$ black rings with fixed Hawking temperature $T_H$.
}
\vspace{0.5cm}

\setlength{\unitlength}{1cm}
\begin{picture}(8,6)
\put(2,0.0){\epsfig{file=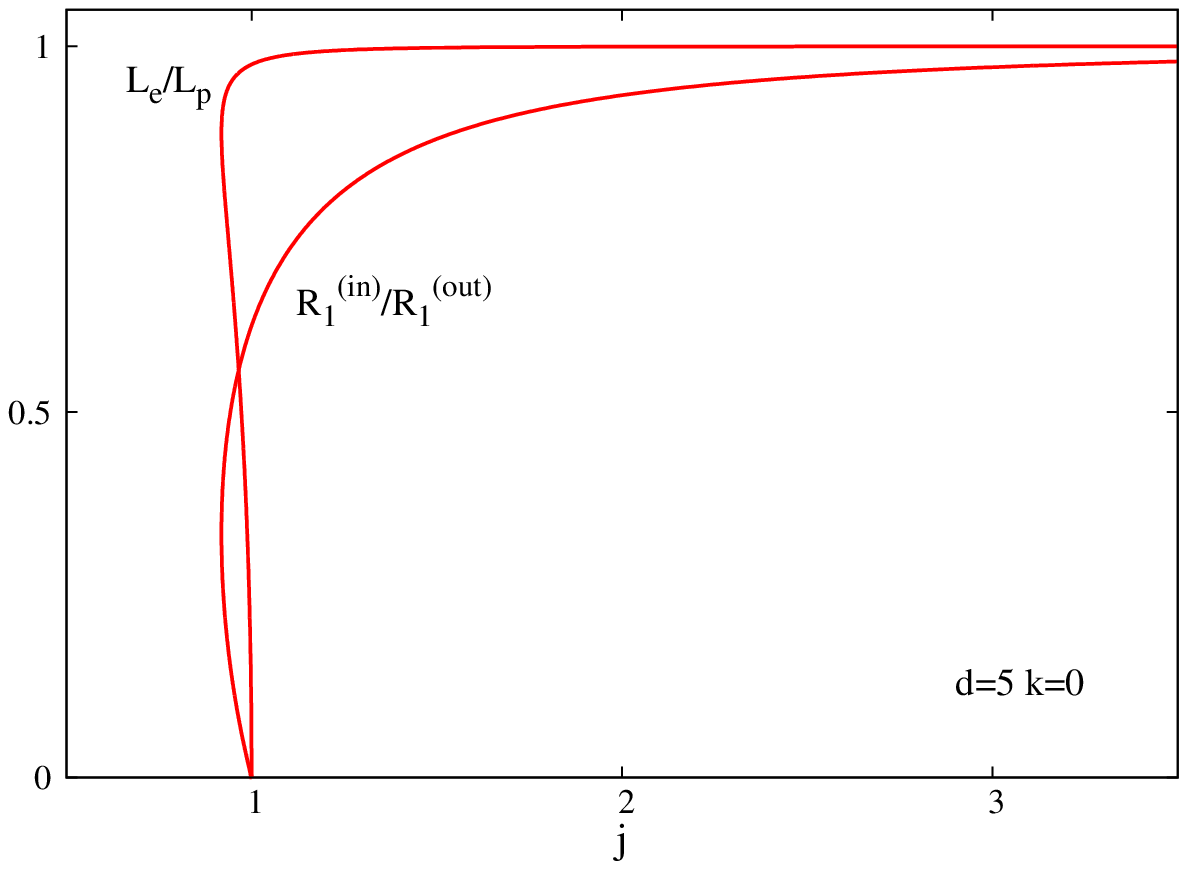,width=8cm}} 
\end{picture}
\\
\\
{\small {\bf Figure 7.}
The ratios $L_e/L_p$ and $R_1^{(in)}/R_1^{(out)}$, which encode the deformation of the horizon,
are shown $vs.$ the reduced angular momentum $j$ for $d = 5$ black ring solutions.}
\vspace{0.5cm}

The behaviour of the solutions in a grand canonical ensemble
is shown in Figure 6 (left).
One can notice the existence
of only one BR and one MP black hole
with the same values of $\Omega_H$ and $T_H$.
Moreover, the solutions
exist for all possible values of these variables.
Also, 
as discussed in \cite{Elvang:2006dd}, in a grand canonical potential
the MP black holes
are always thermodynamically favoured
over the BRs\footnote{This results from the fact that,
for given $T_H,~\Omega_H$, the grand canonical potential
$W$ is minimized by the MP black holes.
}.

As seen in Figure 6 (right),
a different picture is found for 
the same black objects
in a canonical potential.
The MP black holes
exhibit in this case two branches,
with a `swallowtail' structure,
while only one branch of solutions
is found for the BRs.
Note that in the region of co-existence,
the potential $F$
is minimized by a MP solution,
which is therefore thermodynamically preferred.
Also, at $(JT_H^3\simeq 0.00079,~FT_H^2\simeq 0.0106)$ 
the two curves meet and
only BRs exist for large $J$ (at fixed $T_H$).

Finally, in Figure 7 we show
the ratios $L_e/L_p$ and $R_1^{(in)}/R_1^{(out)}$ (as defined by (\ref{Lep})
and (\ref{def2}),
respectively), which encode the deformation of the horizon,
$vs.$ the reduced angular momentum $j$.
There
one can see $e.g.$ that the hole inside the ring shrinks to zero size
while 
the outer radius goes to infinity as the singular solution is approached.

\section{Black objects with non-spherical horizon topology}

\subsection{The blackfold limit}
 
The ultraspinning limit of the black objects 
with $S^{n+1}\times S^{2k+1}$ horizon topology
has been already discussed in the blackfold literature,
see $e.g.$ \cite{Emparan:2009vd}.
The results there imply the following
expressions for the reduced quantities, valid to leading order:
%
\begin{eqnarray}
\label{bf1}
&&
a_H=\frac{1}{2^{\frac{d(2k-1)+6}{(d-2k-4)(d-3)}}}\frac{1}{j^{\frac{2k+1}{d-2k-4}}},~~
t_H=(d-4)2^{\frac{d(2k-1)+6}{(d-2k-4)(d-3))}}j^{\frac{2k+1}{d-2k-4}},~~w_H=\frac{1}{2j}.
\end{eqnarray} 
The corresponding relations for the ultraspinning
MP
black holes with $n>1$ are also of interest:
\begin{eqnarray}
\label{MP-gen1}
a_H=  q_a q_j^{\frac{2(k+1)}{d-2k-5}}\frac{1}{j^{\frac{2(k+1)}{(d-2k-5)}}} ,~~
t_H=\frac{(d-2k-5)q_t}{q_j^{\frac{2(k+1)}{(d-2k-5)}}} j^{\frac{2(k+1)}{(d-2k-5)}},~~
w_H=\frac{q_jq_w}{j},
\end{eqnarray}
with the coefficients $q_a,~q_j,~q_w$ given by (\ref{MP-rel2})
(note also that the product $a_H t_H$ is constant in both cases).

From the above relations one can see that the
area decreases faster for MP black holes than for the BRs/ringoids.
That is, the black objects with a $S^{n+1}\times S^{2k+1}$ horizon topology
dominate
entropically in the ultraspinning regime.

\subsection{Non-perturbative solutions}

\boldmath
\subsubsection{$k=0$: black rings in $d=6$ dimensions}
\unboldmath

Let us start with the simplest case, $k=0$, corresponding to BRs
with rotation on the $S^1$.
The only dimension we have studied so far in a more systematic way is  $d=6$.


\setcounter{figure}{7}
\begin{figure}[t!]
\setlength{\unitlength}{1cm}
\begin{picture}(15,18)
\put(-1.,0){\epsfig{file=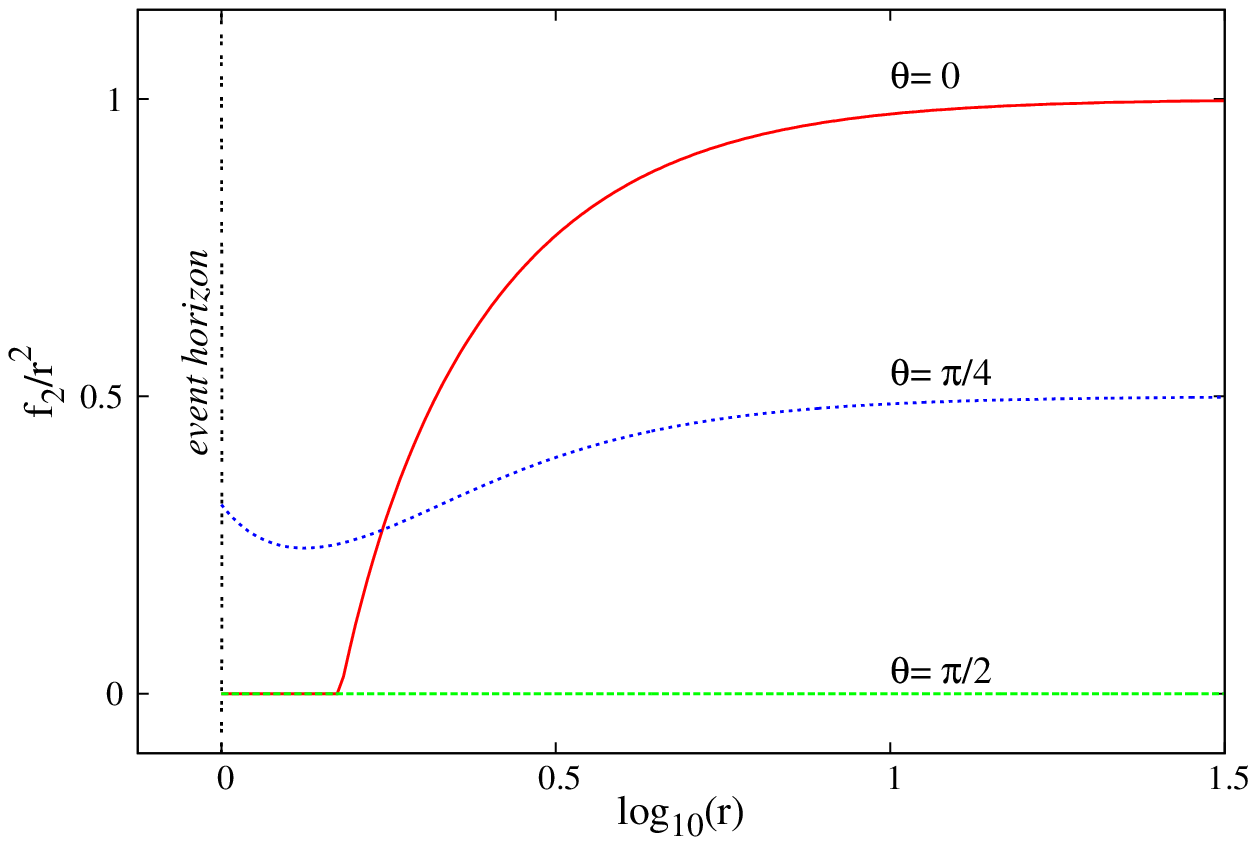,width=8cm}}
\put(7,-0.5){\epsfig{file=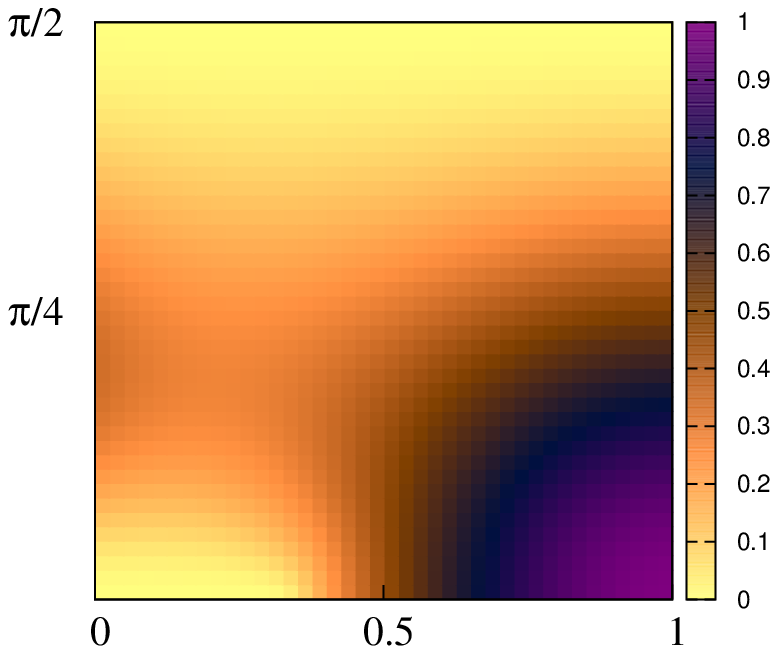,width=10cm}}
\put(-1,6){\epsfig{file=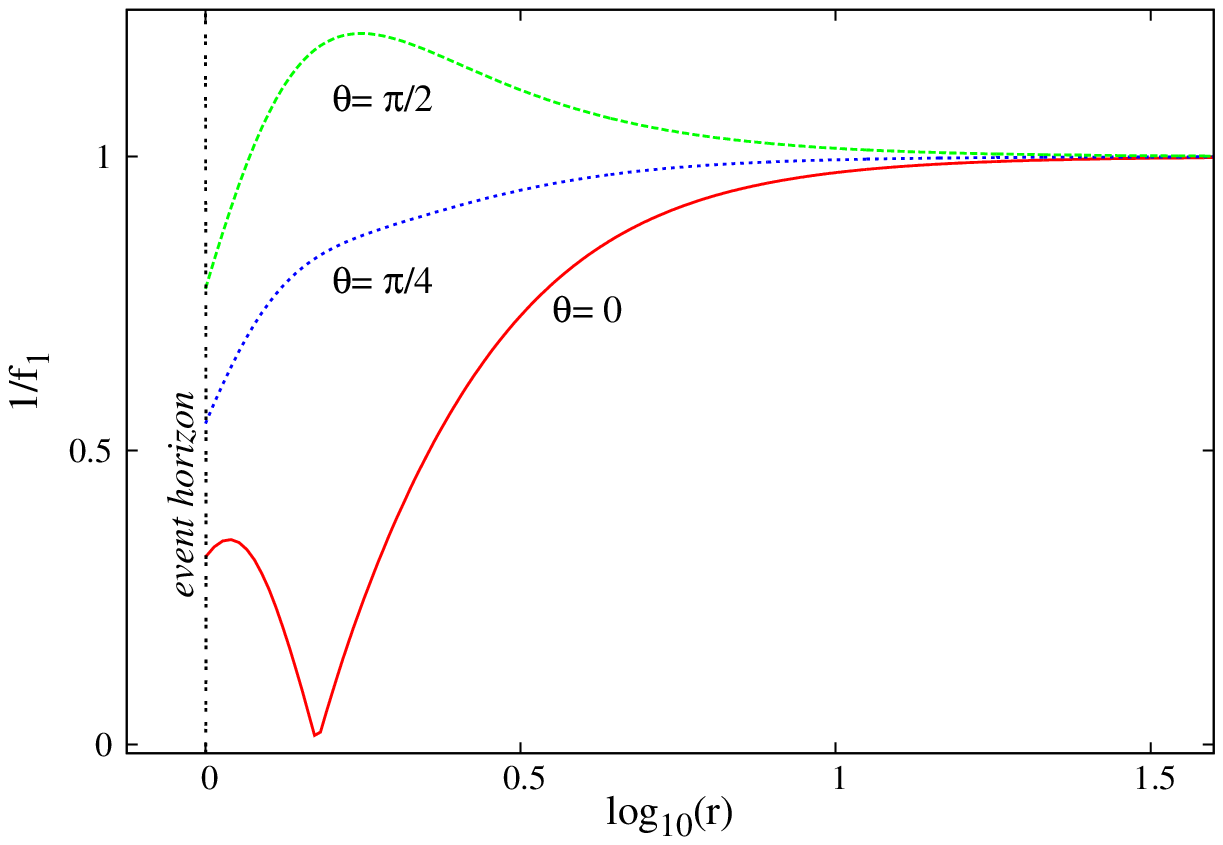,width=8cm}}
\put(7,5.5){\epsfig{file=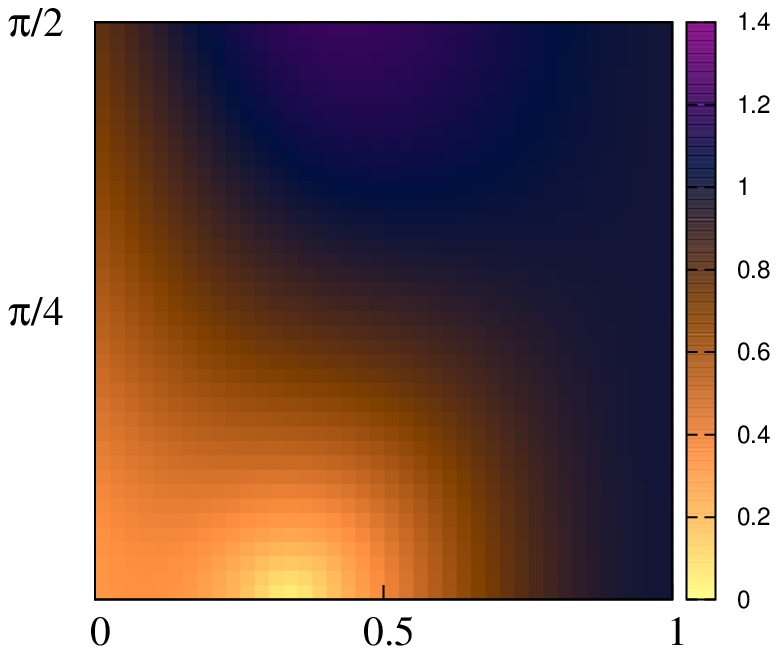,width=10cm}}
\put(-1,12){\epsfig{file=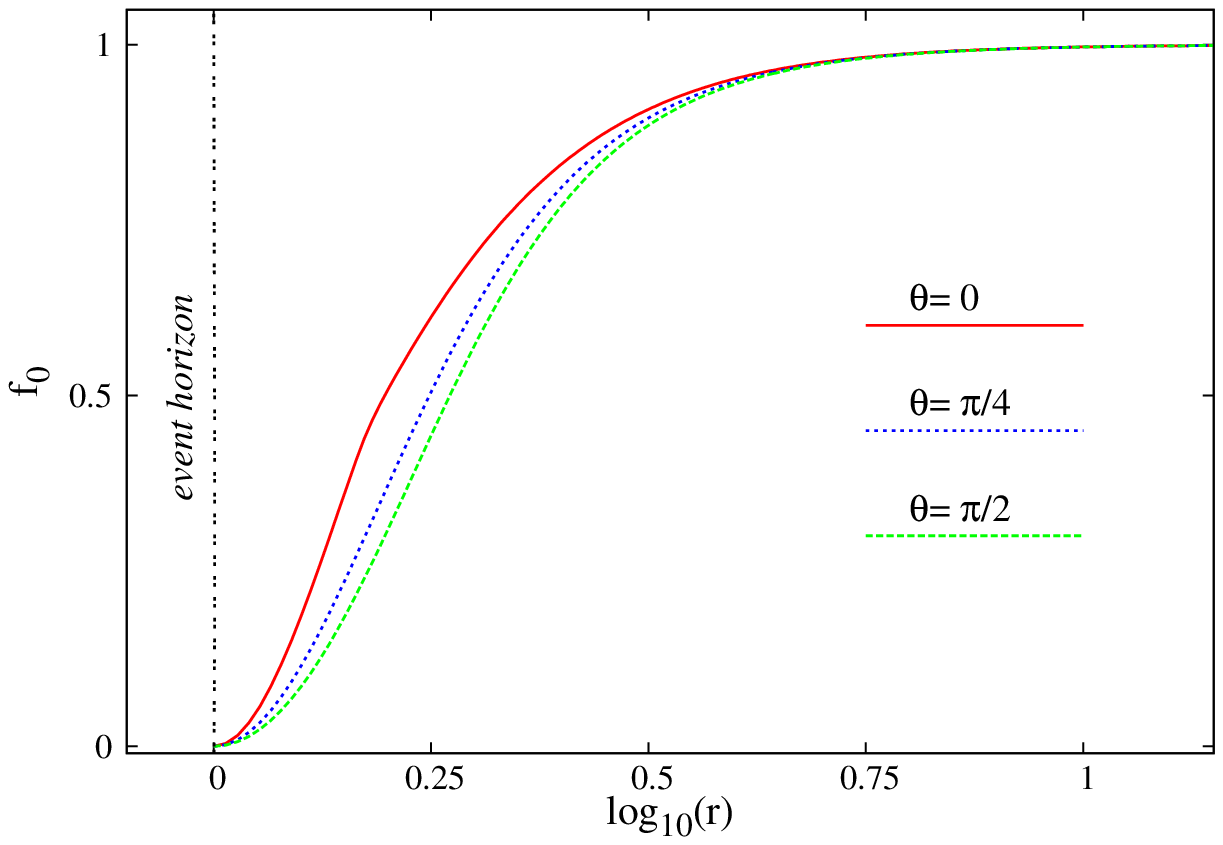,width=8cm}}
\put(7,11.5){\epsfig{file=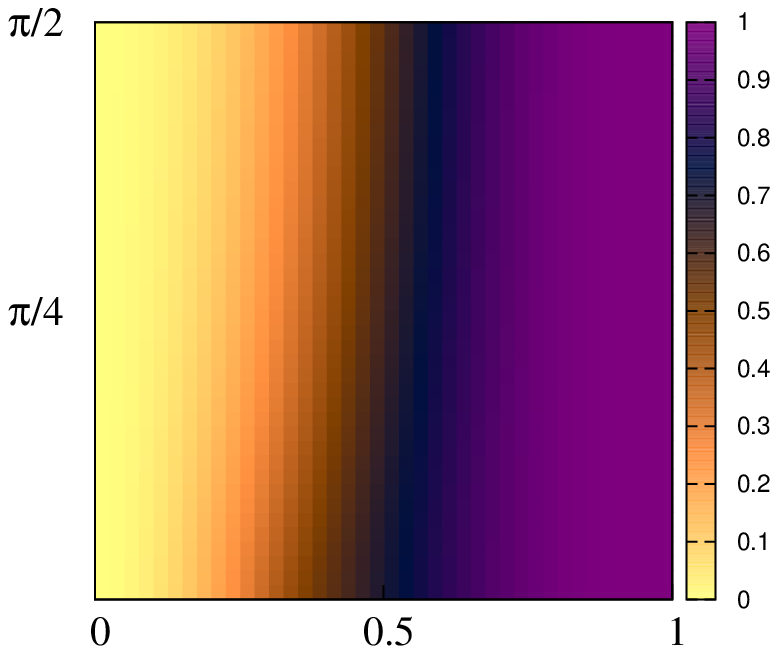,width=10cm}}
\end{picture}
\caption{
The metric functions $f_i,W$ and the Kretschmann scalar $K=R_{\mu\nu\alpha \beta}R^{\mu\nu\alpha \beta}$ 
are shown for a $d=6$ black ring with
the input parameters $r_H=1$, $R=1.504$, $\Omega_H \simeq 0.352$.
}
 \end{figure}

A discussion of the basic properties of these solutions has been given already
in Ref.~\cite{Kleihaus:2012xh};
here we return with a more detailed description. 
Let us also mention that, recently
 these solutions have been constructed independently in \cite{Dias:2014cia}.
Although the results there have been found by using a very different approach\footnote{For example,
the Ref.~\cite{Dias:2014cia} uses 
a ring-like coordinate system.
Also, the Einstein equations are solved
by  using the DeTurck method \cite{Headrick:2009pv}.}  as
 compared
 to the one described above, 
they agree well with those reported in  \cite{Kleihaus:2012xh}.

\setcounter{figure}{7}
\begin{figure}
\setlength{\unitlength}{1cm}
\begin{picture}(15,18)
\put(-1.,0){\epsfig{file=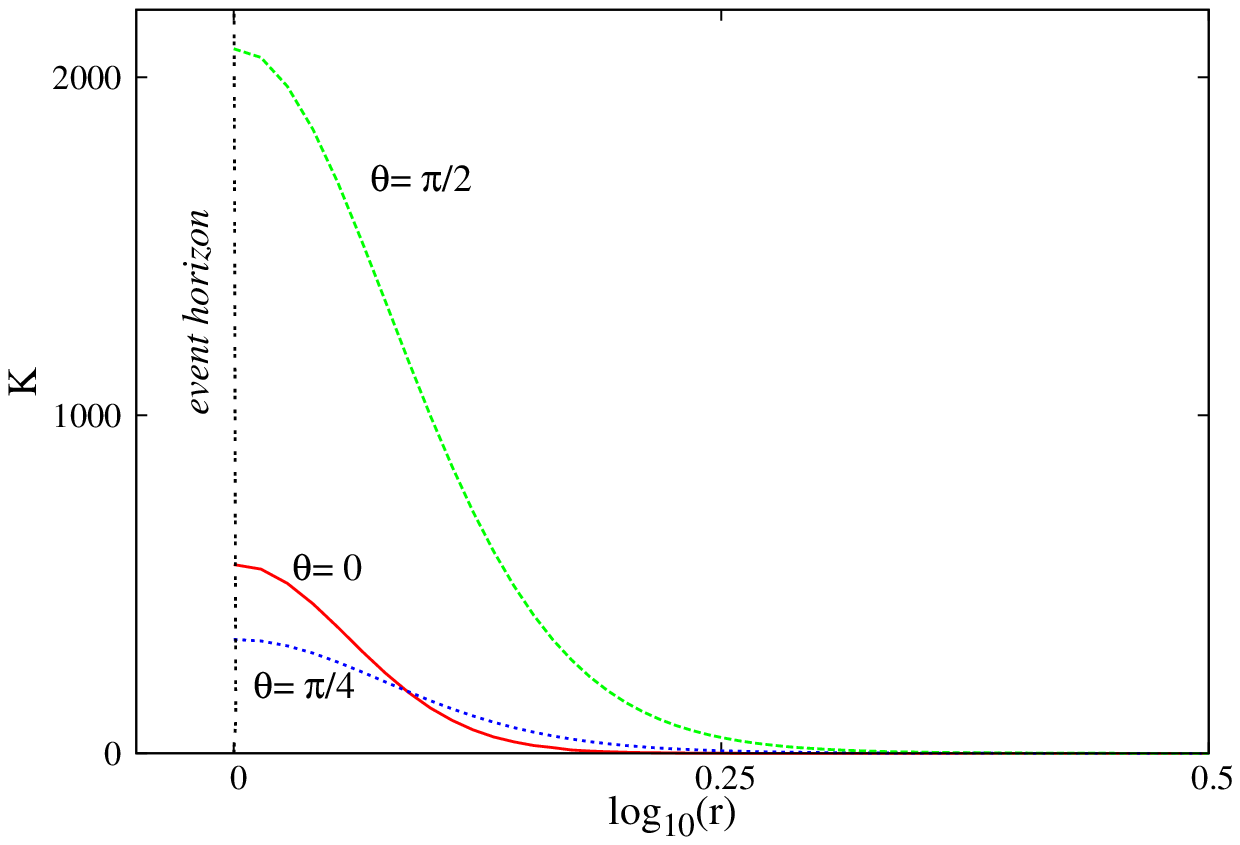,width=8cm}}
\put(7,-0.5){\epsfig{file=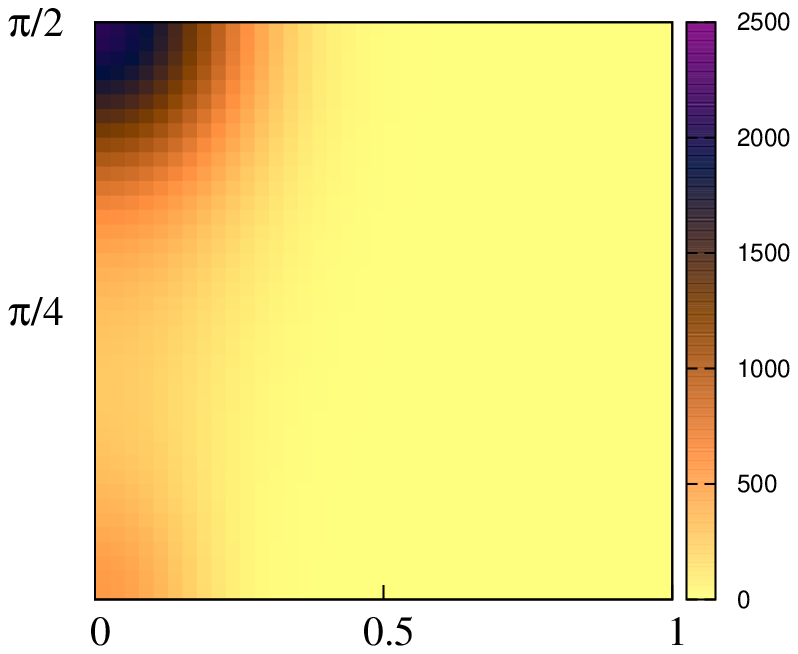,width=10cm}}
\put(-1,6){\epsfig{file=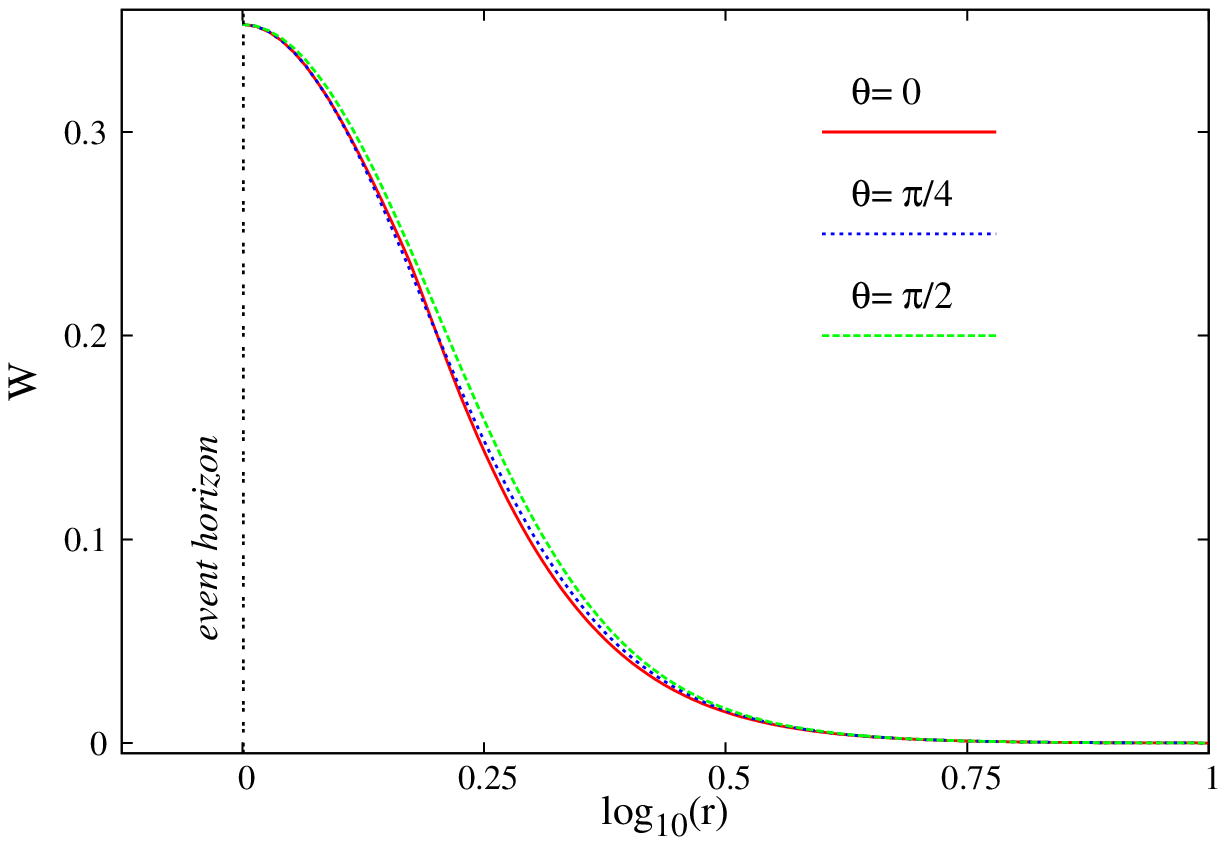,width=8cm}}
\put(7,5.5){\epsfig{file=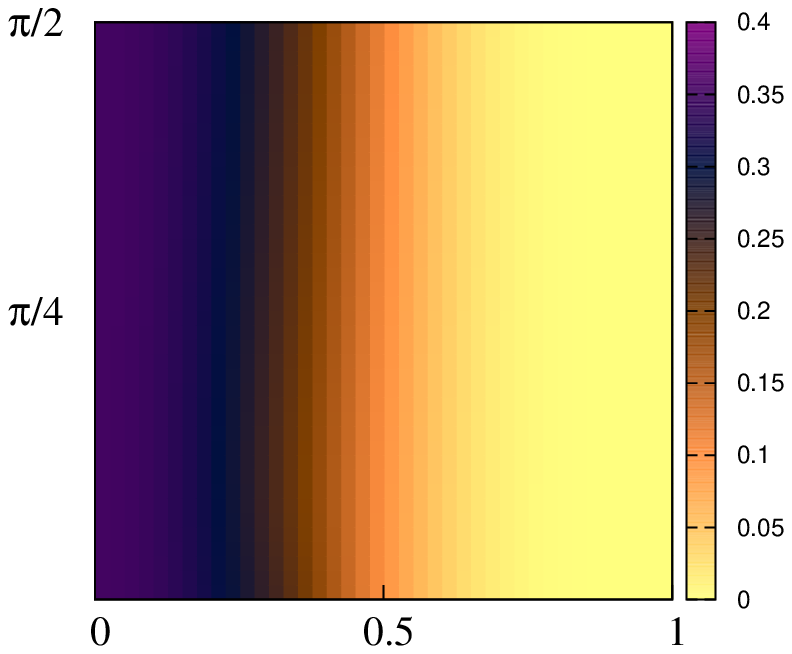,width=10cm}}
\put(-1,12){\epsfig{file=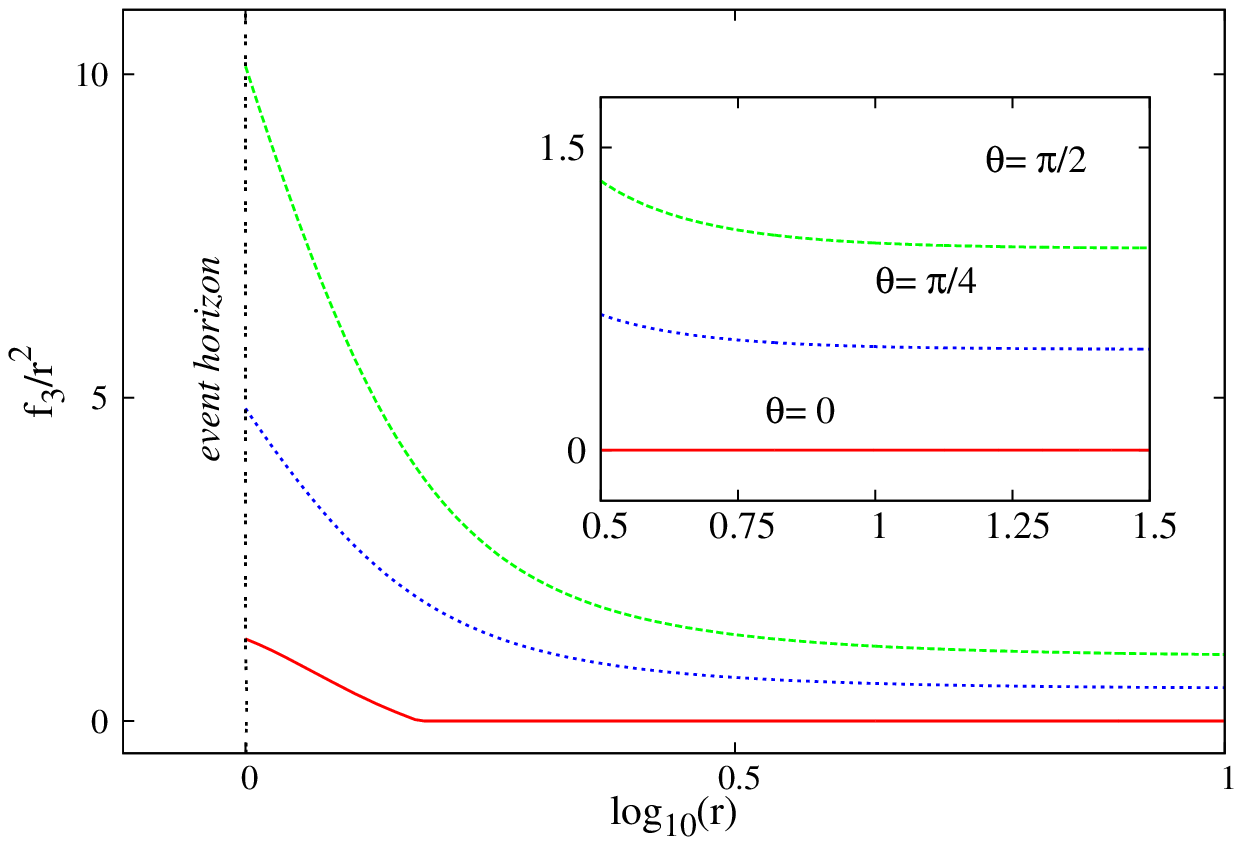,width=8cm}}
\put(7,11.5){\epsfig{file=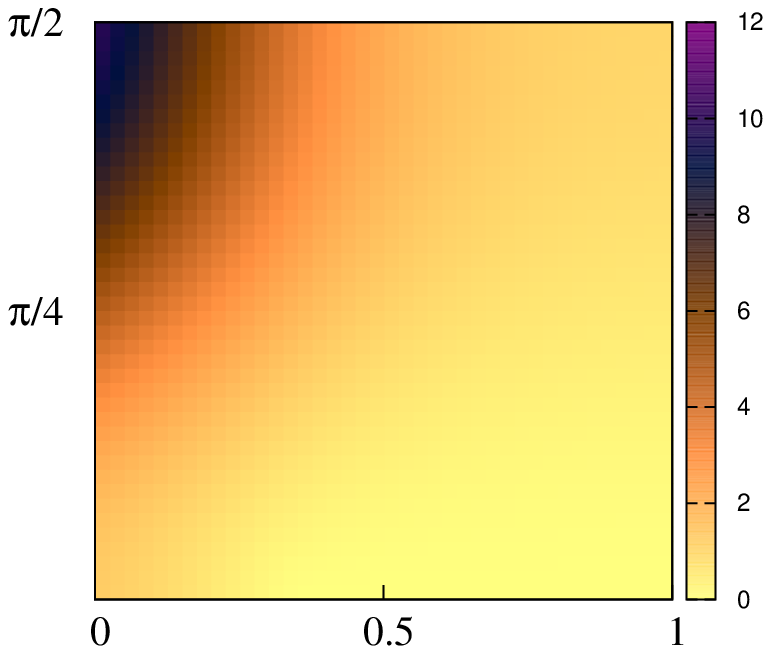,width=10cm}}
\end{picture}
\caption{
continued.
}
\end{figure}

 %
\setlength{\unitlength}{1cm}
\begin{picture}(8,6)
\put(-1,0.0){\epsfig{file=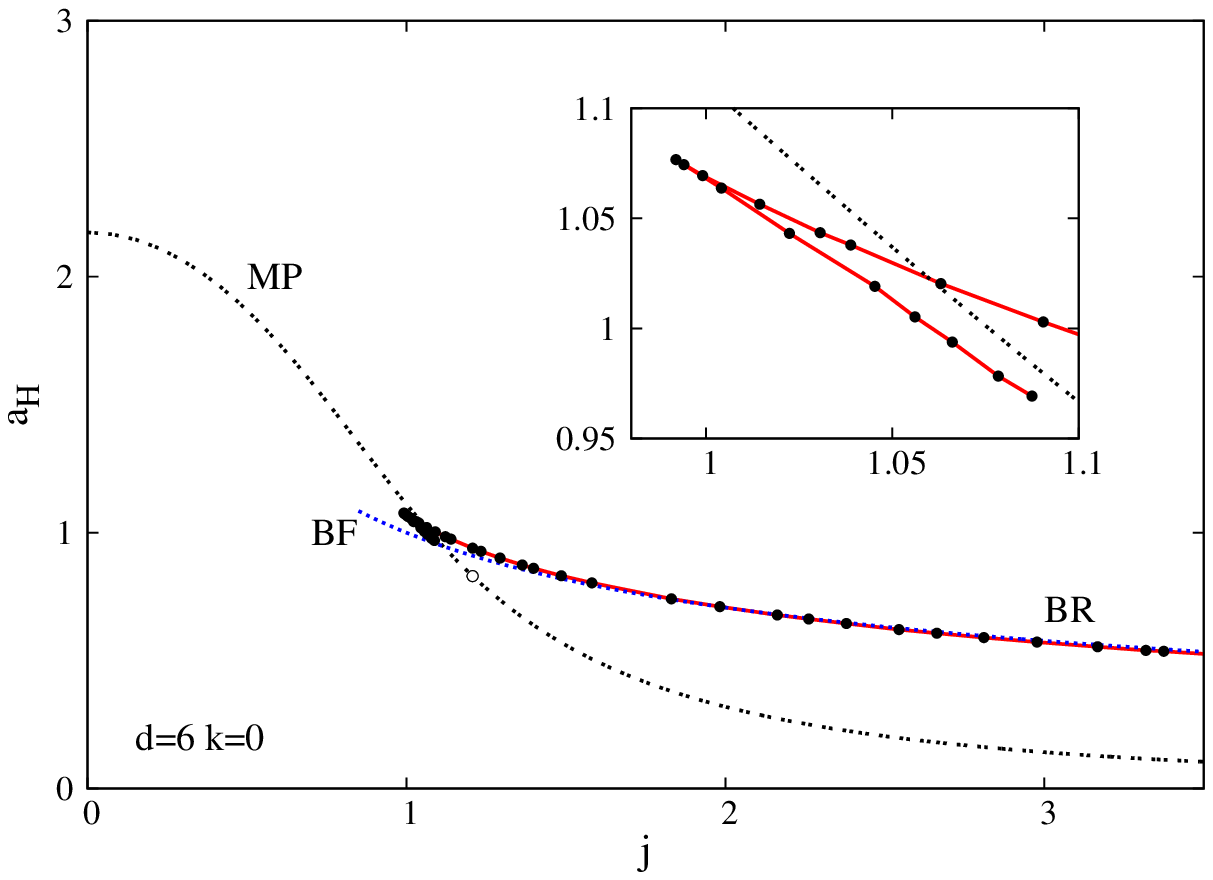,width=8cm}}
\put(7,0.0){\epsfig{file=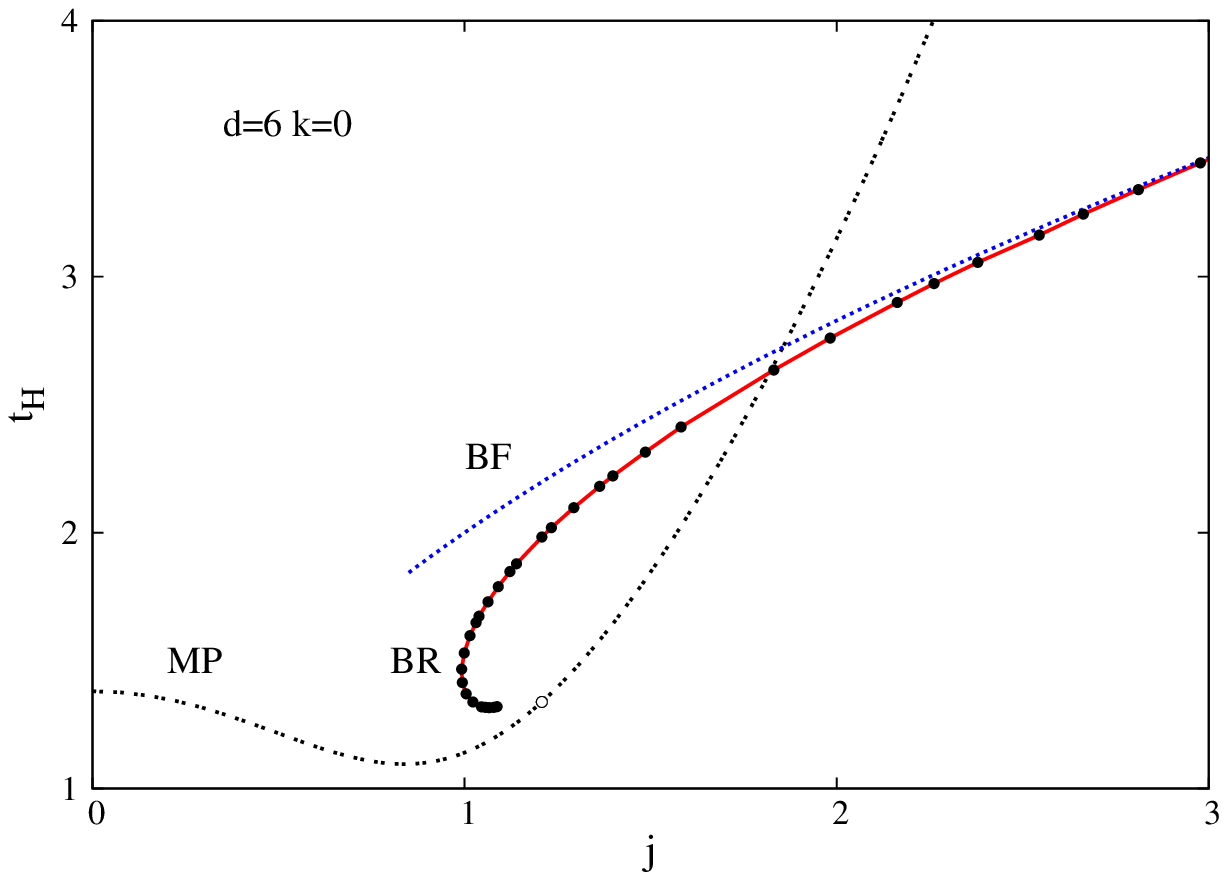,width=8cm}}
\end{picture}
 \vspace{0.5cm}

\ 
\setlength{\unitlength}{1cm}
\begin{picture}(8,6)
\put(2,0.0){\epsfig{file=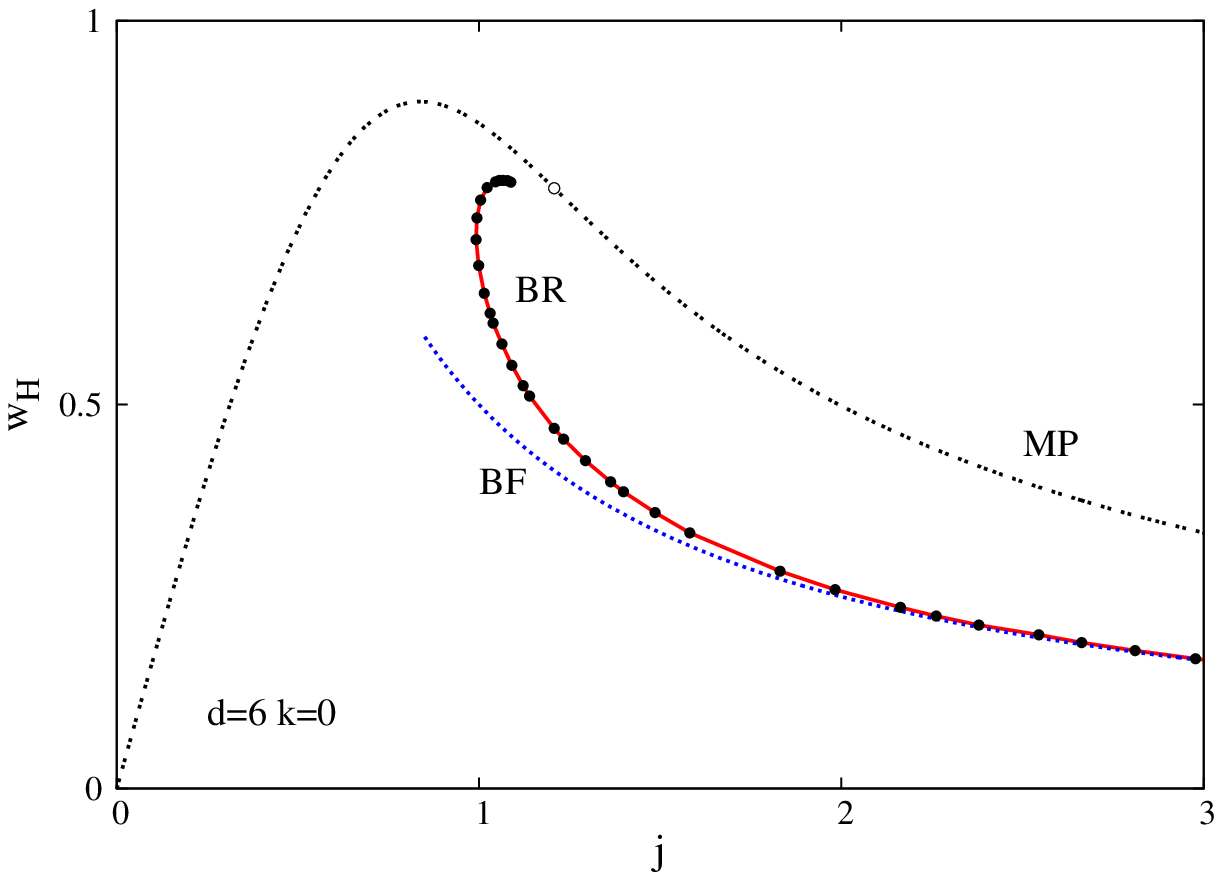,width=8cm}} 
\end{picture}
\\
\\
{\small {\bf Figure 9.}
The  reduced area $a_H$, the reduced temperature $t_H$ and the reduced angular velocity $w_H$  
are shown $vs.$
the reduced angular momentum $j$ for $d=6$
black rings (BR) and Myers-Perry (MP) black holes with a single angular momentum.
The  blackfold (BF) prediction for the black rings
is also shown.
Here and in Figure 10, the
circle on the MP curve indicates the critical solution where the branches of 'pinched'
black holes emerge.
}
\vspace{0.5cm}

The numerical scheme described in Section 3
requires a further adjustment for $d>5$ BRs.
In five dimensions, BRs exist for arbitrary values of $\Omega_H$,
generically possessing conical singularities.

 Only for a critical value of the event horizon velocity a BR becomes balanced
 \cite{Emparan:2001wn}.
We have found that the situation is different for  $d>5$,
since the singularities of the unbalanced configurations
are stronger in this case\footnote{This is not an unexpected feature;
indeed, the analysis in 
 \cite{Emparan:2007wm} predicts the occurrence of  naked singularities for $d>5$
BRs which do not satisfy the   equilibrium condition.}.
That is, for given $R$, $r_H$ and arbitrary $\Omega_H$, 
the numerical algorithm diverges,
which we take as an indication for the occurrence of singularities,  
\setlength{\unitlength}{1cm}
\begin{picture}(8,6)
\put(-1.,0.0){\epsfig{file=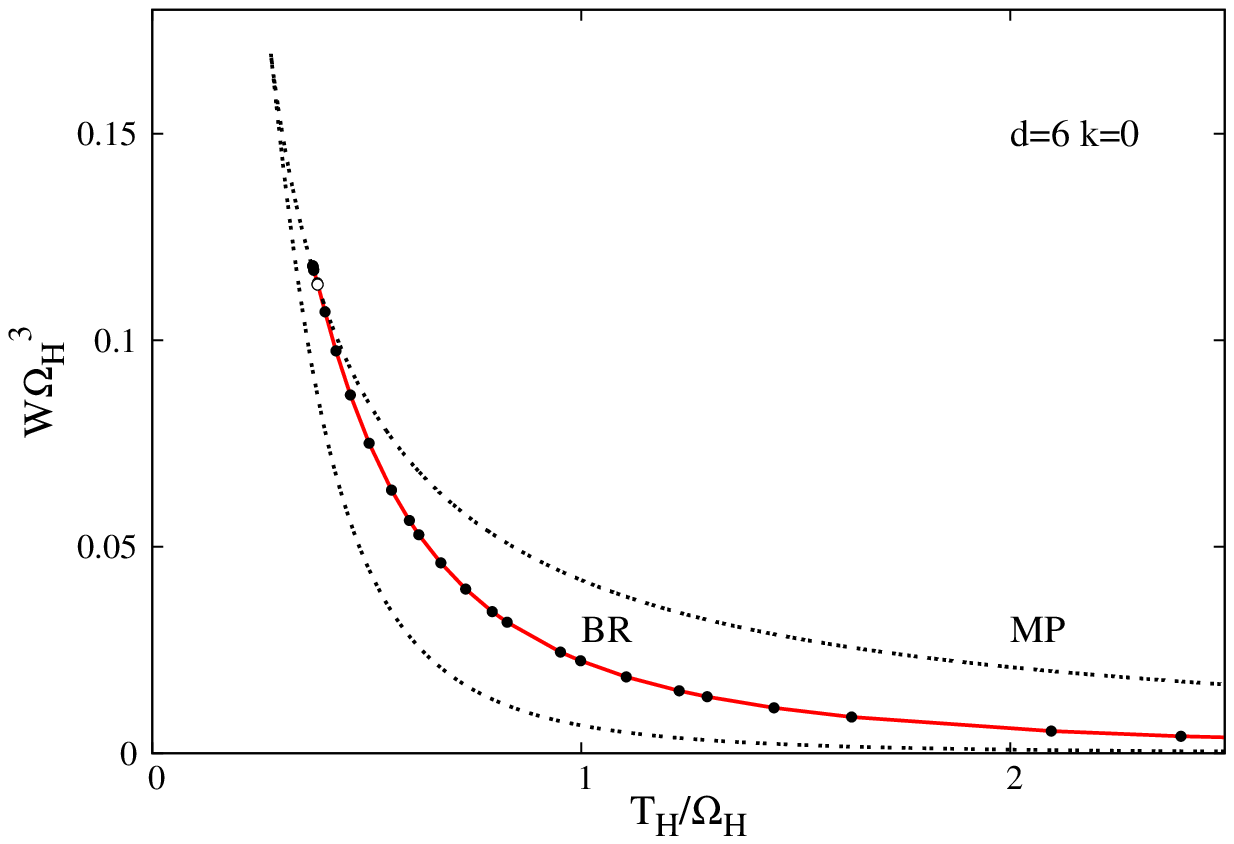,width=8cm}}
\put(7,0.0){\epsfig{file=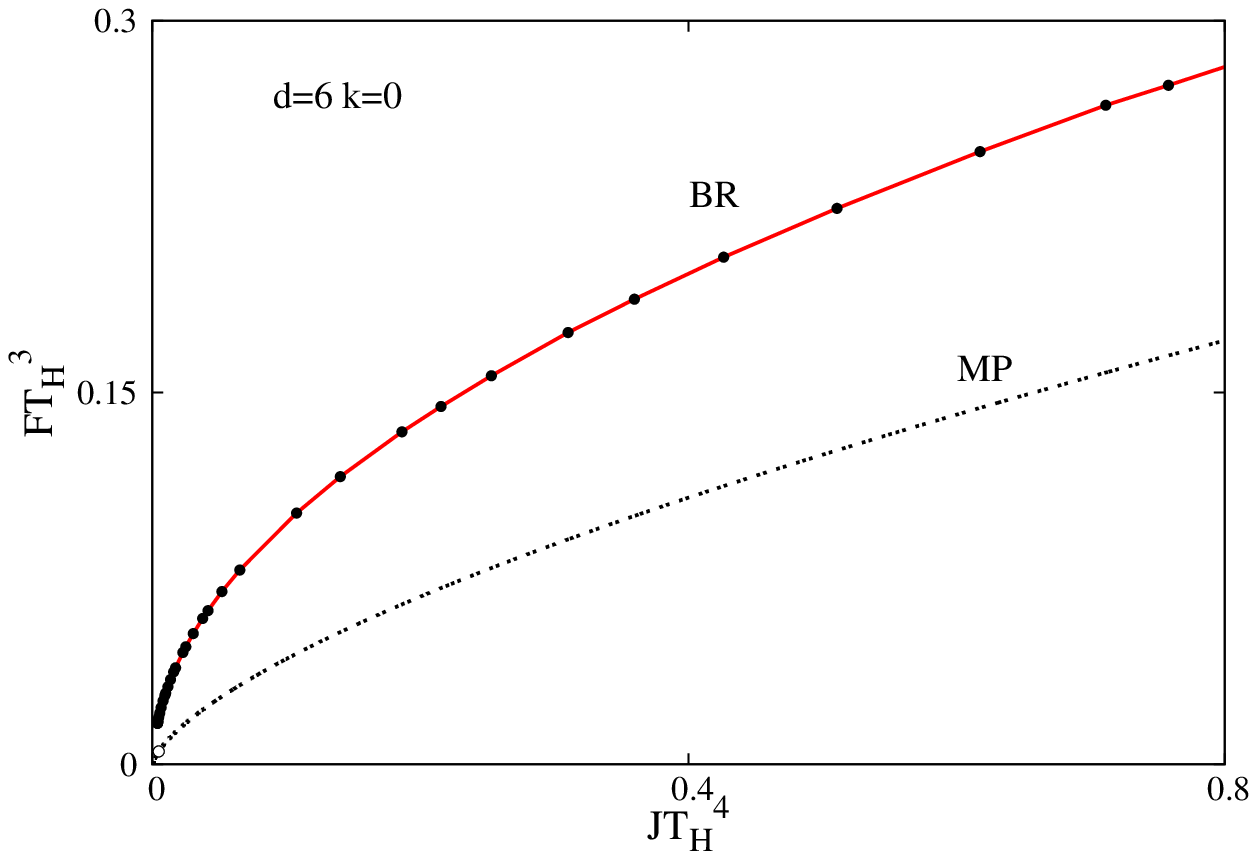,width=8cm}}
\end{picture}
\\
\\
{\small {\bf Figure 10.}
{\it Left}: The grand canonical potential $W$
is shown as a function of the Hawking temperature $T_H$
for $d=6$ black rings (BR) and Myers-Perry (MP) black holes with fixed angular velocity of the horizon $\Omega_H$. 
{\it Right}:
The canonical potential $F$
is shown as a function of the angular momentum $J$
for the same configurations with fixed Hawking temperature $T_H$.}
\vspace{0.5cm}
\\
a situation
that  cannot be dealt with in our scheme.
However, the numerical errors decrease dramatically for some (small)
range of $\Omega_H$
and the solver starts to converge.
The critical value of the event horizon velocity, 
where the ring is precisely balanced,  is found by using a shooting procedure in terms of $\Omega_H$. 
 Then the balanced solution has no singularity on and outside the horizon.
This can be seen by 
computing the Kretschmann scalar which is finite everywhere. 

Therefore, in principle, by varying the value of $R$
for fixed $r_H$
and by adjusting the value of $\Omega_H$ via a shooting algorithm, 
the full spectrum of $d=6$ balanced BRs
can be recovered numerically.

We have studied 
in a systematic way the $d=6$ BR solutions with $1.12 r_H<R<7 r_H$.
 However, we could not obtain BRs closer to the critical point $R=r_H$
with high accuracy, although we have a strong indication for their existence.

The profiles for the metric functions $f_i$, $w$ 
are rather similar to those of the
$d=5$ balanced BR solution,  
a typical configuration
being shown in Fig. 8. 
To illustrate the regular character of the solution,
we plot there also the Kretschmann scalar.

The general picture we have unveiled for $d=6$ BRs 
exhibits a number of
similarities to the well-known $d=5$ case.
Again,  one finds two branches of BR solutions whose 
physical differences are most clearly seen in 
terms of the reduced quantities $a_H$ and $j$ introduced in Section 3.
The $a_H(j)$ diagram of the BRs is shown in Figure 9, where
the singly rotating MP BHs are included as well.
There we show also the dependence of the reduced temperature $t_H$ 
and the reduced horizon angular velocity $w_H$ on the reduced
angular momentum $j$.
The analytical curve corresponding to the blackfold prediction
is also included in those plots.

\setlength{\unitlength}{1cm}
\begin{picture}(8,6)
\put(2,0.0){\epsfig{file=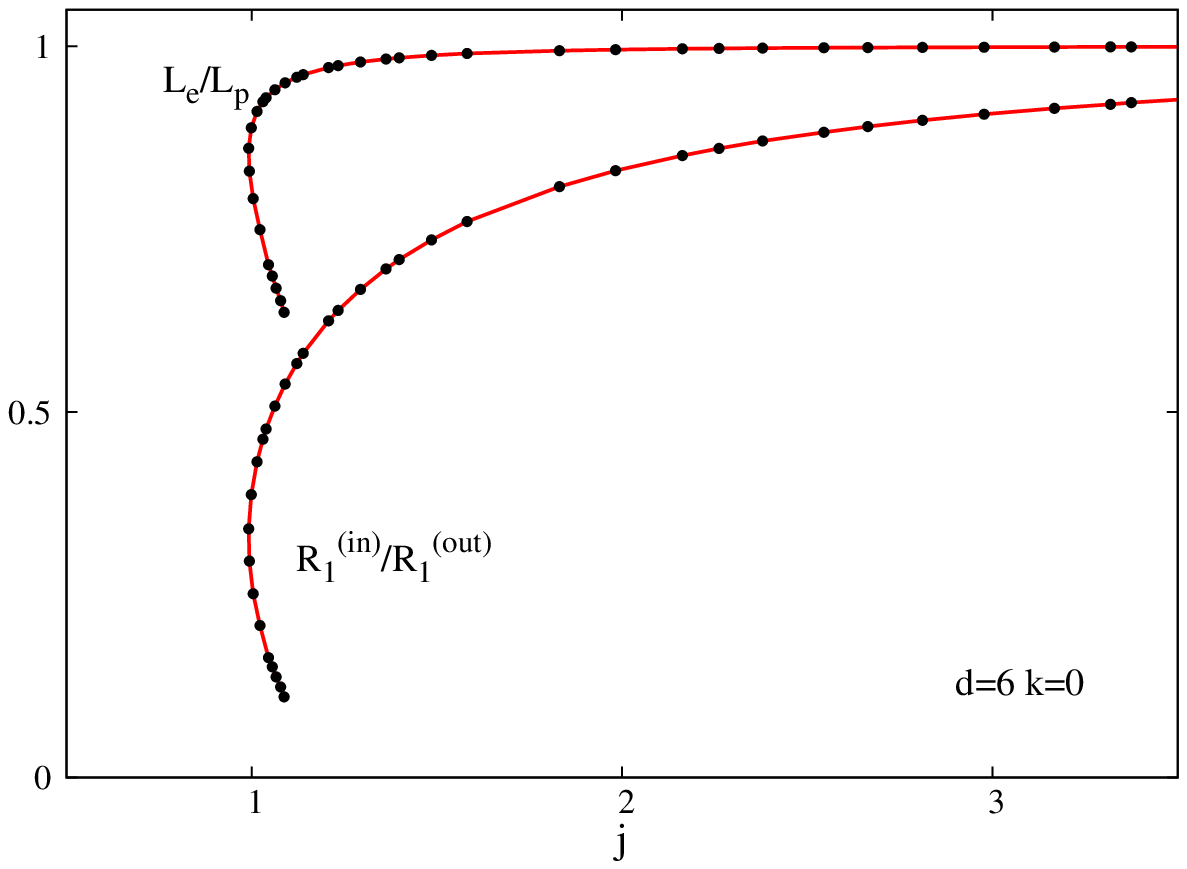,width=8cm}} 
\end{picture}
\\
\\
{\small {\bf Figure 11.}
The ratios $L_e/L_p$ and $R_1^{(in)}/R_1^{(out)}$, which encode the deformation of the horizon,
are shown
$vs.$ the reduced angular momentum $j$ for $d = 6$ black ring solutions.}
\vspace{0.5cm}

These diagrams clearly show
that the nonuniqueness result in five dimensions \cite{Emparan:2001wn}
extends also to the $d=6$ case.
We observe that the reduced area $a_H(j)$ has a cusp 
at a minimal value of $j$, $j_{\rm min}^{(BR)} \simeq 0.991$,
where $a_H$ assumes its maximal value, $a_H\simeq 1.076$.
Starting from this cusp the upper branch of solutions 
extends to $j\to \infty$.
Our results show that 
in the ultraspinning regime, these BRs
 are very well approximated by boosted black strings. 
In fact, 
we have found that the blackfold analytical result provides a good approximation for 
spinning $d=6$
solutions with 
$j \gtrsim 2$
(which include also a set of `not-so fast' spinning rings).

In agreement with the $d=5$ picture,
starting from the cusp there is also a lower branch of BRs,
the branch of `fat' BRs. (Note that this feature is not predicted by the blackfold results.)
Thus, in a certain range of the reduced
angular momentum $j_{\rm min}^{(BR)}< j < j_{\rm max}$
there exist three different solutions with the same global charges. 
 
This lower branch  have a small extent in both $j$ and $a_H$,
  ending in a critical merger configuration
\cite{Emparan:2007wm},
where a branch of `pinched' black holes is approached in a
horizon topology changing transition.
Extrapolations of the present data
together with the results in 
the recent work \cite{Dias:2014cia} 
indicate that the  critical configuration might be in the vicinity of
$j_{max} \simeq 1.14$, $a_H \simeq 0.918$ and $t_H\simeq 1.34$.
The numerical results here
and those in \cite{Dias:2014cia}
clearly show that the critical merger solution has a finite, nonzero area,
while the temperature   stays also finite and thus nonzero.
This critical behaviour of the $d=6$ BRs is in strong contrast with 
the one of the $d=5$ BRs \cite{Emparan:2001wn},
where the branch of `fat' BRs merges with the MP black hole branch 
in a singular solution with $j=1$, $a_H=0$ and $t_H=0$.

The results in \cite{Dias:2014cia} 
show that the  $d=6$ `pinched' black holes 
 branch off 
from a critical MP  black hole solution at $j\simeq 1.27,~a_H \simeq 0.83$, 
along the stationary zero-mode perturbation of
the GL-like instability \cite{Dias:2009iu,Dias:2010maa}.
(In fact, a more complicated picture is unveiled there, 
showing the existence of  two branches of `pinched' black holes,
with only one of them merging with the BRs.
However, these aspects are beyond the purposes of this work.)

\setlength{\unitlength}{1cm}
\begin{picture}(8,6)
\put(-1.,0.0){\epsfig{file=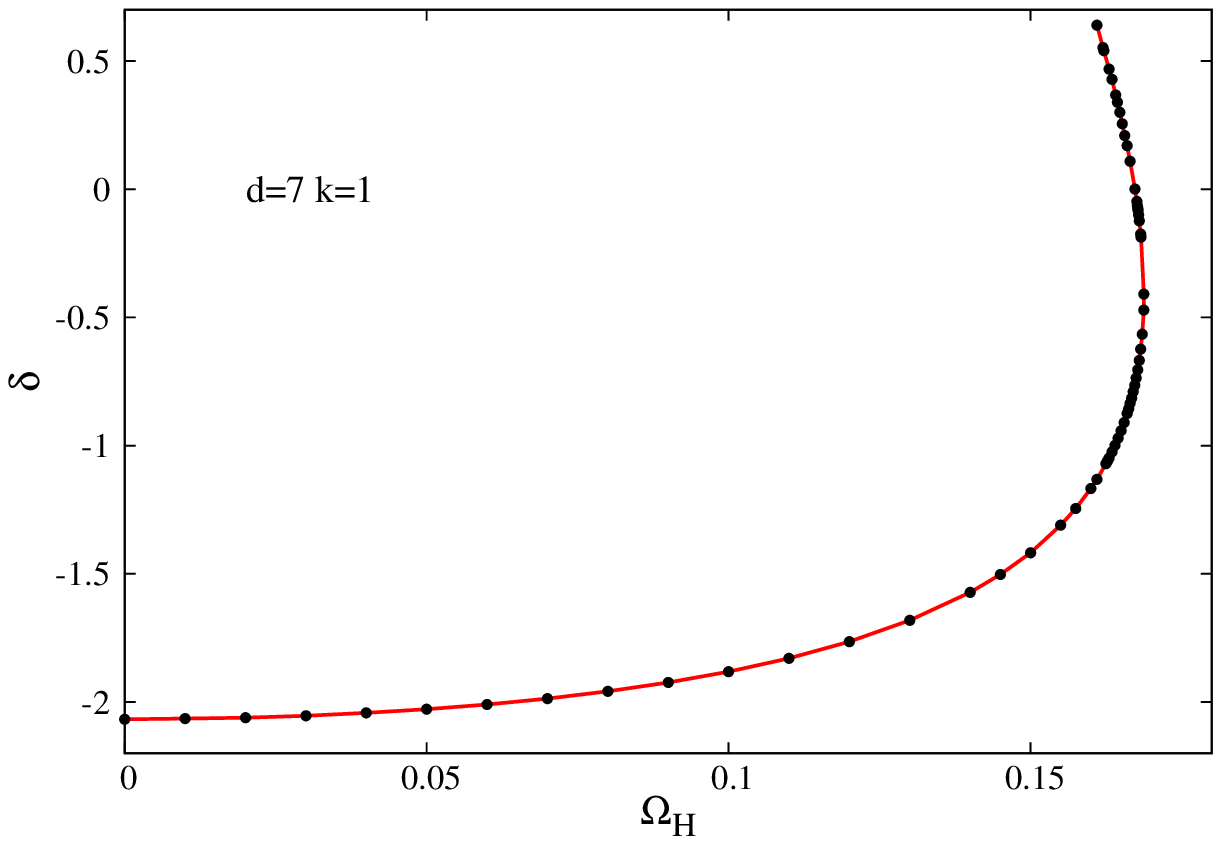,width=8cm}}
\put(7,0.0){\epsfig{file=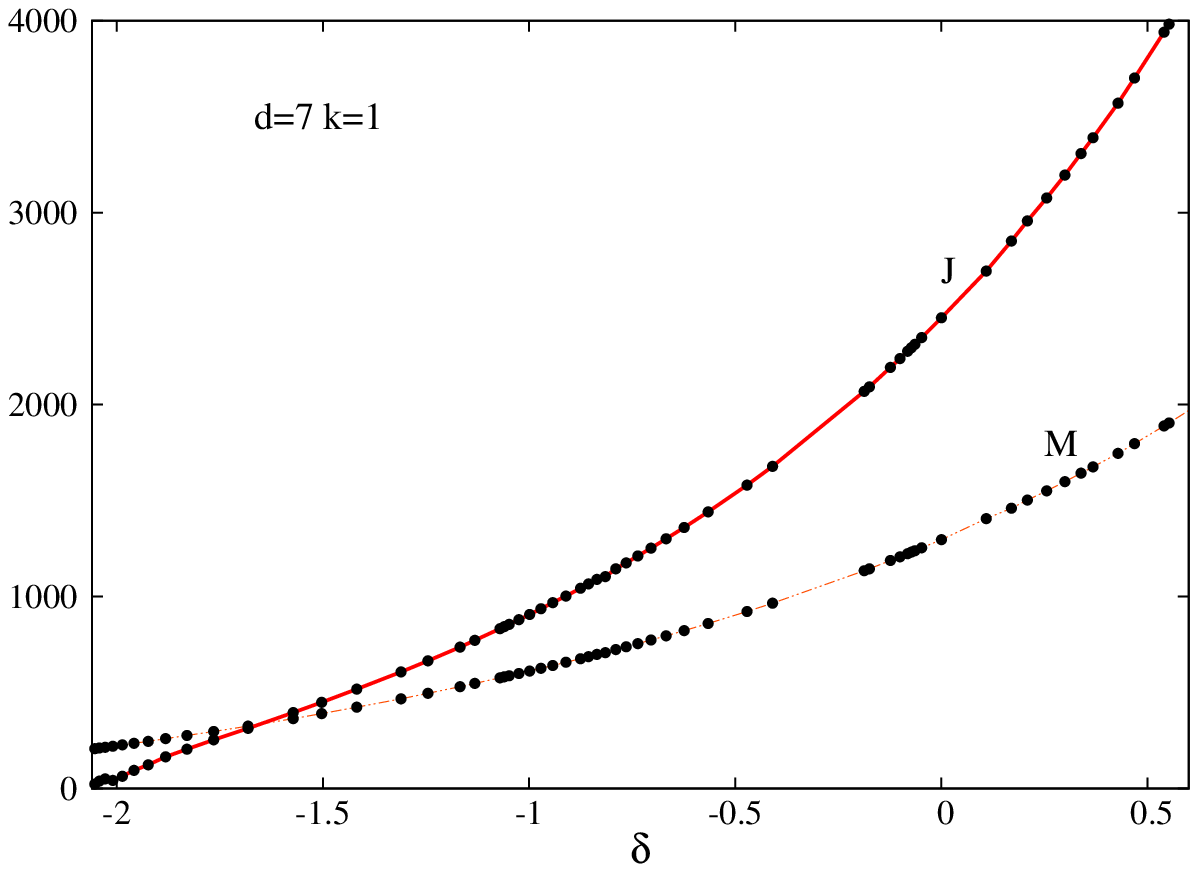,width=8cm}}
\end{picture}
\\
\\
{\small {\bf Figure 12.}
{\it Left:} The  conical deficit/excess $\delta$ of the (unbalanced) $d=7,~k=1$
black ringoid solutions with $r_H=1$, $R=4.6$
is shown as a function of the angular velocity of the horizon $\Omega_H$.
{\it Right:} The  mass $M$ and angular momentum $J$ are shown $vs.$ $\delta$ for the same solutions.
}
\vspace{0.5cm}

In Figure 10 we show the Gibbs potential and the Helmholz free energy
of the BRs together with the corresponding MP black holes.
The situation there looks rather different as compared to the $d=5$
case in Figure 6.
This originates in the different behaviour of the MP black holes
together with the existence of a critical merger solution in $d=6$,
with nonzero values of $W,F$.

Finally, in Figure 11 we exhibit
the deformations of the $S^3$ and $S^1$-components of the horizon, 
as given by the ratios $L_e/L_p$
and $R_1^{(in)}/R_1^{(out)}$, as functions of the reduced angular momentum.
One can see that
as the critical horizon topology changing solution is approached,
both 
$L_e$ and $L_p$ stay finite and nonzero.
Moreover, our results suggest that
this is the case as well  for  $R_1^{(out)}$, whereas
 $R_1^{(in)}\to 0$.

\boldmath
\subsubsection{$k=1$:  black ringoids in $d = 7$  dimensions }
\unboldmath

The $d=7,~k=1$ solutions with a horizon topology $S^{2}\times S^{3}$
have very different properties from those of the $d=6$ BRs
discussed above. 
This is not surprising, since the slowly rotating solutions
can be describes as perturbative deformations of the static configurations
in \cite{Kleihaus:2009wh}. 
The results there show that the
limiting static
solutions necessarily possess a conical singularity which prevents their collapse,
and no other pathologies.
Moreover,
as discussed in 
\cite{Herdeiro:2009vd},
\cite{Herdeiro:2010aq},
 the asymptotically flat black objects with conical singularities still
admit a consistent thermodynamical description.
Also, when working with the appropriate set
of thermodynamical variables, the Bekenstein-Hawking law still holds for such solutions.

As expected, 
their (generic) rotating generalizations 
possess also conical singularities, while being physically acceptable in all other aspects.
Moreover, this pathology
has a rather neutral effect on the numerics, since the solver does not notice it directly. 

\newpage
\setcounter{figure}{12}
\begin{figure}[t!]
\setlength{\unitlength}{1cm}
\begin{picture}(15,18)
\put(-1.,0){\epsfig{file=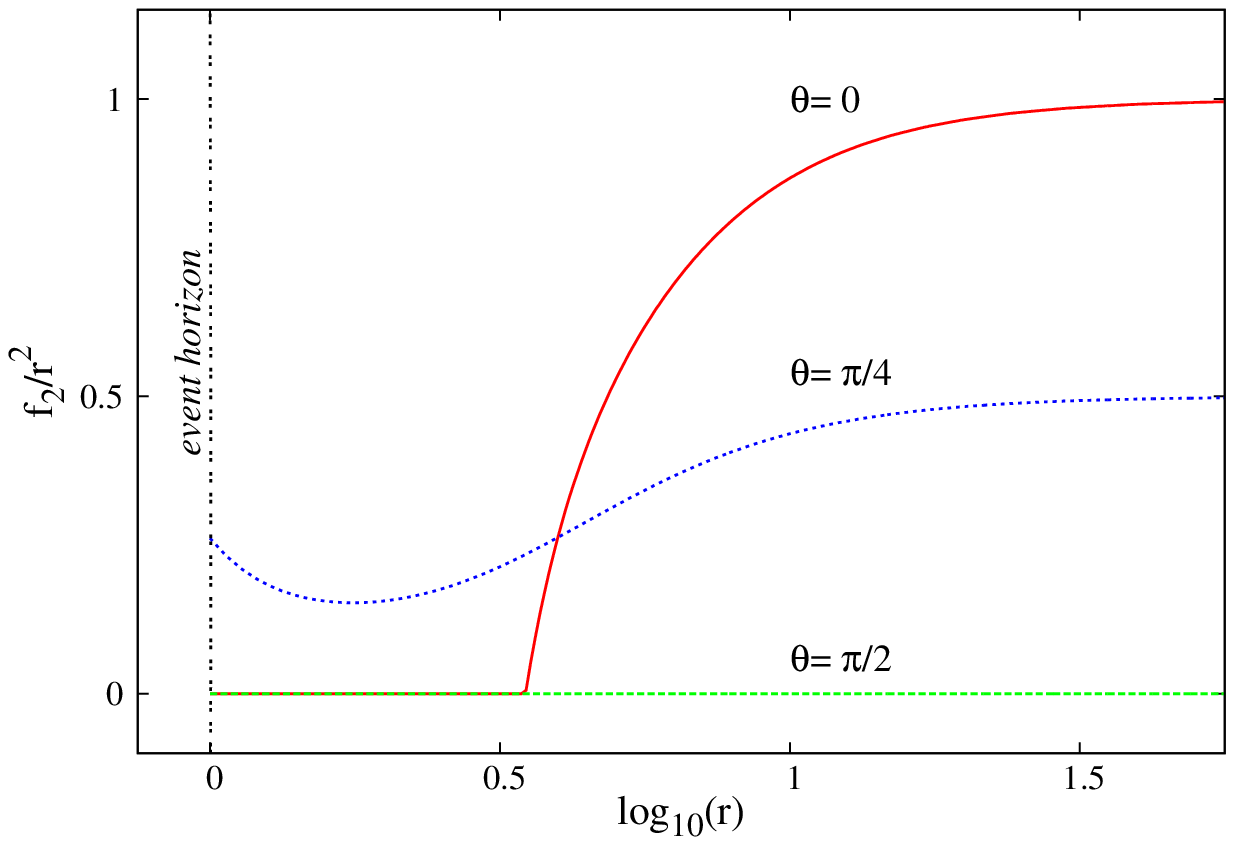,width=8cm}}
\put(7,-0.5){\epsfig{file=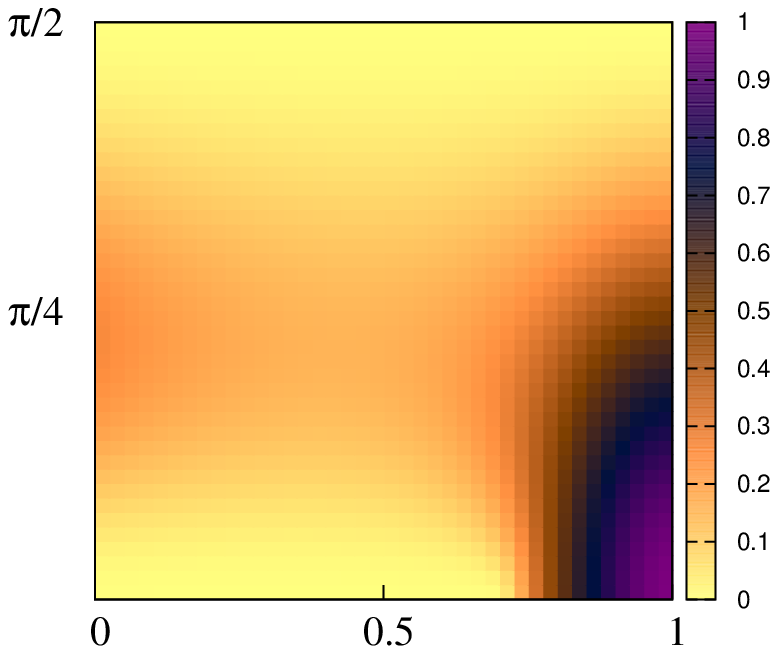,width=10cm}}
\put(-0.9,6){\epsfig{file=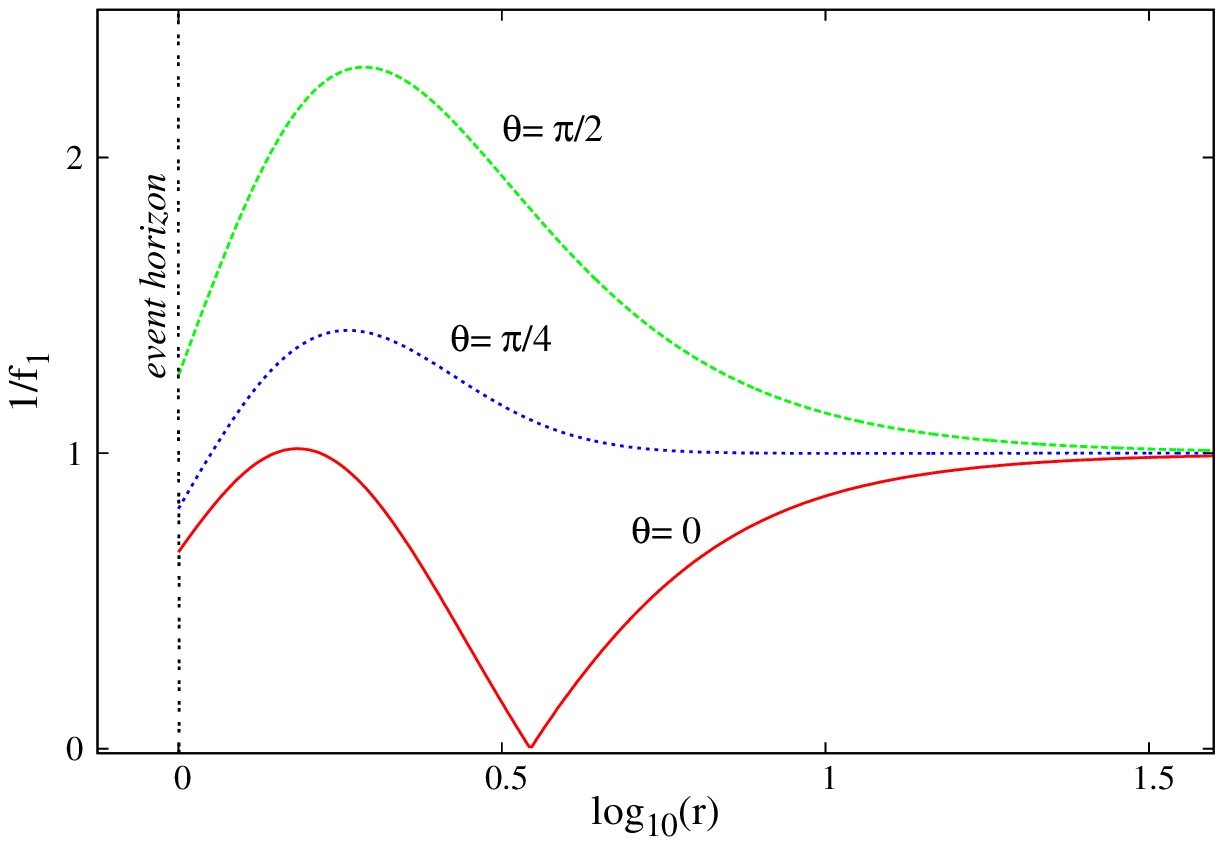,width=7.9cm}}
\put(7,5.5){\epsfig{file=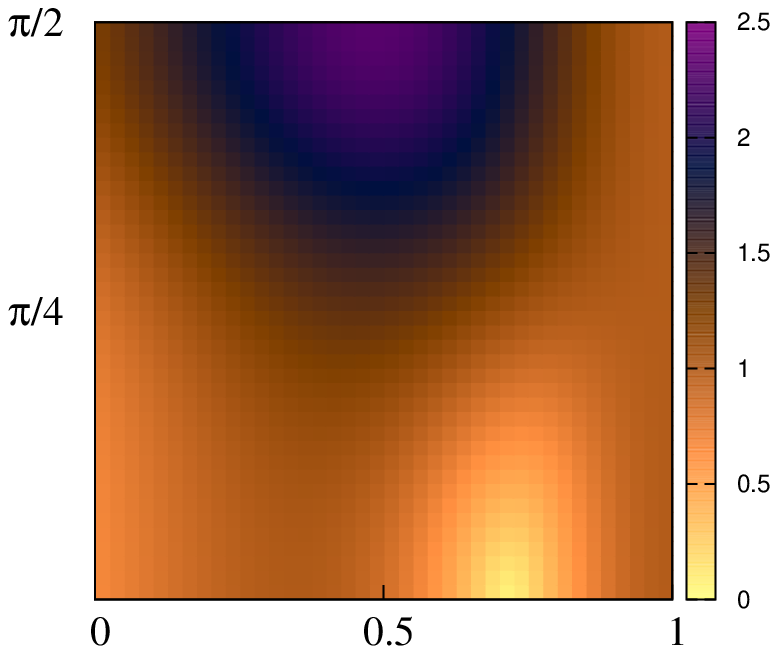,width=10cm}}
\put(-1,12){\epsfig{file=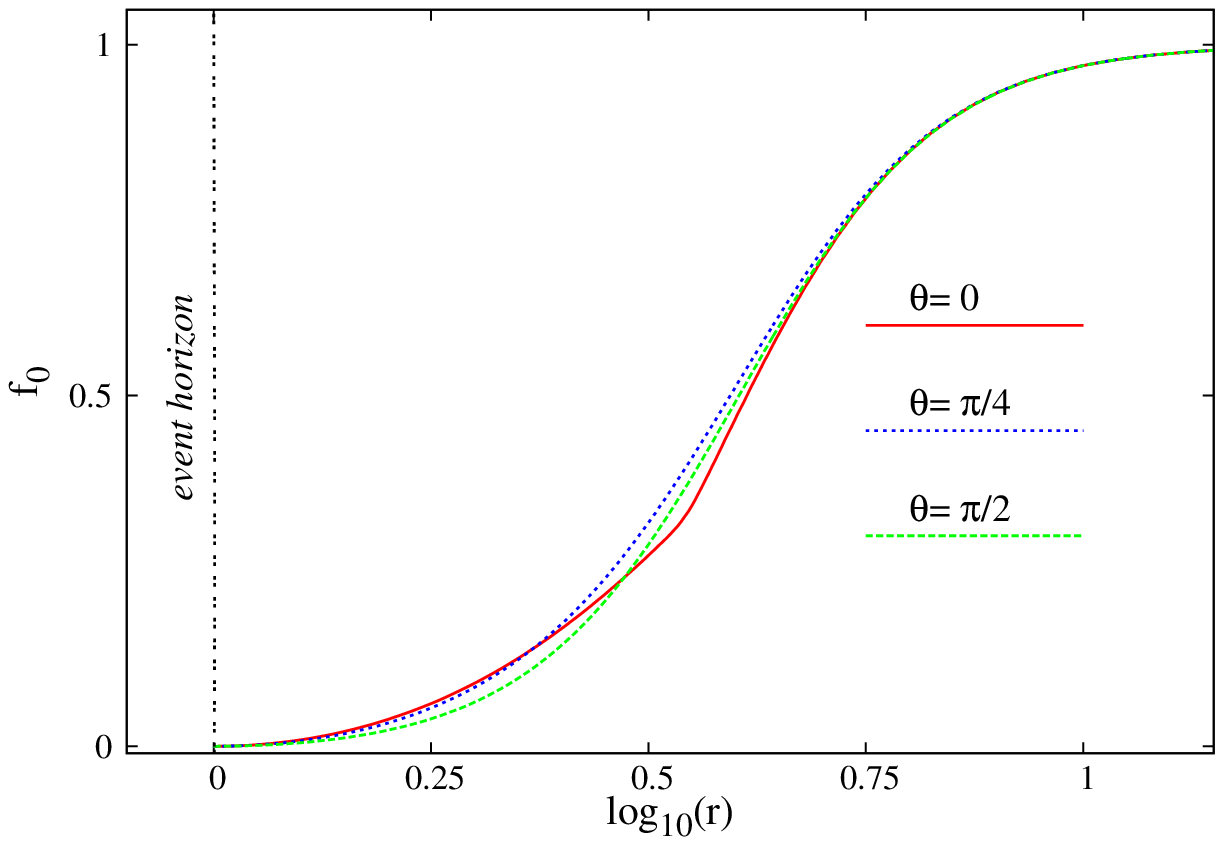,width=8cm}}
\put(7,11.5){\epsfig{file=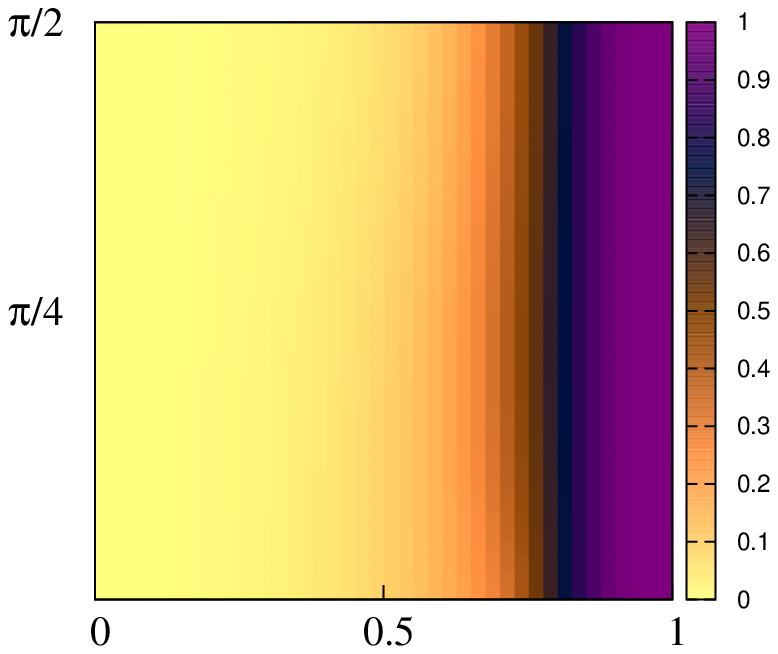,width=10cm}}
\end{picture}
\newpage
\caption{
The metric functions $f_i,W$ and the Kretschmann scalar $K=R_{\mu\nu\alpha \beta}R^{\mu\nu\alpha \beta}$  are shown for a $d=7$ black ringoid
with the input parameters 
$r_H=1$,
$R=3.5$
and
$\Omega_H\simeq 0.199$.
}
 \end{figure}

In our work,
we have chosen to locate the conical singularity 
at  $\theta=0$, $r_H<r<R$, where 
the generic configurations have 
a conical 
deficit/excess

\newpage
\setcounter{figure}{12}
\begin{figure}
\setlength{\unitlength}{1cm}
\begin{picture}(15,18)
\put(-1.,0){\epsfig{file=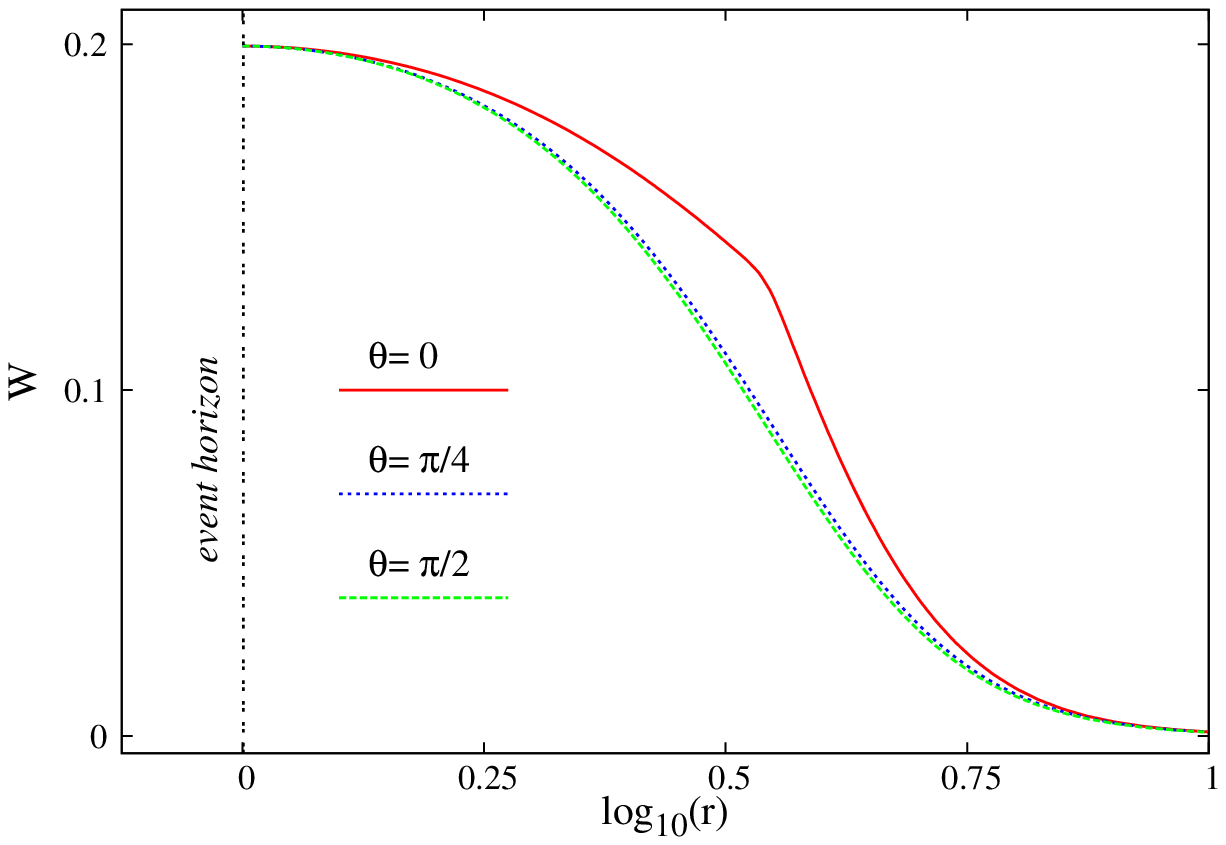,width=8cm}}
\put(7,-0.5){\epsfig{file=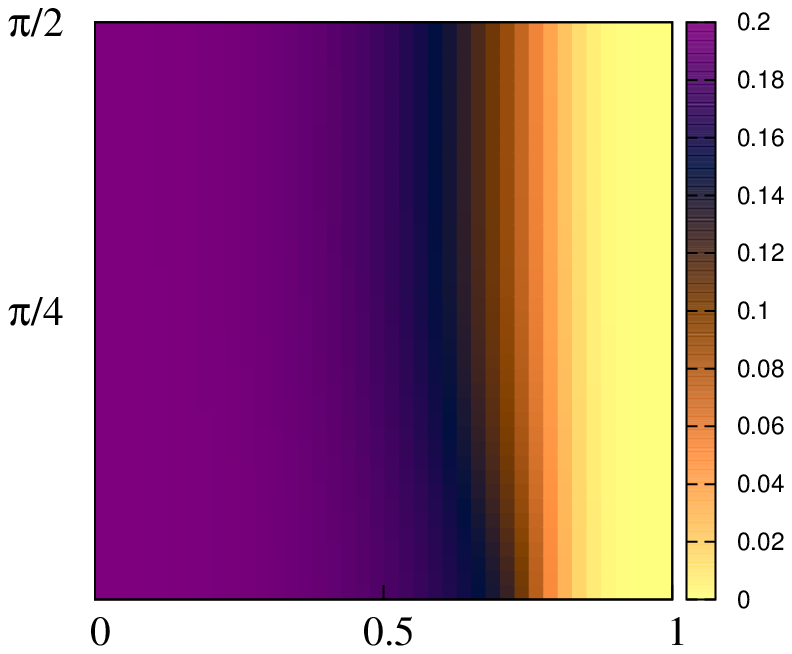,width=10cm}}
\put(-1,6){\epsfig{file=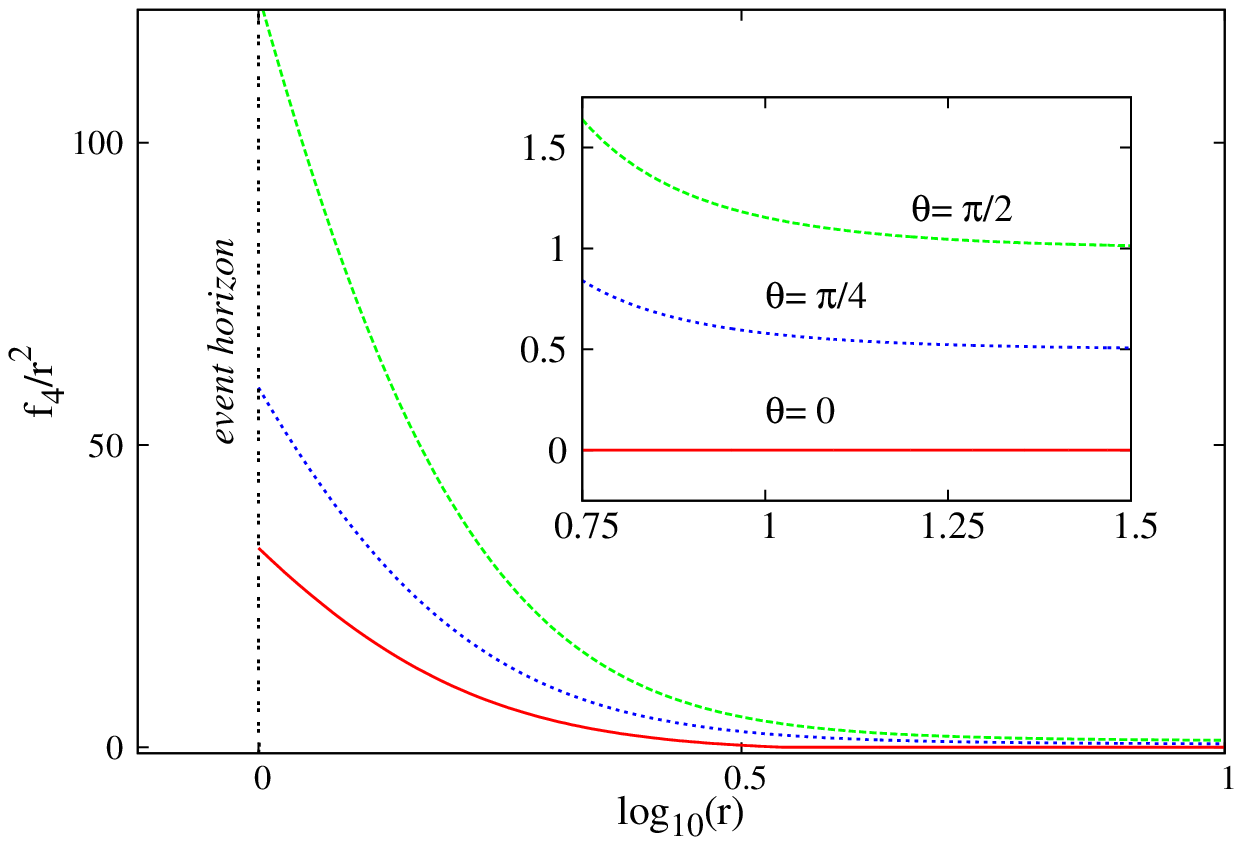,width=8cm}}
\put(7,5.5){\epsfig{file=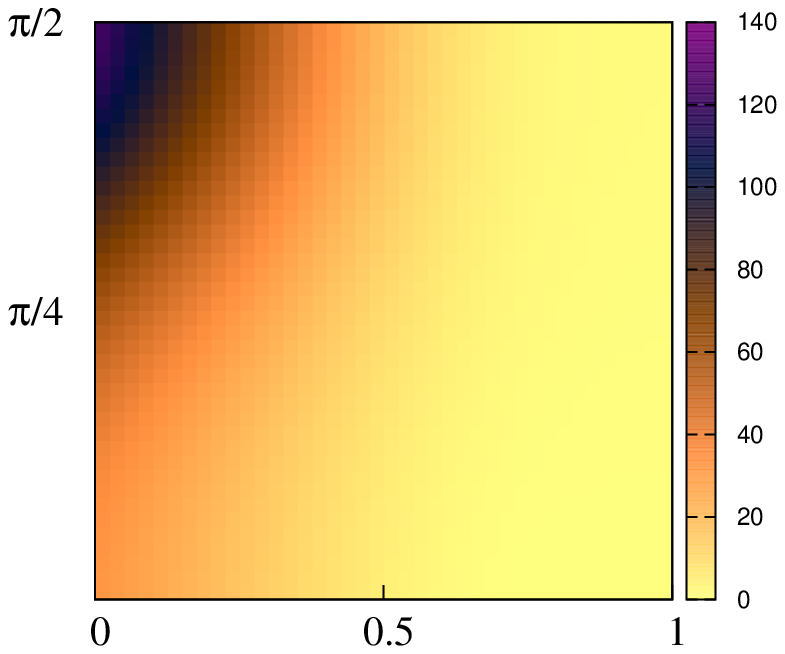,width=10cm}}
\put(-1,12){\epsfig{file=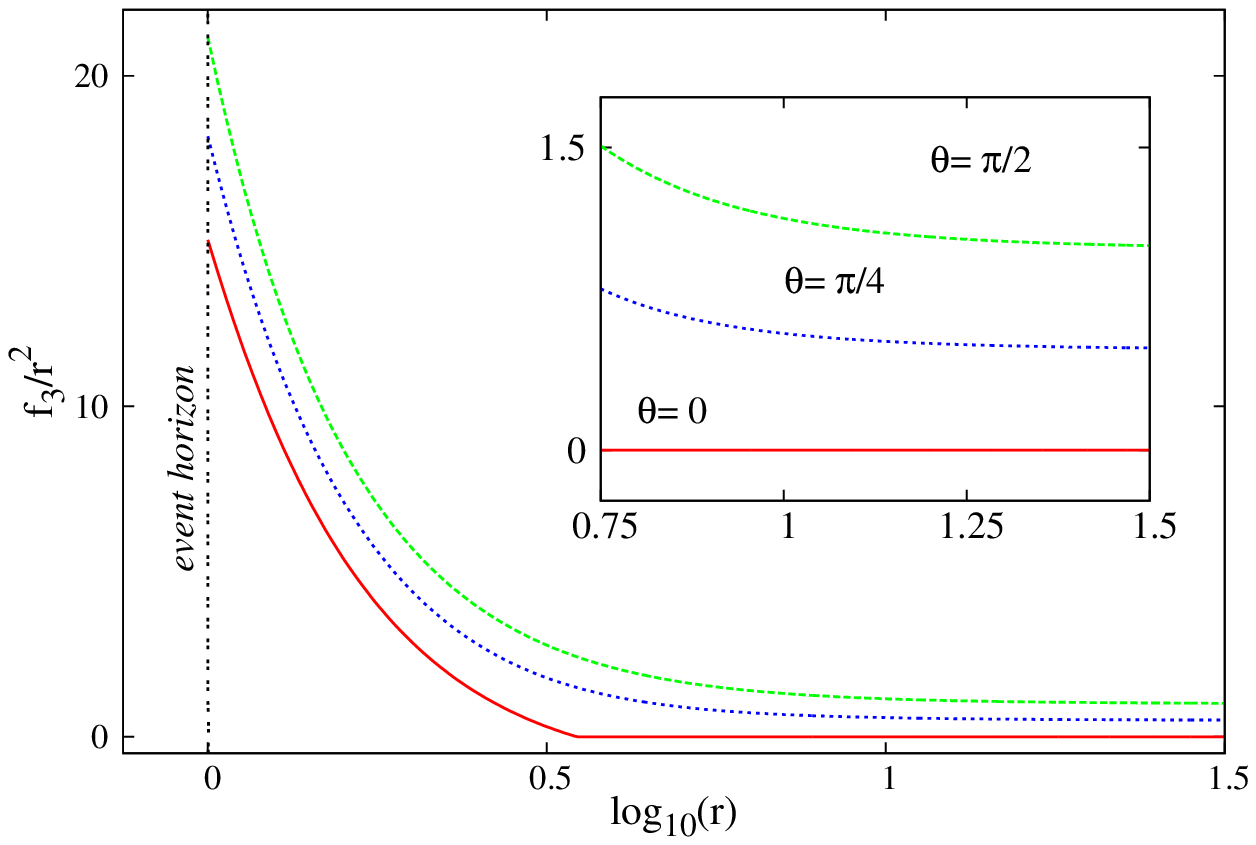,width=8cm}}
\put(7,11.5){\epsfig{file=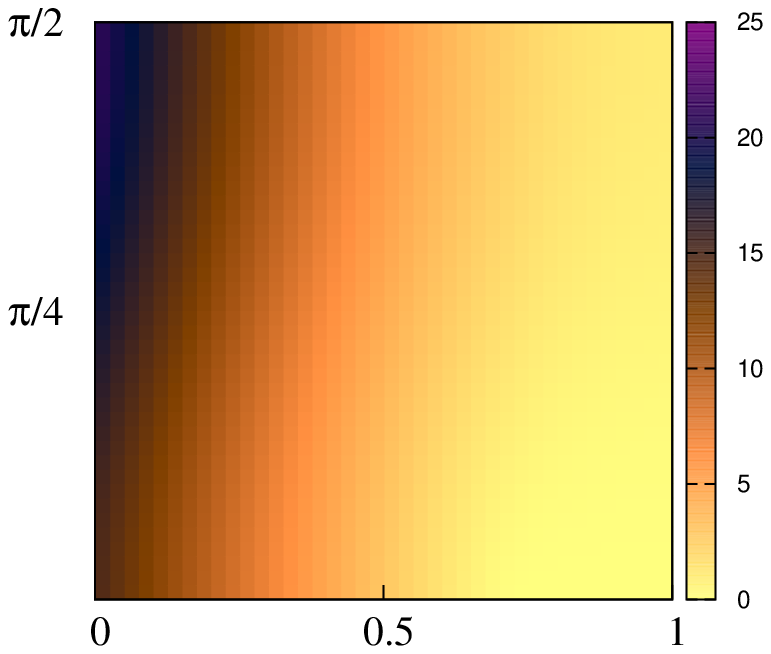,width=10cm}}
\end{picture}
\caption{
continued.
}
\end{figure} 

\begin{eqnarray}
\label{delta-def}
\delta= 2\pi (1-\lim_{\theta\to 0}\frac{f_2}{\theta^2 r^2 f_1 }),
\end{eqnarray}
with $\delta<0$ in the static case. 
\setcounter{figure}{12}
\begin{figure}
\setlength{\unitlength}{1cm}
\begin{picture}(8,6)
\put(-1,0.0){\epsfig{file=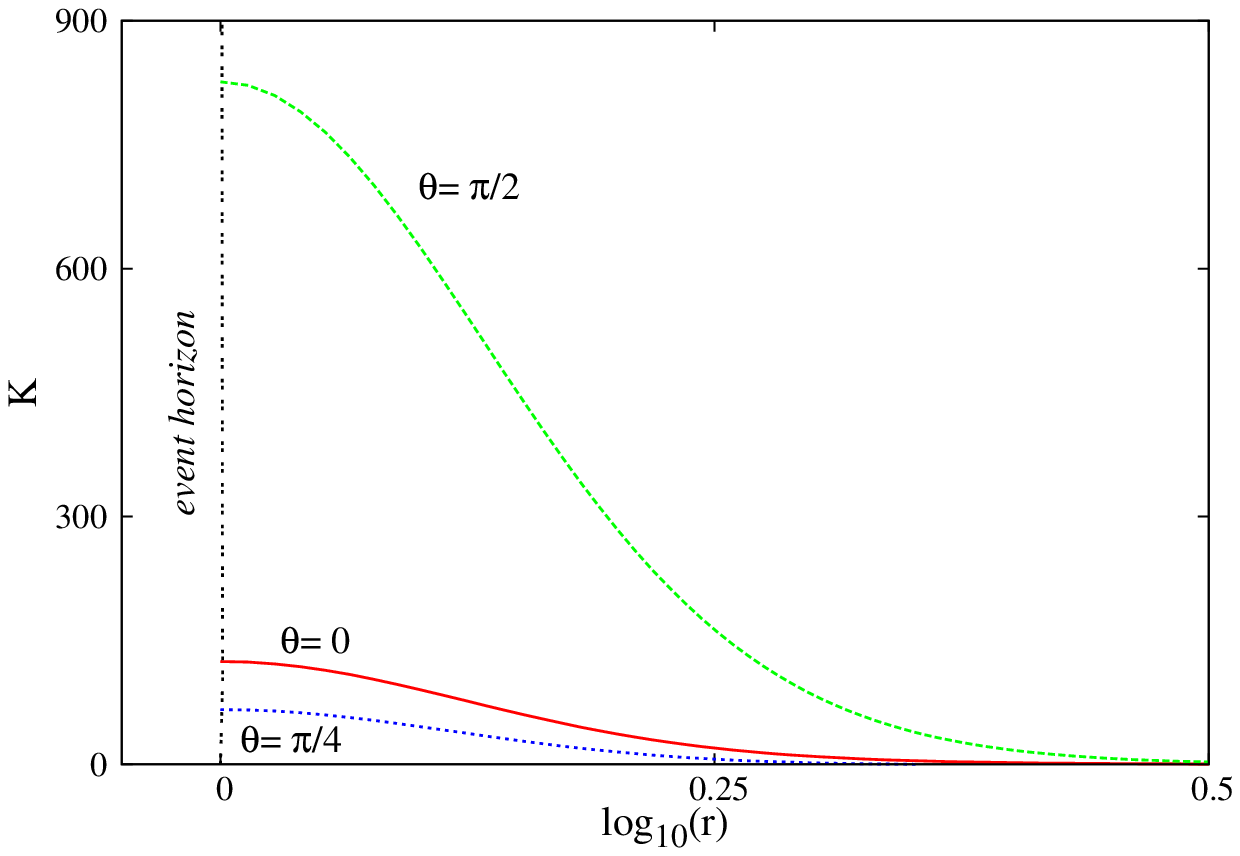,width=8cm}}
\put(7,-0.5){\epsfig{file=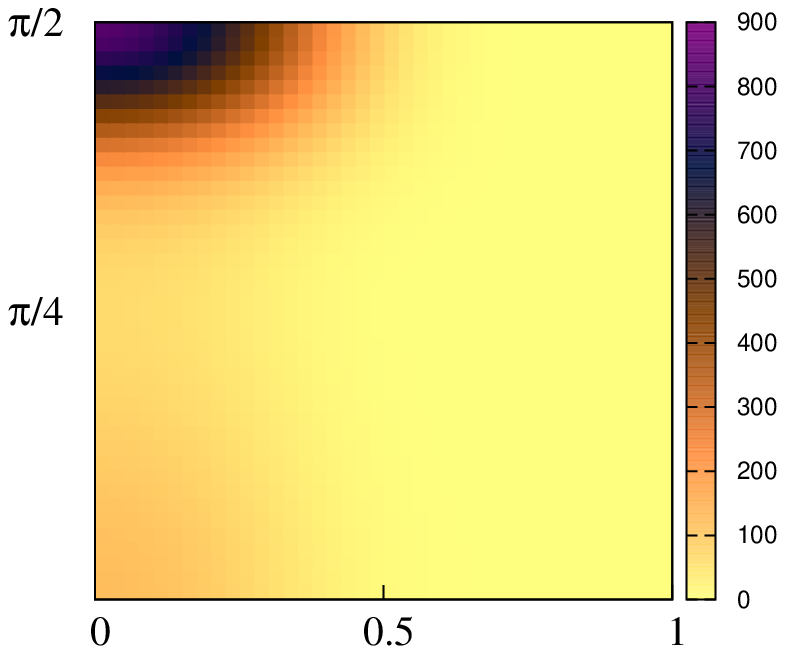,width=10cm}}
\end{picture}%
\caption{
continued.
}
\end{figure} 
(Therefore these solutions satisfy the more general condition 
$f_{22}= r^2f_{10} const.$
in  the $\theta=0$ expansion  (\ref{t04}),
 the case $const.=1$ implying the absence
 of a conical singularity.)
This can be interpreted as the higher dimensional analogue of a `strut' 
preventing the collapse of the configurations.

As expected, adding rotation to a static solution 
decreases the absolute value of $\delta$, such that $\delta=0$
is realized for a critical value of $\Omega_H$.
When the parameter $\Omega_H$ is varied further, one finds instead an over-rotating
black ringoid with a conical excess, $\delta>0$.

Note also that for the solutions studied so far, the global charges 
increase with $\delta$.
These features are illustrated in Figure 12,
for a family of solutions
with fixed $r_H,~R$. 
(Note that this behaviour is qualitatively similar to the one found for $d=5$
unbalanced BRs, see $e.g.$ \cite{Herdeiro:2010aq}.)

However, a systematic study of the generic unbalanced
case is beyond the purposes of this work. 
Therefore for the rest of this section we shall consider the physically most interesting case
of balanced black ringoids. 

  The set of balanced solutions is constructed 
 again by varying the parameter $R$ for fixed $r_H$,
 looking for configurations with $\delta=0$,
 a condition which is realized for a unique value of
 the input parameter $\Omega_H$. 
 To the best of our knowledge,
these solutions represent the first set of balanced nonperturbative solutions 
 obtained for a non-spherical and non-ring horizon topology.
 We show in Figure 13
 the profiles of such a typical balanced black ringoid,
 together with the Kretschmann scalar $K=R_{\mu\nu\alpha \beta}R^{\mu\nu\alpha \beta}$.

  The results of the numerical integration are shown in Figure 14
 for the same set of reduced quantities as in the BR case.
One can see that the blackfold results provide again a very good description of 
the fast spinning solutions.
Also, as expected, 
the angular momentum of the balanced black ringoids
is bounded below, but not above.

\setlength{\unitlength}{1cm}
\begin{picture}(8,6)
\put(-1,0.0){\epsfig{file=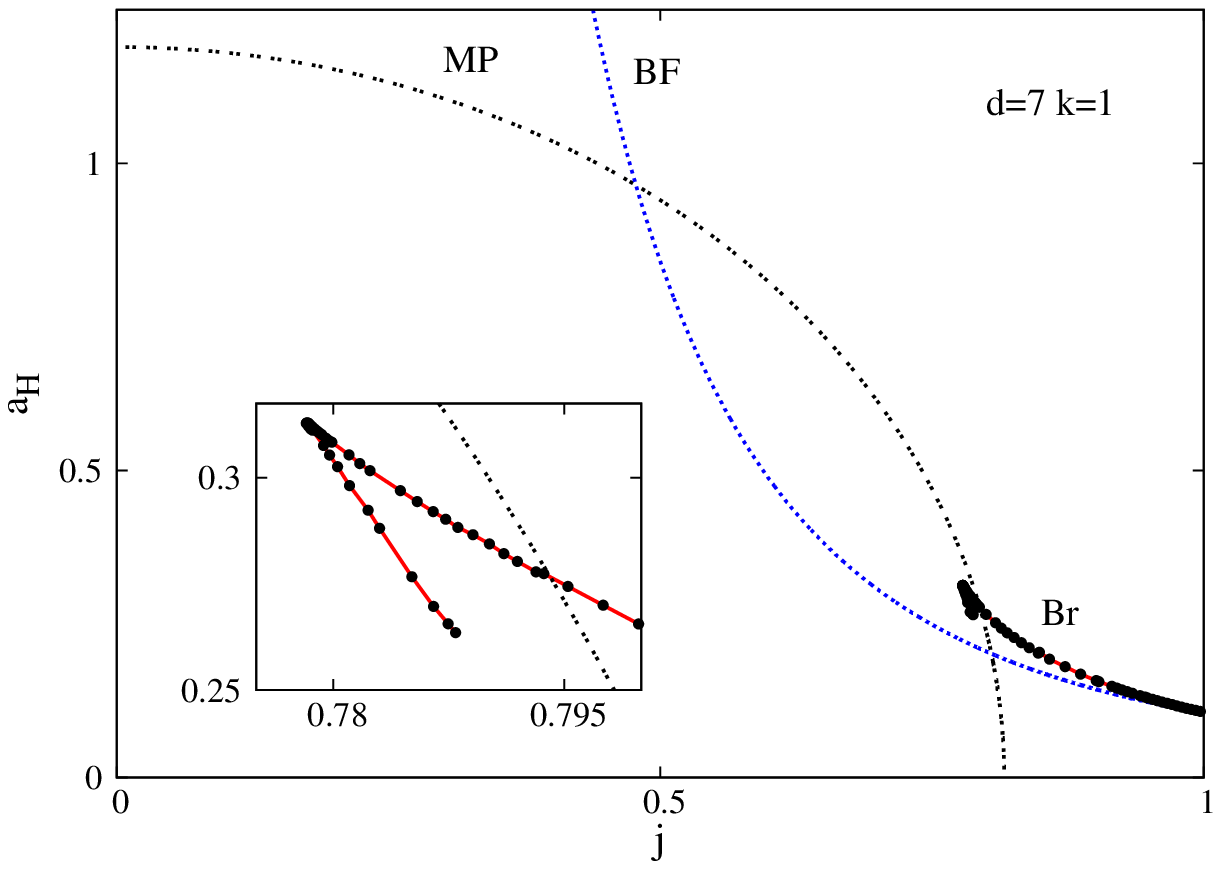,width=8cm}}
\put(7,0.0){\epsfig{file=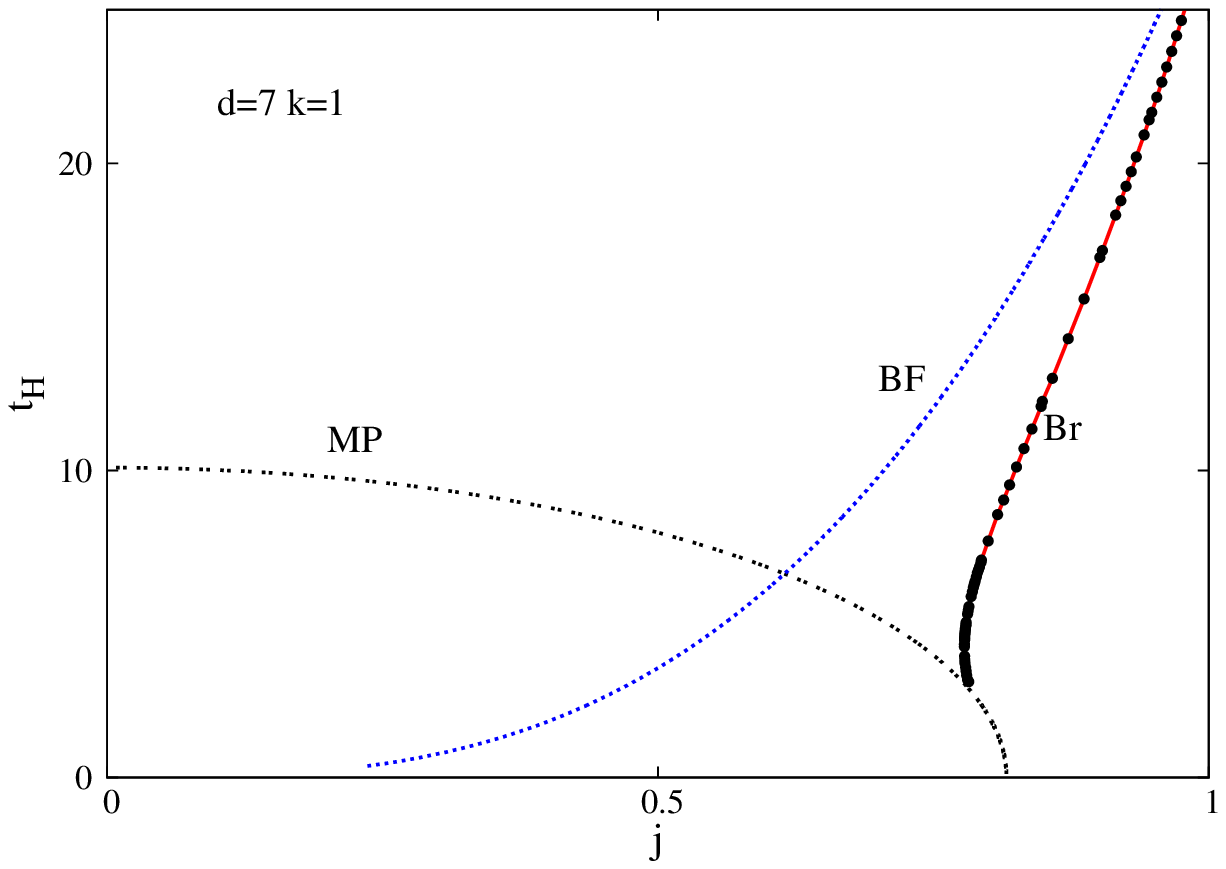,width=8cm}}
\end{picture}
 \vspace{0.5cm}

\setlength{\unitlength}{1cm}
\begin{picture}(8,6)
\put(2,0.0){\epsfig{file=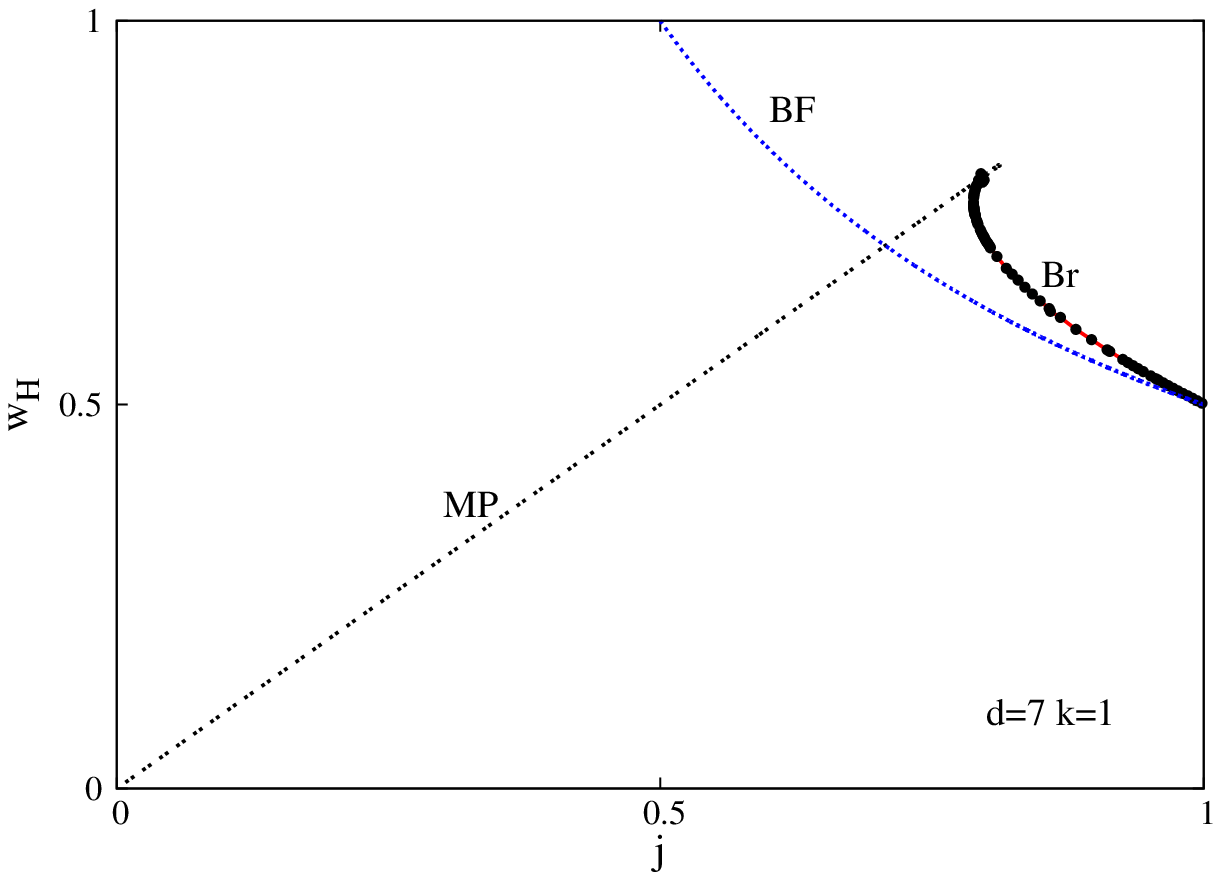,width=8cm}} 
\end{picture}
\\
\\
{\small {\bf Figure 14.}
The  reduced area $a_H$, the reduced temperature $t_H$
and the reduced angular velocity $w_H$  
are shown $vs.$
the reduced angular momentum $j$ for $d=7$
black ringoids (Br) with $S^2\times S^3$ event horizon topology
together with the corresponding results for
Myers-Perry (MP) black holes with two equal angular momenta.
The curves corresponding to the blackfold (BF) prediction
are also shown.
}
\vspace{0.5cm}
 
 Our numerical results show that,
similar to BRs, there are two branches of solutions,
which are dubbed again `thin' and `fat', according to their shape.
These two branches meet in a cusp at 
$j\simeq 0.778$,
$a_H \simeq 0.312$,
where $a_H$ assumes its maximal value, while 
$j$
takes its minimal value.

This minimally spinning solution has a non-degenerate regular horizon,
and thus does not correspond to an extremal black hole.
Also, for some range of $j$,
there are three different solutions with the same global charges:
one MP black hole and two black ringoids.

The numerical results strongly suggest that the branch of `fat' black ringoids
ends in a limiting solution with $a_H=0$ and a nonzero $j$.

\setlength{\unitlength}{1cm}
\begin{picture}(8,6)
\put(-1.,0.0){\epsfig{file=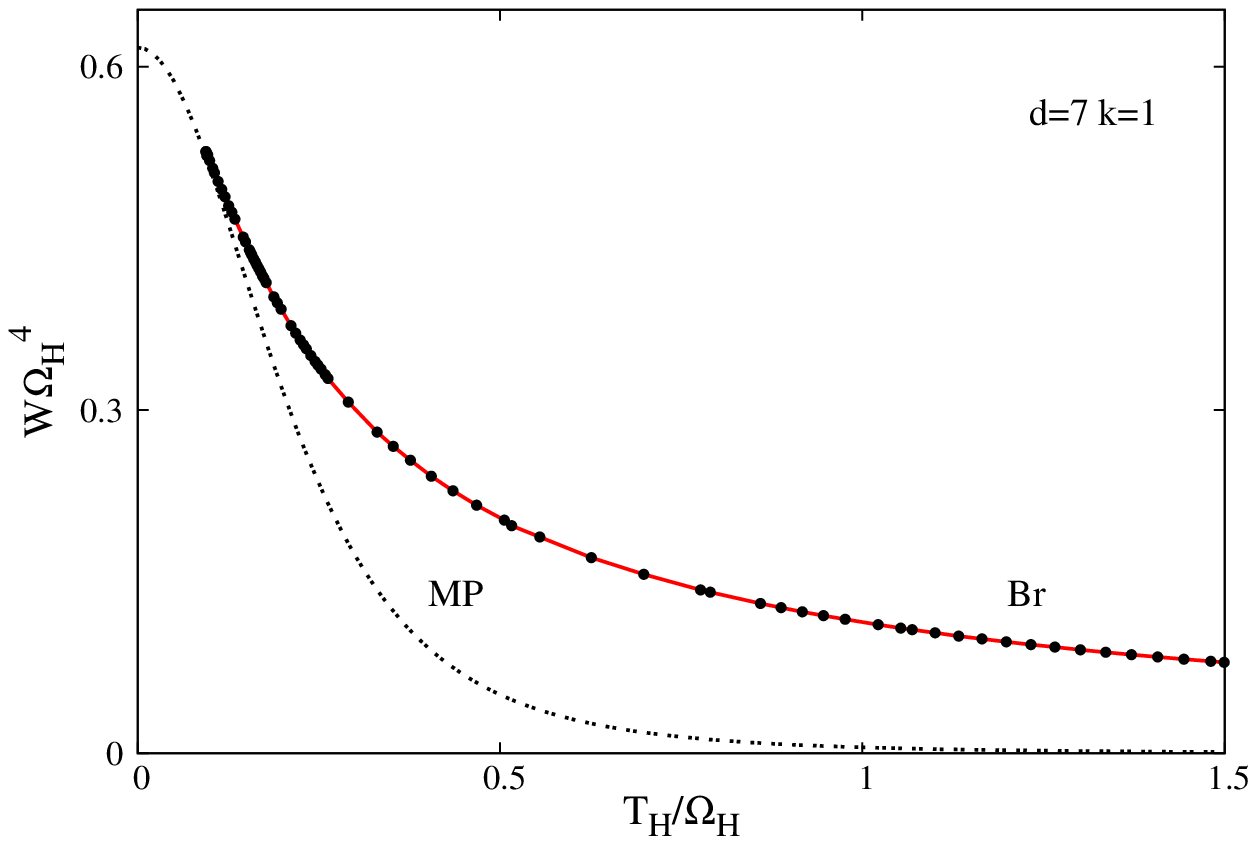,width=8cm}}
\put(7,0.0){\epsfig{file=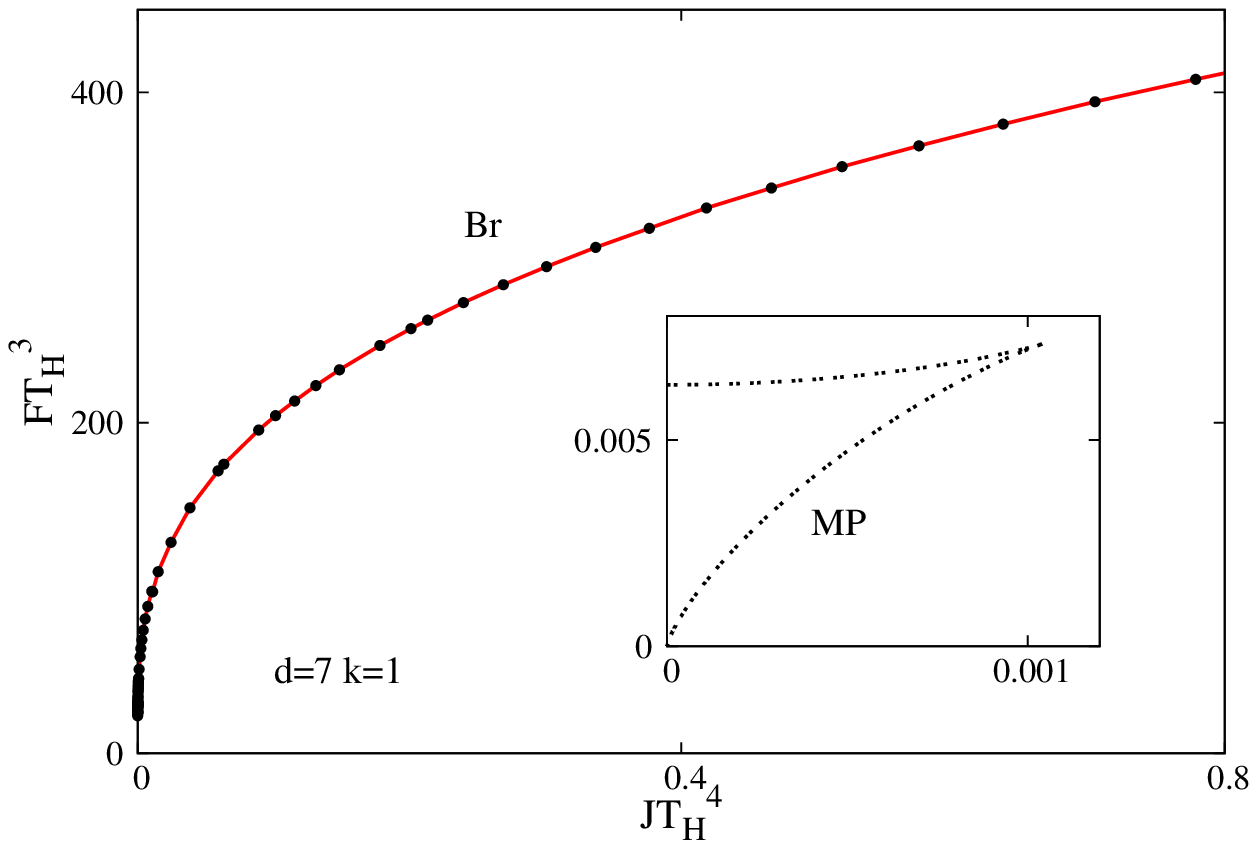,width=8cm}}
\end{picture}
\\
\\
{\small {\bf Figure 15.}
{\it Left}: The grand canonical potential $W$
is shown as a function of the Hawking temperature $T_H$
for $d=7$ black ringoids (Br) and Myers-Perry (MP) 
black holes with fixed angular velocity of the horizon $\Omega_H$.
{\it Right}:
The canonical potential $F$
is shown as a function of the angular momentum $J$
for the same $d=7$ solutions with fixed Hawking temperature $T_H$.
}
\vspace{0.5cm}

\setlength{\unitlength}{1cm}
\begin{picture}(8,6)
\put(2,0.0){\epsfig{file=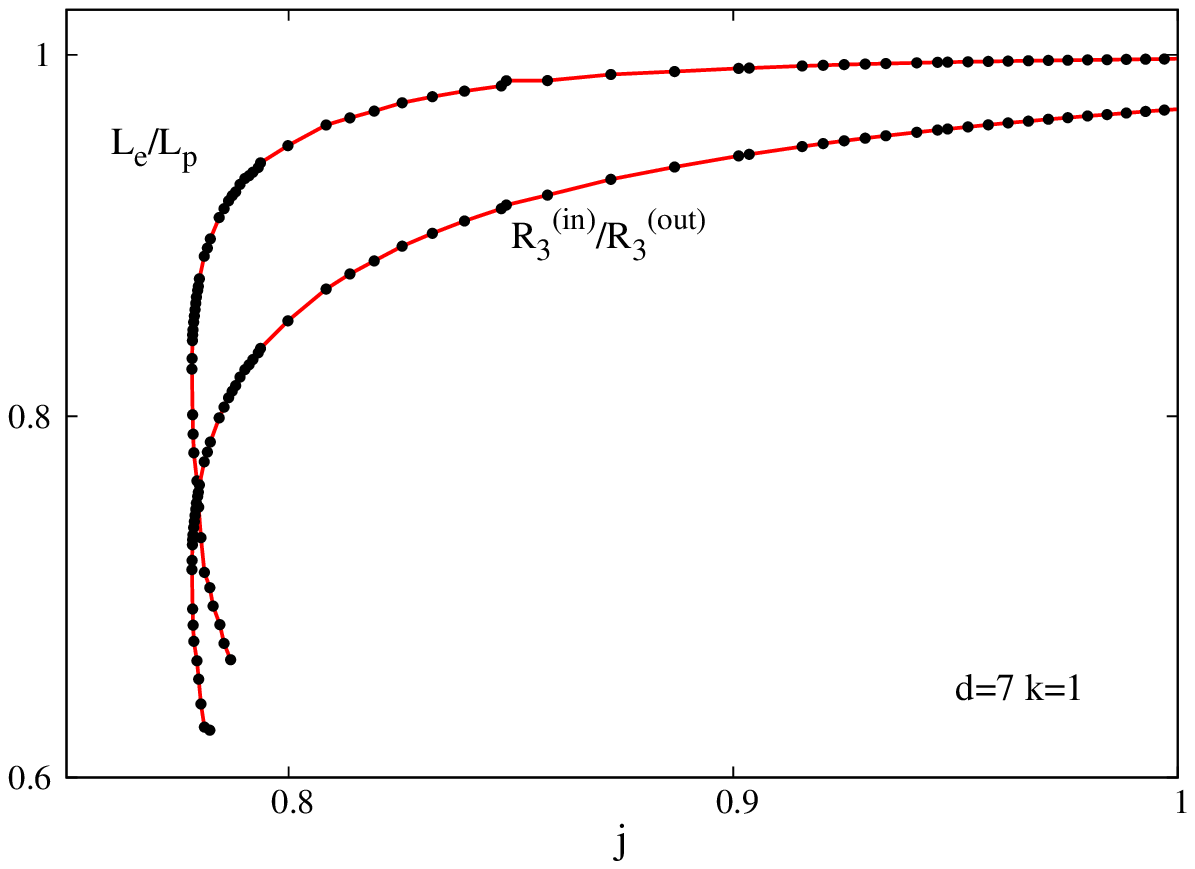,width=8cm}} 
\end{picture}
\\
\\
{\small {\bf Figure 16.}
The ratios $L_e/L_p$ and $R_3^{(in)}/R_3^{(out)}$, which encode the deformation of the horizon,
are shown 
$vs.$ the reduced angular momentum $j$ for $d = 7,~k=1$ black ringoid solutions.}
\vspace{0.5cm}
\\
Unfortunately, the numerical accuracy 
deteriorates before approaching that point,
so that we could not fully clarify this issue\footnote{
So far, we did not manage to construct solutions for $R/r_H<1.7$
with good accuracy,
obtaining large errors (in particular, for the constraint equations)
for smaller values of this ratio.
However, we believe that this is a numerical problem only.
We conjecture that the numerical difficulties
we encounter are related to the singular nature 
of the $R\to r_H$
limiting solution.
For example, an analysis of the $d=5$
BR solution in Appendix B shows that
some metric functions (as well as some global quantities) diverge in this limit.
}.
However, based on an extrapolation of the existing
results, we conjecture that, similar to the $d=5$ case, 
this limiting  solution coincides with the extremal MP black hole,
forming a naked singularity.

A further argument that this is indeed the case is suggested by 
the results in \cite{Kleihaus:2009wh}.
There the $d=7$ static solutions with $S^2\times S^3$
horizon topology have been constructed in a systematic way
(although in a different coordinate system, see also \cite{Kleihaus:2010pr}).
The results there indicate that as $R/r_H\to 1$,
the black ringoids  approach  a solution with $S^5$
horizon topology ($i.e.$ the Schwarzschild-Tangerlini 
black hole). 
Thus it is natural to expect that this continues to hold when
these configurations are spinning.

Moreover, this is also suggested by the 
plots of the thermodynamical potentials in Figure 15, 
which resemble again the $d=5$ case.
Also
the ratios 
$L_e/L_p$
 and 
$R_3^{(in)}/R_3^{(out)}$, 
which encode the deformation of the horizon (see Figure 16),
follow the pattern of Figure 7,
although our last reliable numerical results 
are still far away from the putative limiting 
solution with $L_p\to \infty$,  $R_1^{(in)}\to 0$.

\subsubsection{Further cases and the conjectured picture}
Let us start with the case of BRs in $d=7$.
So far,  we have constructed only solutions
on the branch of thin BRs.
(We mention that
our approach here was similar to the
$d=6$ case. In particular, the generic solutions found
were again singular, 
with singularities 
stronger than the conical ones.) 
However, the recent results in \cite{Dias:2014cia}
(see also \cite{Armas:2014bia})
clearly show that the picture found for the nonperturbative region in $d=6$
holds also in $d=7$.
Again there exist
two branches of solutions;
the branch of `fat' black rings 
connects via a topology-changing merger solution
with a branch of `pinched' black hole solutions.
In particular, the diagrams for the reduced physical properties
look very similar to those in Figure 9.
We   expect
that this picture remains valid for BR solutions in $d>7$
dimensions as well.
 
Returning to the case of black ringoids ($i.e.$ $k>0$), we
mention that we have managed to construct a number of 
solutions for  $d=8$ and $d=9$, possessing $S^3\times S^3$
and $S^2\times S^5$ horizon topologies, respectively.
However, although we could confirm their existence,
we have not yet managed to study their properties in a systematic way.
All configurations we could obtain so far
are well described by the blackfold results.
But the numerical accuracy decreases and the 
solutions are lost well before approaching the respective branches of `fat'  black ringoids.
We believe that a refinement of
the numerical scheme is required in order to succeed in obtaining those branches.

Let us mention that the properties of the generic (unbalanced) solutions
are different for $d=8$ as compared to $d=9$.
The numerical construction of the $d=8$ black ringoids with $S^3\times S^3$
horizon topology
is similar to the one of the $d>5$ BRs.
In particular, all generic configurations appear to be singular and
a `shooting' procedure is required in order to construct regular solutions
(analogous to the $d>5$ BRs discussed above).
Also, since there is no upper bound on the angular momenta 
for the
corresponding $n=2,~k=1$  MP black holes,
we expect the existence of new branches of axisymmetric 
 `pinched'  black holes with
 a spherical horizon topology.
 These black holes would branch  off from the MP solutions along the stationary axisymmetric
 zero-mode perturbation of the Gregory-Laflamme-like instability.
As suggested by the  $k=0$ pattern observed for $d=(6,7)$ BRs,
we expect
the  `pinched'  black holes
to connect
these  $d=8$ black ringoids with the MP black holes.

By contrast, the generic $d=9$
configurations with $S^2\times S^5$
horizon topology 
 are supported against collapse by conical singularities, 
 their (well-defined) static limit being considered in 
 \cite{Kleihaus:2009wh},
 in a more general context. 
Similar to the $d=7$ black ringoids discussed above,
the balanced configurations are found by finetuning
the value of the event horizon velocity $\Omega_H$ 
for given input parameters $(r_H,R)$.
Since the angular momentum of the corresponding $n=1,~k=2$ MP solutions
is bounded from above, 
no branches of `pinched'  black holes are expected to exist in this case.
Therefore we conjecture the picture found for 
$d=5$ black objects to be valid also for these configurations,
the $n=1,~k=2$ black ringoids and MP black holes
meeting in a nakedly singular configuration given by (\ref{extr1}). 

\section{Conclusions and further remarks}

The last decade has witnessed a tremendous progress
in the physics of black objects with a non-spherical
horizon topology.
These developments,
originating in the discovery by Emparan and Reall of the $d=5$
BR,
have revealed the existence of a zoo of higher dimensional
solutions with various topologies of the event horizon.
However, while the ultraspinning regime of some of these objects
is very well  described by the corresponding
blackfolds, the behaviour of the solutions
in other regions of the parameter space is relatively
poor understood.
In particular, only little is known about their limiting
behaviour and topology-changing mergers  
with branches of `pinched' black holes, emerging from
the respective known MP black hole solutions.

In the absence of analytical solutions, one possible approach to
make progress in this direction is to solve the Einstein equations numerically.
In this work, 
by using a special coordinate system,
we have been able to formulate the problem of (a class of) 
$d\geq 5$ balanced black objects
with $S^{n+1}\times S^{2k+1}$
event horizon topology
in a numerically manageable manner.
Thus such solutions can be constructed numerically,
by solving a set of  elliptic
partial differential equations
(with a dependence on two variables only),
with suitable boundary conditions.
These boundary conditions enforce the
topology of the horizon and the asymptotic structure of the spacetime. 
 
A number of  nonperturbative solutions have been constructed in this way.  
Our results for $d=6$ BRs
have confirmed the conjecture of \cite{Emparan:2007wm}
for the phase diagram of such objects.
In particular, the limiting behaviour of the `fat' black ring solutions is very different
as compared to the $d=5$ case,
since they are connected to the MP black hole via a set of 
`pinched' black holes.
We have found a different picture for
$d=7$ black ringoids with two equal angular momenta.
Although we could not fully explore this case,
a number of features (observed in the phase diagrams
for the physical properties and the thermodynamics) strongly suggest that
these configurations share the properties of the 
$d=5$ BRs.
In particular, they are likely to meet the
corresponding MP black holes in
a limiting (nakedly singular) extremal configuration.

Another interesting new result is the
occurrence of a cusp in the $a_H(j)$  black ring(oid) diagram,
where a branch of `fat' black ring(oid) solutions emerges.
Since this is the case for all known solutions,
we expect this to hold generically.
Thus nonuniqueness with respect to the global charges
also appears to be a generic property of these solutions.

As possible avenues for future research,
let us mention that 
the general ansatz in Section 3 as well as the specific solutions
in Section 5 can be extended to study black objects possessing an $S^{n+1}\times S^{2k+1}$ horizon topology, 
with rotation not only on the $S^{2k+1}$ but also on the $S^{n+1}$.
However, in this case, $n$ should be an odd number,
 $n=2p+1$, and  the angular momenta in that sector should all be equal as well
 (although their magnitude could differ
 from the one of the angular momenta on the  $S^{2k+1}$).
 This extension would involve more functions in
the metric ansatz, but no new conceptual difficulty should arise,
the problem remaining of codimension-two.
Moreover, our methods should readily extend to more general situations, 
$e.g.$ to the presence of matter fields
\cite{Kleihaus:2010hd},
\cite{Kleihaus:2013zpa},
 or for anti-de Sitter asymptotics of the spacetime background.
The latter case is of special interest, given the absence 
of exact solutions even for $d=5$
(see, however,
\cite{Caldarelli:2008pz},
\cite{Armas:2010hz}
 for results found within the blackfold approach). 
Also, it should be interesting to consider multi-black hole 
configurations and to extend 
the phase diagram proposed in \cite{Emparan:2010sx}
to the fully nonperturbative regime.

Finally, let us speculate about another possible implication of the results in this work.
Hopefully, the activity in the area of higher dimensional black objects will result in
an encyclopedia of solutions,
with a well-understood phase space.
Then we consider it very likely that, at some stage, some  general structures
would be revealed, leading to a description
of the solutions
in terms of 
a number of (relatively) simple patterns.
Let us exemplify this with the case of the particular classes of MP black holes
and the $S^{n+1}\times S^{2k+1}$ black ring(oid)s
in this work. 
We expect that the well-known picture found for $d=5,~k=0$
is generic for all higher dimensional solutions
with $n=1$, $i.e.$
for black ringoids with $S^{2}\times S^{d-4}$
event horizon topology and the corresponding MP black holes
with $k+1$ equal angular momenta
in $d=2k+5$ dimensions.
Thus we conjecture that
the diagrams in Figures 5-7 should always apply for these $n=1$ solutions,
with the branch of `fat' black ringoids meeting the MP solutions
in the  zero-area configuration (\ref{extr1}).
At the same time, we expect the 
properties of the known $d=6,~7$
BRs (in conjunction with the corresponding MP and `pinched' black holes)
to provide the pattern for all balanced 
black objects with  $S^{n+1}\times S^{2k+1}$ horizon topology and with $n>1$.
Moreover,
it is likely that more complicated black objects will exhibit a generic behaviour as well
($e.g.$ the $d=5$
doubly spinning BRs \cite{Pomeransky:2006bd} would provide the pattern for 
$S^{2}\times S^{2k+1}$ black ringoids with un-equal
angular momenta, possibly spinning also on the $S^2$).

Such a classification should result at some stage in 
{\it a periodic table} of higher dimensional black objects,
 organized on the basis of a small number of characteristic features.

 \vspace*{0.7cm}
\noindent{\textbf{~~~Acknowledgements.--~}}  
We would like to thank C.~Herdeiro,  N.~Obers and  J.~Rocha for helpful
discussions at various stages of this project.
We acknowledge support by the DFG Research Training Group 1620 Models
of Gravity.
E.R.~acknowledges support from the FCT-IF programme.

\appendix
\setcounter{equation}{0}

\section{The approximate form of the solutions
on the boundaries}
In our work,
we have found it useful to construct
an  approximate form of the solutions
 on the boundaries of the domain of integration.
There we suppose the
existence of a power series expansion at the horizon/infinity and
also at $\theta=0,\pi/2$.
Also, we assume
that the metric functions $f_i$ ($i=1,4$)
preserve the behaviour  on the boundaries of the background metric.
Here we recall that for black objects with an
$S^{n+1}\times S^{2k+1}$
horizon topology, the relevant part of the background metric is
given by (\ref{new-form}).
At the same time,
the obvious background for black holes with a spherical
horizon topology is (\ref{Mink}).

\subsection{The horizon}
%
As $r\to r_H$, it is natural to suppose that the non-extremal solutions (which are 
the only type constructed in this work), 
possess a power series expansion of the form  
(here we assume a metric gauge with $\Delta(r)=r^2$;
also, the relations below hold as well for a spherical topology of the horizon):
\begin{eqnarray}
\label{rh}
f_i(r,\theta)=\sum_{j\geq 0}f_{ij}(\theta)(r-r_H)^j,
~~W(r,\theta)=\sum_{j\geq 1}w_{j}(\theta)(r-r_H)^j,
\end{eqnarray}
with nonvanishing $f_{10},f_{20},f_{30},f_{40}$ and 
 a double zero for $f_0$,
 \begin{eqnarray}
 \label{rh-sup}
f_0(r,\theta)=f_{02}(r-r_H)^2+\sum_{j\geq 3}f_{ij}(\theta)(r-r_H)^j.
\end{eqnarray}
Then the Einstein equations are solved order by order in $\epsilon=(r-r_H)$,
which lead to a solution in terms of five functions
\begin{eqnarray}
\label{rh1}
\{
f_{02}(\theta),~
f_{20}(\theta),~
f_{30}(\theta),~
f_{40}(\theta),~
w_{2}(\theta)
 \},
\end{eqnarray}
and two constants
\begin{eqnarray}
\label{rh2}
\{
c_1=\frac{f_{10}(\theta)}{f_{02}(\theta)}, ~~~w_0=\Omega_H
 \},
\end{eqnarray}
which  
fix the Hawking temperature and 
the event horizon velocity of the black objects.

 We have verified that, at least up to order six, all other functions in (\ref{rh})
 vanish or are fixed by (\ref{rh1}), (\ref{rh2}).
 For example, one finds
 \begin{eqnarray}
\label{rh3}
  f_{21}= f_{31}= f_{41}= w_{1}=0,~~
f_{03}=-\frac{f_{02}}{r_H},~f_{11}=-\frac{2f_{10}}{r_H},~ w_{3}=-\frac{w_{2}}{r_H},
\end{eqnarray}
while 
 \begin{eqnarray}
\label{rh4}
&&
f_{04}= \frac{1}{12}
\bigg(
\frac{11f_{02}}{r_H^2}
-\frac{4f_{02}f_{32}}{f_{30}}
-\frac{8kf_{02}f_{42}}{f_{40}}
+8 f_{30}w_2^2+\frac{{\dot f}_{02}^2}{r_H^2f_{02}}
\\
\nonumber
&&{~~~~~~~~~~~~~~~~~}
-\frac{(d-2k-4)}{f_{20}}(4f_{02}f_{22}+\frac{{\dot f}_{02}{\dot f}_{20}}{r_H^2})
-\frac{{\dot f}_{02}{\dot f}_{30}}{r_H^2f_{30}}
-\frac{2k{\dot f}_{02}{\dot f}_{40}}{r_H^2f_{40}}
-\frac{2{\ddot f}_{02} }{r_H^2}
\bigg).
\end{eqnarray}
In a numerical approach, the functions (\ref{rh1})
together with the constant $c_1$ are read from the numerical output.

%
For completeness, we give also the approximate form of the metric
close to the horizon 
\begin{eqnarray}
\label{qsa}
  &&
 ds^2= f_{02}(\theta ) \big( c_1(dr^2+r_H^2  d\theta^2)-(r-r_H)^2 dt^2 \big) 
 + f_{20}(\theta)d\Omega_{n}^2
  \\
  \nonumber
  &&{~~~~~~~~~~~~~~~~~~~~~~~~~~~~}
  +f_{30}(\theta) \big(d\psi+{\cal A}-\Omega_H dt\big)^2+f_{40}(\theta) d\Sigma_{k}^2~.
\end{eqnarray}

\subsection{The $\theta=0$ boundary}
The situation is more complicated in this case, since 
  the black holes with $S^{n+1}\times S^{2k+1}$
horizon topology,
possess a different expansion  for $r_H\leq r<R$ and for $r>R$, respectively.
(This feature can be understood from the study in Section 2 of 
background functions $F_i$.)

For $r_H\leq r<R$ and $\theta\to 0$, the solutions with a non-spherical horizon topology  
can be written as
  \begin{eqnarray}
\label{t01}
f_{i}(r,\theta)=\sum_{j\geq 0}f_{ij}(r)\theta^j,~~
W(r,\theta)=\sum_{j\geq 0}w_{j}(r)\theta^j,
 \end{eqnarray}
 with 
\begin{eqnarray}
\label{t02}
f_{2}(r,\theta)=f_{22}(r)\theta^2+\sum_{j\geq 3}f_{2j}(r)\theta^j.
\end{eqnarray}
The essential functions in this expansion are
\begin{eqnarray}
\label{t03}
\{
f_{00}(r),~
f_{10}(r),~
f_{30}(r),~
f_{40}(r),~
w_{0}(r)
 \},
\end{eqnarray}
all other functions in (\ref{t01})
 vanishing or being fixed by those in (\ref{t03}).
 One finds $e.g.$
\begin{eqnarray}
\label{t04}
f_{01}=f_{11}=f_{31}=f_{41}=f_{23}=w_1=0,~~f_{22}=r^2f_{10},
\end{eqnarray}
and
\begin{eqnarray}
\label{t05}
&&
f_{42}=\frac{r^2}{4(d-2k-3)}
\bigg(
8f_{10}(1+k-\frac{f_{30}}{f_{40}})
-(
\frac{2(d-2k-3)}{r}+\frac{f_{00}'}{f_{00}}
\\
&&
\nonumber
{~~~~~~~~~~~~~~~~~~~~~~}
+\frac{(d-2k-4)f_{10}'}{f_{10}}
+\frac{f_{20}'}{f_{30}}
)f_{40}'
-\frac{2(k-1)}{f_{40}}f_{40}'^2
-2f_{40}''
\bigg).
\end{eqnarray}
Therefore for $r_H<r<R$,  
the approximate form of the line element close to $\theta=0$
reads
\begin{eqnarray}
\label{t06}
  &&
 ds^2=f_{10}(r ) \left(dr^2+r^2 (d\theta^2 +\theta^2 d\Omega_{n}^2)\right) -f_{00}(r) dt^2
  \\
  \nonumber
  &&{~~~~~~~~~~~~~~~~~~~~~~~~~~~~}
  +f_{30}(r) \big(d\psi+{\cal A}-w_0(r) dt\big)^2+f_{40}(r) d\Sigma_{k}^2~.
\end{eqnarray}
%
%
An expansion similar to (\ref{t01}) holds also for $r>R$,
with
\begin{eqnarray}
\label{t07}
f_{3}(r,\theta)=f_{32}(r)\theta^2+\sum_{j\geq 3}f_{3j}(r)\theta^j,~~
f_{4}(r,\theta)=f_{42}(r)\theta^2+\sum_{j\geq 3}f_{4j}(r)\theta^j,
\end{eqnarray}
and (with $i=1,2)$
\begin{eqnarray}
\label{t07-bis}
f_{i}(r,\theta)=\sum_{j\geq 0}f_{ij}(r)\theta^j,~~ 
W(r,\theta)=\sum_{j\geq 0}w_{j}(r)\theta^j,
\end{eqnarray}
in this case.
The essential functions here are
\begin{eqnarray}
\label{t08}
\{
f_{00}(r),~
f_{10}(r),~
f_{20}(r),~
w_{0}(r)
 \},
\end{eqnarray}
while, $e.g.$
\begin{eqnarray}
\label{t09}
f_{01}=f_{11}=f_{21}=f_{33}=f_{43}=w_1=0,~~f_{32}=f_{42}=r^2f_{10},
\end{eqnarray}
such that
the approximate form of the line element close to $\theta=0$ and $ r>R$ 
reads
\begin{eqnarray}
\label{t06n}
\nonumber
  &&
 ds^2=f_{10}(r ) \bigg(dr^2+r^2 \big [ d\theta^2 +\theta^2 \big ( (d\psi+{\cal A}-w_0(r) dt)^2+ d\Sigma_{k}^2 \big )  \big] \bigg)
 +f_{20}(r ) d\Omega_{n}^2 -f_{00}(r) dt^2~.
 \end{eqnarray}
 For  black holes with a spherical horizon topology, one can formally take $R=r_H$,
  such that the relations (\ref{t07})-(\ref{t06n})
  hold  in that case for any $r>r_H$.

\subsection{The $\theta=\pi/2$ boundary}
The corresponding expansion as $\theta \to \pi/2$ reads
\begin{eqnarray}
\label{t01n}
f_{i}(r,\theta)=\sum_{j\geq 0}f_{ij}(r)(\theta-\frac{\pi}{2})^j,~~
W(r,\theta)=\sum_{j\geq 0}w_{j}(r)(\theta-\frac{\pi}{2})^j,
 \end{eqnarray}
 with 
\begin{eqnarray}
\label{t02q}
f_{2}(r,\theta)=f_{22}(r)(\theta-\frac{\pi}{2})^2+\sum_{j\geq 3}f_{2j}(r)(\theta-\frac{\pi}{2})^j,
\end{eqnarray}
the essential functions in this expansion being
\begin{eqnarray}
\label{t03q}
\{
f_{00}(r),~
f_{10}(r),~
f_{30}(r),~
f_{40}(r),~
w_{0}(r)
 \},
\end{eqnarray}
all other functions in (\ref{t01})
 vanishing or being fixed by (\ref{t03q}).
 Note that the relations (\ref{t04}), (\ref{t05}) still hold in this case
 (although the corresponding expressions of the functions are different, 
 of course). 
Then the approximate form of the line element close to $\theta=\pi/2$
reads
\begin{eqnarray}
\label{t06p}
  &&
 ds^2=f_{10}(r ) \left(dr^2+r^2 (d\theta^2 +(\theta-\frac{\pi}{2})^2 d\Omega_{n}^2)\right) -f_{00}(r) dt^2
  \\
  \nonumber
  &&{~~~~~~~~~~~~~~~~~~~~~~~~~~~~}
  +f_{30}(r) \big(d\psi+{\cal A}-w_0(r) dt\big)^2+f_{40}(r) d\Sigma_{k}^2~.
\end{eqnarray}
The above relations hold for both a $S^{d-2}$ and 
$S^{n+1}\times S^{2k+1}$ horizon topology.

\subsection{The expansion as $r\to \infty$}
 Finally, the solutions admit a $1/r$ expansion as $r\to \infty$, with
\begin{eqnarray}
\label{t01n2}
&&
f_{1}(r,\theta)=1+\sum_{j\geq 2} \frac{f_{1j}(\theta)}{r^j},~~
f_{2}(r,\theta)=r^2 \cos^2\theta(1+\sum_{j\geq 2} \frac{f_{2j}(\theta)}{r^j}),
\\
\nonumber
&&
f_{3}(r,\theta)=r^2 \sin^2\theta(1+\sum_{j\geq 2} \frac{f_{3j}(\theta)}{r^j}),~~
f_{4}(r,\theta)=r^2 \sin^2\theta(1+\sum_{j\geq 2} \frac{f_{4j}(\theta)}{r^j}),~~
\\
\nonumber
&&
f_{0}(r,\theta)=1+\sum_{j\geq d-3} \frac{f_{0j}(\theta)}{r^j},~~
W(r,\theta)= \sum_{j\geq d-1} \frac{w_{j}(\theta)}{r^j},~~
 \end{eqnarray}
with
\begin{eqnarray}
\label{t01n3}
 f_{0(d-3)}=C_t,~~w_{d-1}=C_{\psi}
\end{eqnarray}
two constants
which fix the mass and angular momentum, respectively.

\subsection{On the regularity of the solutions}

One can easily verify that the MP black holes
and the $d=5$ Emparan-Reall black ring
are regular on and outside the event horizon. 
However, given their numerical character, this is not obvious
for the other solutions
discussed in this work.
In particular, the numerical scheme employed in this work
uses a special coordinate system 
which implies the existence of a 
singularity at 
$(r=R,~\theta=0)$.
This singularity is already present for the $D=4$ 
flat space case discussed in Section 2.
%
However, one can show that  similar to that case,
the point $(r=R,~\theta=0)$ is just a coordinate singularity,
assuming that the metric functions
satisfy the $\theta\to 0$ expansion discussed above.

To prove that, we recall first 
that within our numerical scheme the functions
$f_i$ $(i>0)$ which enter the general line element (\ref{metric})
are taken as
\begin{eqnarray}
f_i=F_i {\cal F}_i,~~i=1,2,3, ~~~~{\rm and}~~~f_4=F_3 {\cal F}_4,
\end{eqnarray}
with $(F_1,~F_2,~F_3)$ given by (\ref{Fi}),
and the  unknown functions ${\cal F}_1,~{\cal F}_2,~{\cal F}_3,~{\cal F}_4$ which are found numerically.

Next, in order to focus on the region around the point $(r=R,~\theta=0)$,
we change to adapted coordinates
\begin{eqnarray}
r\sin \theta=\frac{1}{2R {\cal F}_1(R,0)}\rho^2 \sin 2\Theta,~~
r\cos \theta=R+\frac{1}{2R {\cal F}_1(R,0)}\rho^2 \cos 2\Theta,~~
\end{eqnarray}
such that $(r=R,~\theta=0)$ now lies at $\rho= 0$, and then expand the metric in powers
of $\rho$.
To leading order terms in $\rho^2$, the expression of the line element is
\begin{eqnarray}
\label{new-m}
&&
ds^2=d\rho^2+\rho^2 d\Theta^2
+\frac{{\cal F}_2(R,0)}{{\cal F}_1(R,0)}\rho^2\cos^2\Theta d\Omega_n^2
-f_0(R,0)dt^2
\\
\nonumber
&&
{~~~~~~~~~}
+\frac{{\cal F}_3(R,0)}{{\cal F}_1(R,0)}\rho^2\sin^2\Theta (d\psi+{\cal A}-W(R,0)dt)^2
+\frac{{\cal F}_4(R,0)}{{\cal F}_1(R,0)}\rho^2\sin^2\Theta d\Sigma_k^2.
\end{eqnarray}
However, one can easily see that the approximate solution constructed in Appendix A2
implies\footnote{This follows from the last relations in (\ref{t04}), (\ref{t09})
together with the expression  (\ref{Fi}) of the functions $F_i$. } ${\cal F}_3(R,0)={\cal F}_4(R,0)={\cal F}_1(R,0)$.
As a result, (\ref{new-m})
is just the flat spacetime metric written in the form (\ref{Mink}), (\ref{sphere}).
Thus
we conclude that the point $(r=R,~\theta=0)$
is a coordinate singularity only.


To further investigate the regularity of the numerical solutions,
we have analyzed both analytically and numerically
their Kretschmann scalar\footnote{Note that a finite Kretschmann scalar 
does not exclude the existence of other, more subtle pathologies, see
$e.g. $
\cite{Dias:2011at}
for a recent example in this direction.}
\begin{eqnarray} 
K=R_{\mu\nu\alpha \beta}R^{\mu\nu\alpha \beta}.
\end{eqnarray}
Unfortunately, we have not been able to find a general expression for $K$
valid for any $(d,k)$,
similar to that found for the Einstein tensor.
Therefore we have restricted our analytical study to
a set of $(d,k)$.
However, for all considered cases, we have found that the Kretschmann scalar 
is finite on the boundary of the domain of integration,
for generic solutions possessing the approximate form discussed above\footnote{Here we assume that
the various functions $f_{ij}$, $W_i$
which enter the approximate solution on the boundaries are finite, together with their first and second derivatives.}.

We have also investigated the expression of the Kretschmann scalar,
as resulting from the numerical integration of the Einstein equations,
for a number of $d=6$
black rings and $d=7$ black ringoids.
As seen in Figures 8, 13, the scalar $K$ is finite everywhere
and approaches its maximal value on the horizon at $\theta=\pi/2$
(a similar behaviour can be noticed for the considered
$d=5$ Emparan-Reall black ring
and also for the $d=7$ MP black hole, see Figures 4, 3).
One can see also the absence of any special
features at $(r=R$, $\theta=0)$,
the Kretschmann scalar being always finite there.

\section{Five-dimensional black rings }
\subsection{The solution}

 The expression of the metric functions 
 which enter the  line element (\ref{ER1}) reads\footnote{A slightly more complicated form of these 
 functions has been given in \cite{Kleihaus:2010pr}.}
\begin{eqnarray}
\nonumber
&&
f_0 (r,\theta)=\frac{Q_2(r,\theta)}{Q_1(r,\theta)}U_1(r,\theta)U_2(r,\theta),
~~~
f_1(r,\theta)=\frac{r_H^2R^4}{(R^4-r_H^4)^2 }\frac{U_1(r,\theta)Q_3(r,\theta)}{S(r,\theta)},
\\
&&
\label{ER4}
f_2(r,\theta)=\left(1+\frac{r_H^2}{r^2} \right)^2\frac{r^2\sin^2 2\theta}{2U_2(r,\theta)},
~~f_3(r,\theta)=\frac{r^2\left(1-\frac{r_H^2}{r^2}\right)^2}{2\left(1+\frac{r_H^2}{r^2} \right)^2}\frac{Q_1(r,\theta)}{Q_2(r,\theta)U_1(r,\theta)},
\\
&&
\nonumber
W(r,\theta)=\frac{4\sqrt{2}(r_H^2+R^2)\sqrt{R^4+r_H^4}}{R(R^2-r_H^2)}
\frac{\left(1-\frac{r_H^2}{r^2} \right)^2}{r^2\left(1+\frac{r_H^2}{r^2} \right)^2}
\frac{Q_2(r,\theta)Q_4(r,\theta)}{Q_1(r,\theta)}.
\end{eqnarray}
In the above relations, we have defined a number of auxiliary functions
\begin{eqnarray}
\label{ER2}
&&
U_1(r,\theta)=\frac{(r_H^2+R^4)}{r_H^2R^2}
\left(
1+\frac{4r_H^2}{r^2}+\frac{r_H^4}{r^4}
\right) 
+\frac{4r_H^2}{r^2}\cos2\theta-2S(r,\theta),
\\
&&
\nonumber
U_2(r,\theta)=\frac{r_H^2+R^4}{r^2R^2}
-(1+\frac{r_H^4}{r^4})\cos 2\theta+S(r,\theta),
\end{eqnarray}
together with
\begin{eqnarray}
\nonumber
&&
Q_1(r,\theta)=U_1^2(r,\theta)U_2(r,\theta)
-\frac{4(r_H^2+R^2)^2(r_H^4+R^4)}{r^2R^2(R^2-r_H^2)^2}\frac{\left(1+\frac{r_H^2}{r^2} \right)^2}{\left(1-\frac{r_H^2}{r^2}\right)^2}
\times
\\
\nonumber
&&
\label{ER23}
{~~~~~~~~~~~~~~~}
\bigg[
U_1(r,\theta)-
\bigg(
\frac{(r_H^2-R^2)^2}{r_H^2R^2}
+\frac{r_H^2(r_H^2-R^2)^2}{r^4R^2}
+\frac{2(r_H^2+R^2)^2}{r^2 R^2}
+\frac{4(r_H^4+R^4)}{r^2R^2}
\bigg)
\bigg]^2,
\\
\nonumber
&&
Q_2(r,\theta)=U_1(r,\theta)-\frac{8(r_H^4+R^4)}{r^2R^2},
\\
\nonumber
&&
Q_3(r,\theta)=-U_1(r,\theta)+(1+\frac{r_H^2}{r^2})^2\frac{2(r_H^4+R^4)}{r_H^2R^2},
\\
\nonumber
&&
Q_4(r,\theta)=U_2(r,\theta)-2(1-\frac{r_H^2}{r^2})^2\sin^2\theta,
\end{eqnarray}
where
\begin{eqnarray}
\label{ER3}
S(r,\theta)=\sqrt{
\bigg(
1+\frac{R^4}{r^4}
-\frac{2R^2}{r^2}\cos 2\theta
\bigg)
\bigg(
1+\frac{r_H^8}{r^4R^4}
-\frac{2r_H^4}{r^2R^2}\cos 2\theta
\bigg)
}.
\end{eqnarray}

\subsection{The expansion
of the metric functions on the boundaries}

 To make contact with the generic expressions in Appendix A,
it is useful to give the form of these metric functions on the boundaries of the domain of integration.
Thus, for $r\to \infty$ one finds
\begin{eqnarray}
\label{ER5}
&&
f_0(r,\theta)=1-\frac{8r_H^2(R^4+r_H^4)}{(R^2-r_H^2)^2}\frac{1}{r^2}+O(1/r^4),
\\
\nonumber
&&
f_1(r,\theta)=1+
\frac{1}{R^2(R^2-r_H^2)^2}\bigg(
4R^2r_H^2(R^4+r_H^4)+
(
(R^4+r_H^4)^2
\\
\nonumber
&&{~~~~~~~~~~~~~~~~~~~~~~~~~~~~~~~~~~~~}
-2R^2r_H^2(R^2-r_H^2)^2
+4R^4r_H^4
)
\cos 2\theta
\bigg)
\frac{1}{r^2}
+O(1/r^4),
\\
\nonumber
&&
f_2(r,\theta)=r^2 \cos^2\theta 
\bigg(1-
\frac{(R^2-r_H^2)^2-2R^2r_H^2}{R^2}
\bigg)
+O(1/r^2),
\\
\nonumber
&&
f_3(r,\theta)=r^2 \sin^2\theta 
\bigg(1+
\frac{(R^4+r_H^4)^2+2R^2r_H^2 \big((R^2+r_H^2)^2+2R^2r_H^2 \big)}{R^2(R^2-r_H^2)^2}
\bigg)
+O(1/r^2),
\\
\nonumber
&&
W(r,\theta)=
\frac{4\sqrt{2}r_H^2(R^2+r_H^2)^3 \sqrt{R^4+r_H^4}}{R(R^2-r_H^2)^3}\frac{1}{r^4}
+O(1/r^6).
\end{eqnarray}
The corresponding expression close to  the horizon, $r=r_H$ is 
\begin{eqnarray}
\label{ER6}
&&
f_0(r,\theta)=\frac{(R^2-r_H^2)^2(R^4+r_H^4+2R^2r_H^2\cos 2\theta)}{2r_H^2(R^2+r_H^2)^2 (R^4+r_H^4)}(r-r_H)^2+O(r-r_H)^4,
\\
\nonumber
&&
f_1(r,\theta)=\frac{16R^2r_H^2(R^4+r_H^4+2R^2r_H^2\cos 2\theta)}{(R^4-r_H^4)^2}+O(r-r_H),
\\
\nonumber
&&
f_2(r,\theta)=\frac{4R^2r_H^4\sin^2 2\theta)}{R^4+r_H^4-2 R^2r_H^2\cos 2\theta}+O(r-r_H)^2,
\\
\nonumber
&&
f_3(r,\theta)=\frac{2(R^2+r_H^2)(R^4+r_H^4)}{R^2(R^2-r_H^2)^2}\frac{(R^4+r_H^4-2 R^2r_H^2\cos 2\theta)}{(R^4+r_H^4+2 R^2r_H^2\cos 2\theta)}+O(r-r_H)^2,
\\
\nonumber
&&
W(r,\theta)=\frac{R(R^2-r_H^2)}{\sqrt{2}(R^2+r_H^2)\sqrt{R^4+r_H^4}}+O(r-r_H)^2.
 \end{eqnarray}
The expansion at $\theta=0$ for $r_H\leq r<R$ reads:
\begin{eqnarray}
\nonumber
&&
f_0(r,\theta)=\frac{(r^2-r_H^2)^2(R^2-r_H^2)^2}{r^4(R^2-r_H^2)^2
+r_H^4(R^2-r_H^2)^2+2r^2r_H^2(3R^4+2R^2r_H^2+3 r_H^4)}
+O(\theta)^2,
\\
\label{ER7}
&&
f_1(r,\theta)=\frac{(r^2+r_H^2)^4}{r^8\left(\frac{R^4+r_H^4}{r^2R^2})-\frac{r_H^4}{r^4}-1 \right)}
+O(\theta)^2,
\end{eqnarray}
\begin{eqnarray}
\nonumber
&&
f_2(r,\theta)=\frac{R^2(r^2+r_H^2)^4 }{r^4(R^4+r_H^4)-r^2R^2r_H^4-r^6R^2 }\theta^2
+O(\theta)^4,
\\
\nonumber
&&
f_3(r,\theta)=
\bigg(r^2(R^4+r_H^4)-R^2r_H^4-r^4R^2 \bigg)
\bigg(r^4(R^2-r_H^2)^2+r_H^4(R^2-r_H^2)^2
\\
\nonumber
&&
{~~~~~~~~~~~~~~~~~~}
+2r^2r_H^2 (3R^4+2R^2r_H^2+3r_H^4)
\bigg)
\frac{1}{r^2R^2(r^2+r_H^2)^2(R^2-r_H^2)^2}
+O(\theta)^2,
\\
\nonumber
&&
W(r,\theta)=\frac{4\sqrt{2}r^2 R r_H^2(R^2-r_H^2)\sqrt{R^4+r_H^4}}
{(R^2+r_H^2)
\bigg(
r^4(R^2-r_H^2)^2+r_H^4(R^2-r_H^2)^2+3r^2r_H^2(3R^4+2R^2r_H^2+3r_H^4)
\bigg)
}
+O(\theta)^2.
\end{eqnarray}
A different expansion holds at $\theta=0$ for $r\geq R$:
\begin{eqnarray}
\label{ER8}
&&
f_0(r,\theta)=\frac{(r^2-r_H^2)^2(R^2-r_H^2)^2 }
{r^4(R^2-r_H^2)^2
+r_H^4(R^2-r_H^2)^2
+2r^2r_H^2(3R^4+2R^2r_H^2+3r_H^4)
}
+O(\theta)^2,
\\
\nonumber
&&
f_1(r,\theta)=
\frac{R^2(r^2-r_H^2)^2
\big(
(r^4+r_H^4)(R^2-r_H^2)^2+2r^2r_H^2(3R^4+2R^2r_H^2+3r_H^4)
\big)
}
{r^4(R^2-r_H^2)^2(R^2(r^4+r_H^4)-r^2(R^4+r_H^4))}
+O(\theta)^2,
\\
\nonumber
&&
f_2(r,\theta)=
\frac{(r^2+r_H^2)^2
\big(
R^2(r^4+r_H^4)
-r^2(R^4+r_H^4)
\big)
}{r^2R^2(r^2-r_H^2)^2}
+O(\theta)^2,
\\
\nonumber
&&
f_3(r,\theta)=\frac{R^2(r^2-r_H^2)^2( (R^2-r_H^2)^2(r^4+r_H^4)+2r^2r_H^2(3R^4+2R^2r_H^2+3r_H^4))\theta^2}
{r^2(R^2-r_H^2)^2(R^2(r^4+ r_H^4)-r^2(R^4+r_H^4))}+O(\theta)^4,
\\
\nonumber
&&
W(r,\theta)=\frac{4\sqrt{2}R (R^2-r_H^2)(R^2+r_H^2)r^2r_H^2(r^2+r_H^2)^2\sqrt{R^4+r_H^4}}
{r^4(R^2-r_H^2)^2+r_H^4(R^2-r_H^2)^2+r^2(6R^4r_H^2+4R^2r_H^4+6r_H^6))^2}+O(\theta)^2.
\end{eqnarray}
The expansion   at $\theta=\pi/2$ which holds for any values of $r$ is
\begin{eqnarray}
\nonumber
&&
f_0(r,\theta)=\frac{ 
 r^4(R^2-r_H^2)^2+r_H^4(R^2-r_H^2)^2-2r^2r_H^2(3R^4+2R^2r_H^2+4r_H^4) 
}
{ 
-\frac{32r^6(R^2+r_H^2)^6(R^4+r_H^4)}{(r^2-r_H^2)^2}
+\frac{(R^2-r_H^2)^6(r^2+r_H^2)^2(r^2R^2+R^2r_H^4+r^2(R^4+r_H^4))}{r_H^4}
}~~~~~~~~~~~~~~~{~~~~~~~~~~~~~~}
\\
\nonumber
&&
{~~~~~~~~~~~~~~~~~~~} \times 
\frac{r^4 R^2(R^2-r_H^2)^4(1+\frac{r_H^4}{r^4}+\frac{R^4+r_H^4} {r^2R^2})}{r_H^4} 
+O(\theta-\frac{\pi}{2})^2,
\\
\label{ERpi2}
&&
f_1(r,\theta)=\frac{(r^2+r_H^2)^4}{r^8(1+\frac{r_H^4}{r^4}+\frac{R^4+r_H^4}{r^2R^2})}+O(\theta-\frac{\pi}{2})^2,
\nonumber
\\
&&
f_2(r,\theta)=
\frac{R^2(r^2+r_H^2)^4}
{r^2(r^2+R^2)(r^2R^2+r_H^4)}(\theta-\frac{\pi}{2})^2+O(\theta-\frac{\pi}{2})^4,
\end{eqnarray}

\begin{eqnarray}
\nonumber
&&
f_3(r,\theta)=\frac{1}{r^2R^2(R^2-r_H^2)^4(r^2+r_H^2)^2}
\bigg(
r^8R^2(R^2-r_H^2)^4+R^2r_H^8(R^2-r_H^2)^4+
\\
\nonumber
&&
{~~~~~~~~~~~~}
r^6(R^2-r_H^2)^2(R^2+r_H^2)^4+r^2r_H^4(R^2-r_H^2)^2(R^2+r_H^2)^4
+r^4(6 R^{12} r_H^2+26 R^{10} r_H^4 
\\
\nonumber
&&
{~~~~~~~~~~~~}
+66 R^8r_H^6+60 R^6r_H^{8}
+26 R^2r_H^{12}+6 r_H^{14})
\bigg)+O(\theta-\frac{\pi}{2})^2,
\\
\nonumber
&&
W(r,\theta)=
4\sqrt{2} r^4 R r_H^2 (R^2-r_H^2)(R^2+r_H^2)^3\sqrt{R^4+r_H^4} 
\\
\nonumber
&&
{~~~~~~~~~~~~}\times 
\bigg(
r^8R^2(R^2-r_H^2)^4+
R^2r_H^8(R^2-r_H^2)^4
\\
\nonumber
&&
{~~~~~~~~~~~~~~~}
+
r^6(R^2-r_H^2)^4
+r^2r_H^4(R^2-r_H^2)^2(R^2+r_H^2)^4
+r^4(6 R^{12} r_H^2
+26 R^{10} r_H^4 
\\
\nonumber
&&
{~~~~~~~~~~~~~~~}
 +66 R^8r_H^6
 +60 R^6r_H^{8}
+26 R^2r_H^{12}
+6 r_H^{14})
\bigg)^{-1}
+O(\theta-\frac{\pi}{2})^2.
\end{eqnarray} 

 \begin{small}

 \end{small}

\end{document}